\documentclass[transmag]{IEEEtran}
\usepackage{latexsym}
\usepackage[noadjust]{cite}
\usepackage{amsmath,amsthm,graphicx,cite,booktabs,amssymb,gensymb, color, xcolor}
\usepackage{fancyhdr,float,perpage,tikz}
\usepackage[linesnumbered,lined,ruled]{algorithm2e}
\usepackage{algorithmic,array,enumerate,rotating}
\usepackage{subfigure,epsfig,dblfloatfix}
\usepackage{multirow}
\usepackage{float,perpage}
\usepackage[affil-it]{authblk}
\usepackage{hyperref}
\hypersetup{
	colorlinks   = true,
	linkcolor    = red,
	citecolor    = green,
	filecolor    = cyan,      
    urlcolor     = magenta,
}

\makeatletter
\renewcommand{\@algocf@capt@plain}{above}

\newcommand{\SD}[1]{\textcolor[rgb]{0.00,0.00,0.00}{#1}}

\makeatother
\def\bsx{{\boldsymbol{x}}}
\def\bpsi{{\boldsymbol{\psi}}}
\def\balpha{{\boldsymbol{\alpha}}}

\title{Quantum mechanics-based signal and image representation: application to denoising}

\author{Sayantan Dutta$^{1,2}$, Adrian Basarab$^{1}$, Bertrand Georgeot$^{2}$, and Denis Kouam\'e$^{1}$

\thanks{$^{1}$Institut de Recherche en Informatique de Toulouse, UMR CNRS 5505, Universit\'e de Toulouse, Toulouse, France}

\thanks{$^{2}$Laboratoire de Physique Th\'eorique, IRSAMC, Universit\'e de Toulouse, CNRS, UPS, Toulouse, France}

\thanks{E-Mails: sayantan.dutta@irit.fr; adrian.basarab@irit.fr; georgeot@irsamc. ups-tlse.fr; denis.kouame@irit.fr .}}

\begin{document}


\IEEEtitleabstractindextext{\begin{abstract}
Decomposition of digital signals and images into other basis or dictionaries than time or space domains is a very common approach in signal and image processing and analysis. Such a decomposition is commonly obtained using fixed transforms (e.g., Fourier or wavelet) or dictionaries learned from example databases or from the signal or image itself. In this work, we investigate in detail a new approach of constructing such a signal or image-dependent bases inspired by quantum mechanics tools, i.e., by considering the signal or image as a potential in the discretized Schroedinger equation. To illustrate the potential of the proposed decomposition, denoising results are reported in the case of Gaussian, Poisson, and speckle noise and compared to the state of the art algorithms based on wavelet shrinkage, total variation regularization or patch-wise sparse coding in learned dictionaries, non-local means image denoising, and graph signal processing.
\end{abstract}
\begin{IEEEkeywords}
adaptive signal and image representation, denoising, quantum mechanics, adaptive transformation.
\end{IEEEkeywords}
}
\maketitle

\section{Introduction}
\label{sec:intro}

In number of applications, processing or analyzing signals and images require the use of other representations than time or space. While the most famous transformation still remains the Fourier transform, other representations have been proposed to overcome the non-localization in time or space of the Fourier basis vectors. The most used time-frequency representations are the short time Fourier and the wavelet transforms \cite{donoho1994ideal,donoho1995wavelet}. Most often (see, e.g., image compression, restoration, reconstruction, denoising or compressed sensing), such transforms are associated with the concept of sparsity, i.e., their ability to concentrate most of the signal or image energy in a few coefficients. To reinforce the sparsity, overcomplete dictionaries have also been explored over the last decades, such as the wavelet frames or more recently patch-based or convolutional dictionaries learned from a set of training signals or images \cite{Aharon06}. The latter has been shown to be of particular interest in image denoising \cite{Elad06}.

In this paper, we investigate a novel signal and image representation, through a dedicated basis extracted from the signal or image itself, using concepts from quantum mechanics. First preliminary results were published in \cite{smith2018adaptive}. Compared to fixed basis such as Fourier, discrete cosinus, wavelets, curvelets, etc, or dictionary learning that generally needs a training database, the proposed approach has the advantage of computing a transform adapted to the signal or image of interest. 

Several attempts of translating quantum principles in image or signal processing applications have been proposed in the literature. One may note the seminal work in \cite{eldar2002quantum}, or, more recently, the interest of quantum mechanics in image segmentation \cite{Gabbouj2013, youssry2015quantum} or in pulse-shaped signal analysis \cite{Laleg2013,LalegKirati16}. Note that a separate domain also exists on designing image processing algorithms adapted to quantum computers, but is of a different purpose \cite{iliyasu2013towards,zhang2013neqr}.  

More related to our work, we note that there was a recent attempt to use quantum mechanics in the same context in \cite{kaisserli2014image,chahid2018new}. Although there are similarities between the two approaches, there are also some important differences. The authors in  \cite{kaisserli2014image,chahid2018new} start from a  continuous mathematical representation of the signal, and the discretization only occurs at the end of the process. The processing of a large image in these papers is done by decomposing it into lines and columns to get 1D signals, while the proposed work is applied block-wisely, which offers a more efficient solution for image denoising given that the correlation between neighbouring pixels is preserved. Additionally, unlike \cite{kaisserli2014image,chahid2018new}, our method fully takes into account the quantum localization phenomenon, a subtle effect due to quantum interference which makes the distribution of the eigenfunctions of the Schroedinger operator strongly dependent on noise, and has important effects on the denoising process. We also use the physics of the problem to identify the optimal domain of applicability of such methods.

The proposed framework reposes on the discrete version of the Schroedinger equation for a quantum particle in a potential. In our case, the potential is represented by the signal samples or the pixel values. The bases used to decompose the signal or the image are directly computed from the signal and image itself and correspond to the wave function representing the stationary solutions of the Schroedinger equation. These wave functions have interesting properties of temporal or spatial localization and of frequency dependence on the value of the potential. In particular, they present higher frequencies for low potential values, thus allowing an original signal or image decomposition.

The proposed method has a certain formal similarity with graph signal processing methods \cite{meyer2014perturbation, Ortega2018Graph, Cheung2018Graph, Pang2017Graph, Shekkizhar2020Efficient}, which use a graph Laplacian constructed from the signal to build an adaptive basis. However, graph signal processing constructs the graph Laplacian to  emphasize the similarities between neighbouring pixel values, while in the proposed method the adaptive basis is solely related to the individual pixel values, resulting into very different adaptive bases with different properties.

\begin{figure*}[h!]
	\centering
	\includegraphics[width=1 \textwidth]{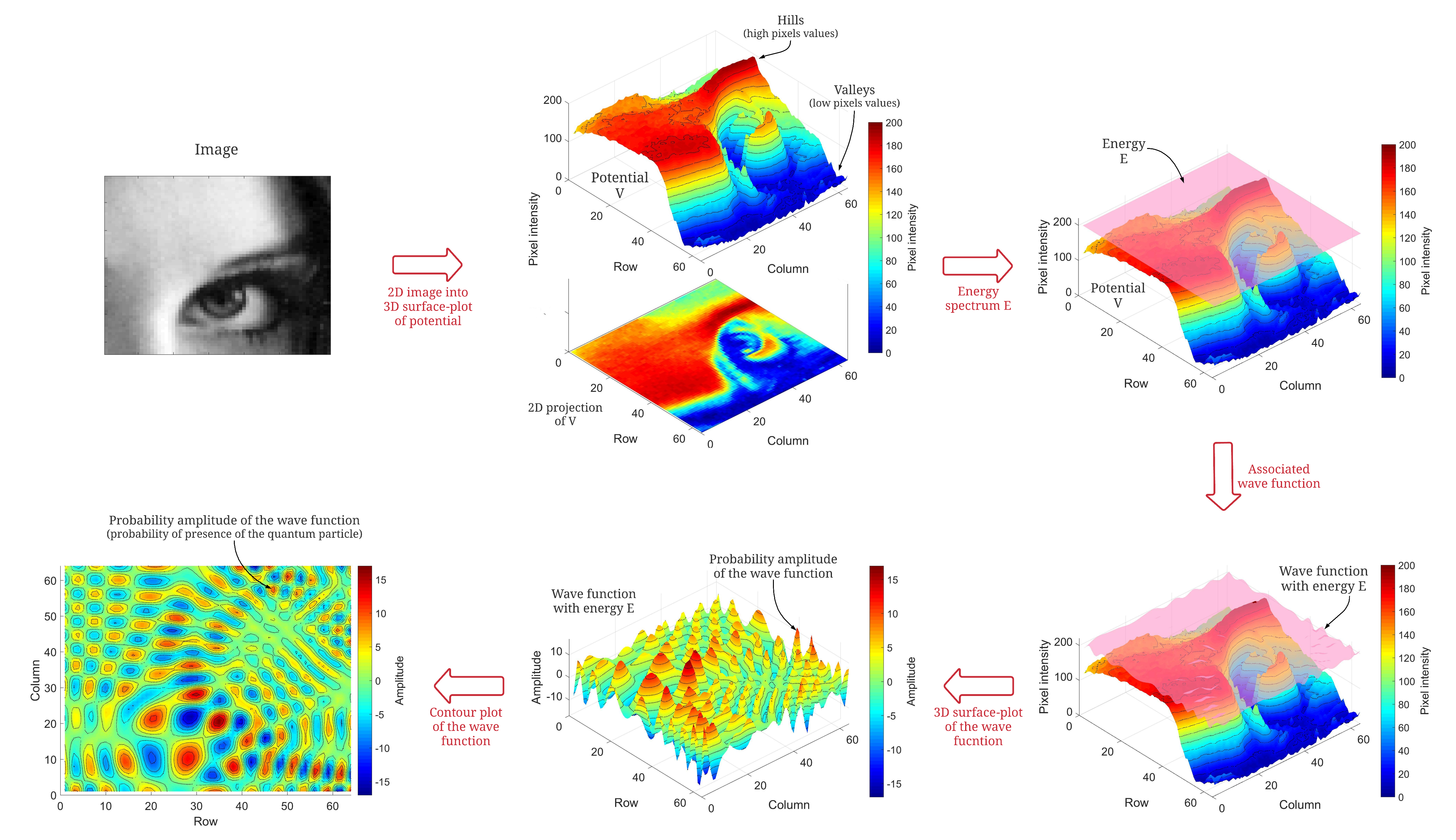}
		
	\caption{Relationship between quantum mechanics and image representation: example on Lena image.}
	\label{fig:relation}
\end{figure*}

Within the proposed framework, the frequency and localization properties of the basis can be controlled through several parameters, thus ensuring flexibility in applications such as denoising. A detailed description of the behavior of the proposed transform and denoising method with respect to the choice of these parameters is provided, allowing to gain insight about the practical consequences in signal and image processing of the quantum mechanical principles involved. Furthermore, the proposed transform embedded in a denoising algorithm shows promising results in different noise scenarios (additive Gaussian, Poisson or speckle noise). 
Finally using different signals and images, comparisons with several state-of-the-art methods are performed.

The remainder of the paper is organized as follows. Section~\ref{sec:adaptive} and \ref{sec:promethdeapp} respectively give the details of the adaptive transform design and its application to denoising. Results and comparisons are provided in Section \ref{sec:results} and concluding remarks are finally reported in Section \ref{sec:conclusion}.

\section{Adaptative basis from quantum mechanics}
\label{sec:adaptive}
\subsection{General framework}
\label{sec:mainidea}
The main idea of the proposed method is to describe a signal or an image onto a specific basis which is constructed through the resolution of a related problem of quantum mechanics: the probability of presence of a quantum particle in a potential related to the signal or image.
While the motivation of using quantum mechanics in this particular context is not straightforward, its main purpose is to produce a basis of oscillating functions with the following properties: 1) the oscillation frequency increases with a parameter of the basis corresponding to the energy, 2) for a given basis function, the oscillation frequency is higher for low values of the signal. The adopted strategy will then be to threshold a noisy signal in energy once expanded in this basis: this will automatically keep higher frequencies for low pixel values, and lower frequencies for high pixel values. Intuitively, one could expect that this method is especially efficient for signal-dependent noise, stronger for high signal values, such as, for instance, Poisson noise.

\begin{figure*}[h!]
	\centering
	\includegraphics[width=1 \textwidth]{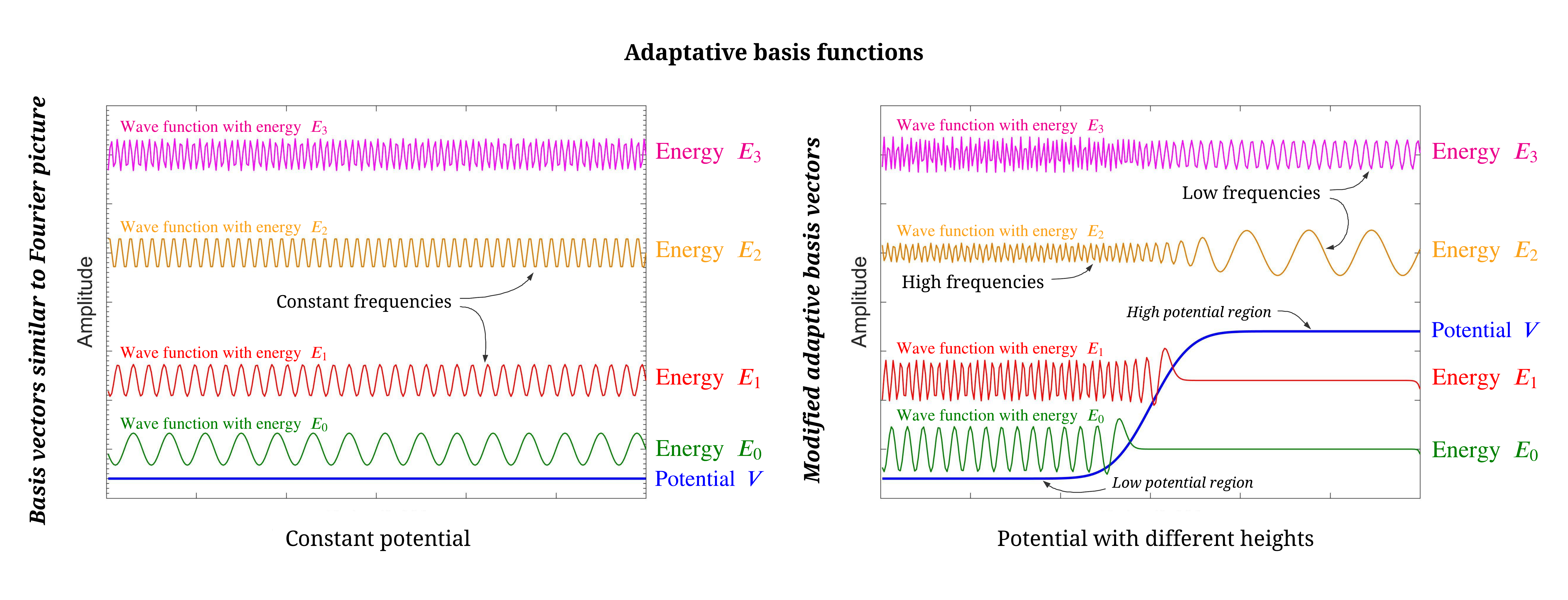}
		
	\caption{Relationship between the frequency of the adaptive basis functions and the height of the potential.}
	\label{fig:difpote}
\end{figure*}

\subsection{Adaptive transform for signals or images}
\label{sec:adtrans}

Our method uses quantum mechanics as a tool for building an adaptative basis suitable for denoising applications. We will only expose here the basics of quantum mechanics which are useful for our purpose, refering the interested reader to more extensive introductions to this vast field of physics \cite{feynman,landau,cohen}.
Our formalism is based on the resolution of the Schroedinger equation of non-relativistic quantum mechanics. This equation determines the wave function $\psi(y)$ which belongs to the Hilbert space of $L^2$-integrable functions, $y$ being e.g. a spatial coordinate. The function $|\psi (y) |^2$ gives the probability of presence of the particle,  which implies that $\int |\psi(y)|^2 \mbox{dy} =1$.

The basic idea of the proposed method is to consider the signal or image as a potential $V(y)$ for a quantum system, as illustrated in Fig.~\ref{fig:relation}. The 3D surface plot of a 2D image is shown in Fig.~\ref{fig:relation}. This surface will act as the potential of the system, where we consider pixel intensity as the height of the potential (\textit{i.e.,} along the z-axis). It is clearly visible that there are many hills and valleys in the potential which are associated with the high and low pixel values respectively. If a quantum particle with energy $E$ probes this surface, then the probability of presence of this quantum particle at some position on the surface will be determined by the wave function $\psi(y)$. Unlike the classical picture where one can precisely determine the position of a classical particle, the quantum theory only gives a probability of finding a quantum particle at some point.
The stationary Schroedinger equation corresponds to the probability of presence of a stationary quantum particle with energy $E$ in a potential $V(y)$, the associated wave function $\psi(y)$ satisfying \cite{schrodinger1926undulatory}:
\begin{equation}
 - \frac{\hbar ^2}{2m} \nabla ^2 \psi = - V(y)  \psi + E \psi,
\label{eq:Schroedinger}
\end{equation}
\noindent with $m$ the mass of the quantum particle, $\hbar$ the Planck constant that are both parameters of the problem, and $\nabla^2$ the Laplacian operator.

To illustrate the nature of the solution of \eqref{eq:Schroedinger}, let us consider a simple case corresponding to a constant potential $V$ and to the wave function $\psi(y)$ following a periodic boundary condition, \textit{i.e.,} $\psi(y + L) = \psi(y)$, where $L$ is the periodicity.
Solving equation \eqref{eq:Schroedinger} under the above conditions is trivial, all solutions having the form:
\begin{equation}
 \psi(y) = A ~ e^{ i \frac{\sqrt{2m(E-V)}}{\hbar} y },
\label{eq:wavefun}
\end{equation}
where $A$ is a given amplitude. Each solution $\psi$ is associated with a specific value of $E$, with $E$ taking  discrete values, all higher than $V$. If space is discretized in $N$ values, there will be $N$ solutions and $E$ takes only $N$ different values. 


In the case of a more intricate 1D potential, where $V$ is no more a simple constant and depends on position, \eqref{eq:Schroedinger} implies that the relation \eqref{eq:wavefun} will still hold locally, with an amplitude and phase depending on position. This means that the stationary solutions of \eqref{eq:Schroedinger} are locally oscillatory functions with an oscillation frequency dependent on the local value of $V$ for a given energy $E$, with a frequency proportional to $\sqrt{E-V}$. This is illustrated in Fig.~\ref{fig:difpote}, where two different potentials are taken into account in the Schroedinger equation \eqref{eq:Schroedinger}: a constant potential and a potential with non-uniform heights.
For the constant potential, the solutions are just plane waves satisfying \eqref{eq:wavefun}. All solutions are indexed by the values of the associated energy, and higher energy translates in higher frequency of oscillations. This frequency is the same for all positions for a given wavefunction. For a non constant potential which depends on position (right panel of Fig.~\ref{fig:difpote}), the oscillation frequency still increases with higher values of $E$, but at the same time a given stationary solution of \eqref{eq:Schroedinger}, which corresponds to the physical  wavefunction, contains different local oscillation frequencies according to the local value of $V$. Thus, although at each local position the frequency increases with $E$, it does so in a different way from place to place according to the local value of $V$. 
In other words, for a given energy $E$ the wave function $\psi(y)$ associated with a quantum particle will use a higher frequency to probe a low potential region in comparison with a high potential region. In the regions where $E-V$ is negative, \eqref{eq:wavefun} leads to exponentially decreasing functions which quickly become constant (see e.g. the solutions for $E_0$ and $E_1$ in the right side of Fig.~\ref{fig:difpote}).

In 2D, \eqref{eq:wavefun} is not exactly valid, but the solutions of \eqref{eq:Schroedinger} will still have typically an oscillation frequency proportional to $\sqrt{E-V}$. This is illustrated in Fig.~\ref{fig:relation} (bottom panels) where the wave function frequency of oscillation is clearly seen to increase in the regions of low potential.

To summarize, the global properties of the wavefunctions which form the proposed adaptive basis are the following:
\begin{enumerate}
	\item they are oscillating functions indexed by the energy $E$,
	\item the local frequency is typically proportional to $\sqrt{E-V}$, thus increasing with $E$ while differing locally for the same wavefunction depending on the local value of $E-V$,
	\item the precise dependence on the frequency of oscillation with respect to $E-V$ is tuned by the parameter $\hbar ^2/2m$.
\end{enumerate}

In the application addressed herein, the Schroedinger equation \eqref{eq:Schroedinger} is just a way to obtain an adaptive basis possessing these properties, which can further be used independently of its quantum mechanical nature as a tool for signal or image processing.

The basis of eigenvectors of \eqref{eq:Schroedinger} naturally describes with different frequencies the different parts of the signal or image, in contrast to e.g the Fourier or wavelet bases. As said above, the precise relation between the local frequency of the eigenvectors and the value of the signal or image pixel is governed by the parameter $\hbar^2/2m$. In the physical problem of quantum mechanics, this quantity is linked to Planck's constant and the particle mass, but in our framework it is a free parameter. It should be chosen with care, as extreme values are clearly inadequate. Indeed, as the problem is discretized there is a maximal frequency in the problem, linked to the inverse of the discretization step. If $\hbar^2/2m$ is very small, the local frequencies $\sqrt{2m(E-V)}/\hbar$ become very high even for low values of the energy, the maximal energy becomes very low, and the basis does not explore properly high values of the signal or pixels of the
image. On the other side for very large values of $\hbar^2/2m$, the local frequencies become smaller and smaller at fixed energy, the maximal energy becomes larger and larger, and eventually when $\hbar^2/2m$ tends to infinity most vectors of the adaptive basis are so high above the signal or image pixel that they do not discriminate between low and high values, becoming closer and closer to the standard Fourier basis vectors. Therefore it is crucial to tune the free parameter $\hbar^2/2m$ in the right way.


\section{Proposed method for denoising applications}
\label{sec:promethdeapp}

\subsection{Explicit construction of the adaptive basis}
\label{sec:techconst}

In operator notation, \eqref{eq:Schroedinger} corresponds to $H \psi =  E \psi$ with $H=- \frac{\hbar ^2}{2m} \nabla ^2 +V$ the Hamiltonian operator.  The energy $E$ of the particle in \eqref{eq:Schroedinger} labels the solutions of the problem. Solutions of this stationary Schroedinger equation in a bounded domain correspond to a discrete set of energy levels, from a minimal energy to infinity.

Solutions of \eqref{eq:Schroedinger} form a basis of the Hilbert space to which the wavefunctions belong. This space is infinite-dimensional for continuous values of the position $y$ in \eqref{eq:Schroedinger}. However, we are interested in signal or image processing applications, where the space is discretized in a finite number of points. Specifically, we assume that the potentiel $V$ is represented by the value of signal sample or image pixel $\bsx$. In the case of a discretized problem, the operators become finite matrices and the resolution of \eqref{eq:Schroedinger} is equivalent to diagonalizing the Hamiltonian matrix.

	
Specifically, one has, following \eqref{eq:Schroedinger},	
	 $ H = - \frac{\hbar ^2}{2m} \nabla ^2 + \bsx $, with:
	\begin{itemize}
		\item the potential $V$ represented by $\bsx$ (the signal or the image),
		\item  if $\bsx$ is a signal of size $N$, then the size of $H$ is $N\times N$,
		\item if $\bsx$ is an image of size $N\times N$, it is transformed into a vector (in lexicographical order) of size $N^2$ and $H$ is a $N^2 \times N^2$ matrix,
		\item in both cases ($\bsx$ is a signal or an image), $\bsx$ is considered in a vector form.
	\end{itemize}
    For a 1D signal, we have:
	\begin{itemize}
	    \item numerical derivatives of $\psi$: $(\nabla \psi)_i = \psi (i+1) - \psi (i)$,
	    \item numerical Laplacian of $\psi$: $ (\nabla ^2 \psi)_i = \psi (i+1) - 2 \psi(i) + \psi(i-1)$.
	\end{itemize}
	Thus, $(H \psi)_i = - \frac{\hbar ^2}{2m} (\psi (i+1) - 2 \psi(i) + \psi(i-1)) +\bsx(i) \psi(i) $\\
	 $\implies (H \psi)_i = \left( \bsx(i) + 2 \frac{\hbar ^2}{2m} \right)  \psi(i) - \frac{\hbar ^2}{2m} (\psi (i+1) - \frac{\hbar ^2}{2m} \psi(i-1)) $.\\
	Therefore, $(H \psi)_i = \sum_{j = i-1}^{i+1} H(i,j) \psi(j) $, for $ i = 1,2,3, \cdots,N $.\\
	where,
\begin{eqnarray}
H(i,j)= \left \{
   \begin{array}{r c l}
      \bsx(i)+ 2 \frac{\hbar ^2}{2m} &  & for \; j = i,\\
       -\frac{\hbar ^2}{2m} & & for \; j = i \pm 1,\\
      0 & & otherwise.
   \end{array}
   \right.
\end{eqnarray}
where $\bsx(i)$ represents the $i$-th component of a signal and $H(i,j)$ is the $(i,j)$-th element of the Hamiltonian matrix.

The resolution of \eqref{eq:Schroedinger} is thus equivalent to finding eigenvectors and eigenvalues of the discretized Hamiltonian matrix $H \in \mathbb{R}^{N \times N}$ written as:
$$ H = 
\begin{bmatrix} 

\bsx(1) + 2 \frac{\hbar ^2}{2m} & -\frac{\hbar ^2}{2m} & & \\

-\frac{\hbar ^2}{2m} & & & \\
& \ddots & \ddots & \ddots & \\
& & & & -\frac{\hbar ^2}{2m}\\

& & & -\frac{\hbar ^2}{2m} & \bsx(N) + 2 \frac{\hbar ^2}{2m} 

\end{bmatrix} $$

For a 2D image $\bsx \in \mathbb{R}^{N\times N}$ the methodology is similar.
In \eqref{eq:Schroedinger}, the Laplacian operator should be replaced by its discrete version, following the standard numerical definitions of the gradient operator:
\begin{align}
\nabla_{\rm{h}} \bsx(i,j) &= \bsx(i+1,j) - \bsx(i,j) \; &\textrm{if} \; i < N 
\nonumber
\\
\nabla_{\rm{v}} \bsx(i,j) &= \bsx(i,j+1) - \bsx(i,j) \; &\textrm{if} \; j < N
\nonumber
\end{align}

\noindent where $\nabla_{\rm{h}}$ and $\nabla_{\rm{v}}$ are associated to the horizontal and vertical gradients. The boundary conditions correspond simply to a zero padding of the image.

The Hamiltonian matrix is thus:
\begin{eqnarray}
\label{eq:H}
H(i,j)= \left \{
   \begin{array}{r c l}
      \bsx(i)+ 4 \frac{\hbar ^2}{2m} &  & for \; i=j,\\
       -\frac{\hbar ^2}{2m} & & for \; i = j \pm 1,\\
        -\frac{\hbar ^2}{2m} & & for \; i = j \pm N,\\
      0 & & otherwise,
   \end{array}
   \right.
\end{eqnarray}
where $\bsx(i)$ represents the $i$-th component of a vectorized image $\bsx$ in the lexicographical order.

As the boundary conditions correspond to zero padding of the image, a few individual coefficients of the matrix $H$ follow specific rules. Indeed, $H(i,j) = \bsx(i)+ 2 \frac{\hbar ^2}{2m}$ for $i=j$ and $i \in \{1,N,N^2-N+1,N^2\}$, $H(i,j) = \bsx(i)+ 3 \frac{\hbar ^2}{2m}$ for $i=j$ and $i \in \{ 2,3,...,N-1,N^2-N+2,N^2-N+3,...,N^2-1\}$, $H(i,j) = \bsx(i)+ 3 \frac{\hbar ^2}{2m}$ for $i=j$ and other than the previous set with $i ~ \mod ~ N\in \{0,1\}$ and $H(i,i+1)=H(i+1,i)=0$ for any $i$ multiple of $N$ apart from $N^2$. 

In the specific case of $N = 4$, \textit{i.e.} for an image of size $4\times4$ the discretized Hamiltonian is of size $16\times16$. This Hamiltonian matrix is explicitly shown in Table \ref{Hmatrix01}.



\begin{table*}[h!]
\begin{tiny}
\setlength\tabcolsep{-3pt}
\begin{center}
\caption{The Hamiltonian matrix of size $16\times16$ corresponding to an image of size $4\times4$.}
\label{Hmatrix01}

\begin{tabular}{c c c c c c c c c c c c c c c c}
\hline
1 & 2 & 3 & 4 & 5 & 6 & 7 & 8 & 9 & 10 & 11 & 12 & 13 & 14 & 15 & 16\\
\hline
~\\

$\SD{\bsx(1)} + 2 \dfrac{\hbar^{2}}{2m}$ & $-\dfrac{\hbar^{2}}{2m}$ & 0 & 0 & $-\dfrac{\hbar^{2}}{2m}$ & 0 & 0 & 0 & 0 & 0 & 0 & 0 & 0 & 0 & 0 & 0\\

$-\dfrac{\hbar^{2}}{2m}$ & $\SD{\bsx(2)}+ 3 \dfrac{\hbar^{2}}{2m}$ & $-\dfrac{\hbar^{2}}{2m}$ & 0 & 0 & $-\dfrac{\hbar^{2}}{2m}$ & 0 & 0 & 0 & 0 & 0 & 0 & 0 & 0 & 0 & 0\\

0 &  $-\dfrac{\hbar^{2}}{2m}$ & $\SD{\bsx(3)}+ 3 \dfrac{\hbar^{2}}{2m}$ & $-\dfrac{\hbar^{2}}{2m}$ & 0 & 0 & $-\dfrac{\hbar^{2}}{2m}$ & 0 & 0 & 0 & 0 & 0 & 0 & 0 & 0 & 0 \\

0 & 0 &  $-\dfrac{\hbar^{2}}{2m}$ & $\SD{\bsx(4)}+ 2 \dfrac{\hbar^{2}}{2m}$ & 0 & 0 & 0 & $-\dfrac{\hbar^{2}}{2m}$ & 0 & 0 & 0 & 0 & 0 & 0 & 0 & 0 \\

$-\dfrac{\hbar^{2}}{2m}$ & 0 & 0 & 0 & $\SD{\bsx(5)}+ 3 \dfrac{\hbar^{2}}{2m}$ & $-\dfrac{\hbar^{2}}{2m} $ & 0 & 0 & $-\dfrac{\hbar^{2}}{2m}$ & 0 & 0 & 0 & 0 & 0 & 0 & 0 \\

0 & $-\dfrac{\hbar^{2}}{2m} $ & 0 & 0 & $-\dfrac{\hbar^{2}}{2m} $ & $\SD{\bsx(6)}+ 4 \dfrac{\hbar^{2}}{2m}$ & $-\dfrac{\hbar^{2}}{2m}$ & 0 & 0 & $-\dfrac{\hbar^{2}}{2m}$ & 0 & 0 & 0 & 0 & 0 & 0 \\

0 & 0 & $-\dfrac{\hbar^{2}}{2m} $ & 0 & 0 & $-\dfrac{\hbar^{2}}{2m} $ & $\SD{\bsx(7)}+ 4 \dfrac{\hbar^{2}}{2m}$ & $-\dfrac{\hbar^{2}}{2m}$ & 0 & 0 & $-\dfrac{\hbar^{2}}{2m}$ & 0 & 0 & 0 & 0 & 0 \\

0 & 0 & 0 & $-\dfrac{\hbar^{2}}{2m} $ & 0 & 0 & $-\dfrac{\hbar^{2}}{2m} $ & $\SD{\bsx(8)}+ 3 \dfrac{\hbar^{2}}{2m}$ & 0 & 0 & 0 & $-\dfrac{\hbar^{2}}{2m}$ & 0 & 0 & 0 & 0 \\

0 & 0 & 0 & 0 & $-\dfrac{\hbar^{2}}{2m} $ & 0 & 0 & 0 & $\SD{\bsx(9)}+ 4 \dfrac{\hbar^{2}}{2m}$ & $-\dfrac{\hbar^{2}}{2m}$ & 0 & 0 & $-\dfrac{\hbar^{2}}{2m}$ & 0 & 0 & 0 \\

0 & 0 & 0 & 0 & 0 & $-\dfrac{\hbar^{2}}{2m} $ & 0 & 0 & $-\dfrac{\hbar^{2}}{2m} $ & $\SD{\bsx(10)}+ 4 \dfrac{\hbar^{2}}{2m}$ & $-\dfrac{\hbar^{2}}{2m}$ & 0 & 0 & $-\dfrac{\hbar^{2}}{2m}$ & 0 & 0 \\

0 & 0 & 0 & 0 & 0 & 0 & $-\dfrac{\hbar^{2}}{2m}$ & 0 & 0 & $-\dfrac{\hbar^{2}}{2m}$ & $\SD{\bsx(11)}+ 4 \dfrac{\hbar^{2}}{2m}$ & $-\dfrac{\hbar^{2}}{2m}$ & 0 & 0 & $-\dfrac{\hbar^{2}}{2m}$ & 0\\

0 & 0 & 0 & 0 & 0 & 0 & 0 & $-\dfrac{\hbar^{2}}{2m}$ & 0 & 0 & $-\dfrac{\hbar^{2}}{2m}$ & $\SD{\bsx(12)}+ 3 \dfrac{\hbar^{2}}{2m}$ & 0 & 0 & 0 & $-\dfrac{\hbar^{2}}{2m}$\\

0 & 0 & 0 & 0 & 0 & 0 & 0 & 0 & $-\dfrac{\hbar^{2}}{2m}$ & 0 & 0 & 0 & $\SD{\bsx(13)}+ 2 \dfrac{\hbar^{2}}{2m}$ & $-\dfrac{\hbar^{2}}{2m}$ & 0 & 0\\

0 & 0 & 0 & 0 & 0 & 0 & 0 & 0 & 0 & $-\dfrac{\hbar^{2}}{2m}$ & 0 & 0 & $-\dfrac{\hbar^{2}}{2m}$ & $\SD{\bsx(14)}+ 3 \dfrac{\hbar^{2}}{2m}$ & $-\dfrac{\hbar^{2}}{2m}$ & 0\\

0 & 0 & 0 & 0 & 0 & 0 & 0 & 0 & 0 & 0 & $-\dfrac{\hbar^{2}}{2m}$ & 0 & 0 & $-\dfrac{\hbar^{2}}{2m}$ & $\SD{\bsx(15)}+ 3 \dfrac{\hbar^{2}}{2m}$ & $-\dfrac{\hbar^{2}}{2m}$\\

0 & 0 & 0 & 0 & 0 & 0 & 0 & 0 & 0 & 0 & 0 & $-\dfrac{\hbar^{2}}{2m}$ & 0 & 0 & $-\dfrac{\hbar^{2}}{2m}$ & $\SD{\bsx(16)}+ 2 \dfrac{\hbar^{2}}{2m}$\\

~\\

\hline
\end{tabular}
\end{center}
\end{tiny}

\end{table*}



The set of eigenvectors gives a basis of the Hilbert space, with each eigenvector associated to an energy $E$, which is the corresponding eigenvalue of the Hamiltonian operator. The $N^2$ eigenvectors, denoted by $\bpsi_{i} \in \mathbb{R}^{N^2 \times 1}$, each with components $\bpsi_{i}^j$ with $j=1, \cdots, N^2$, are the main tool for the proposed adaptive transform in this work. Indeed, our method consists in projecting the signal or image on this particular basis and use the energy associated to each eigenfunction as a parameter on which we perform the thresholding of these coefficients.

\subsection{A technical problem for noisy signals or images: the problem of quantum localization}
\label{sec:techprob}

In order to use the adaptive basis for various problems of signal or image processing, including denoising, the procedure should be adapted for noisy signals and images. A technical problem then arises, linked to the phenomenon of quantum localization. Quantum localization is a property of wave functions in a disordered potential which makes the adaptive basis localized in position space, which in turn makes it less useful for our purpose. In this subsection, we propose a way to mitigate this technical problem which will be implemented throughout the paper.

Indeed, it is known in quantum mechanics that a disordered potential localizes the wavefunctions in one and two dimensions. Due to destructive interference, the different wave functions are exponentially localized at different positions of the potential, an effect known as Anderson localization, which earned the Nobel prize in 1977 to its discoverer \cite{Anderson1958}. If the signal or image are not smooth, which certainly arises in the case of a noisy signal or image, we expect the vectors of the basis to be localized, with a localization length which will be smaller and smaller for increasing noise intensity.

Let us start from a wave function defined as eigenfunction of \eqref{eq:H}, $\bpsi_i$ $ \in \mathbb{R}^{N^2 \times 1}$ with components  $\bpsi_{i}^j$.  The level of localization is measured by computing the inverse participation ratio (IPR) of the wave functions, mathematically defined for a given wave function as:
\begin{eqnarray}
\label{eq:IPR}
\mbox{IPR} (\bpsi_i)= \dfrac{\sum _{j = 1}^{N^2} |\bpsi_{i}^j|^2}{\sum _{j = 1}^{N^2} |\bpsi_{i}^j|^4},
\end{eqnarray}
where $N^2$ is the dimension of the Hilbert space. For a vector uniformly spread over $P$ indices and zero elsewhere, this quantity is exactly $P$. For an exponentially localized vector such as the wavefunctions in a disordered potential, it is proportional to the localization length for each vector in the adaptive basis. In this case, these vectors will still be oscillating functions, but will no longer have different frequencies at different locations since they are localized in a specific part of the potential.

\begin{figure}[h!]
  \centering
  \includegraphics[width=0.22\textwidth]{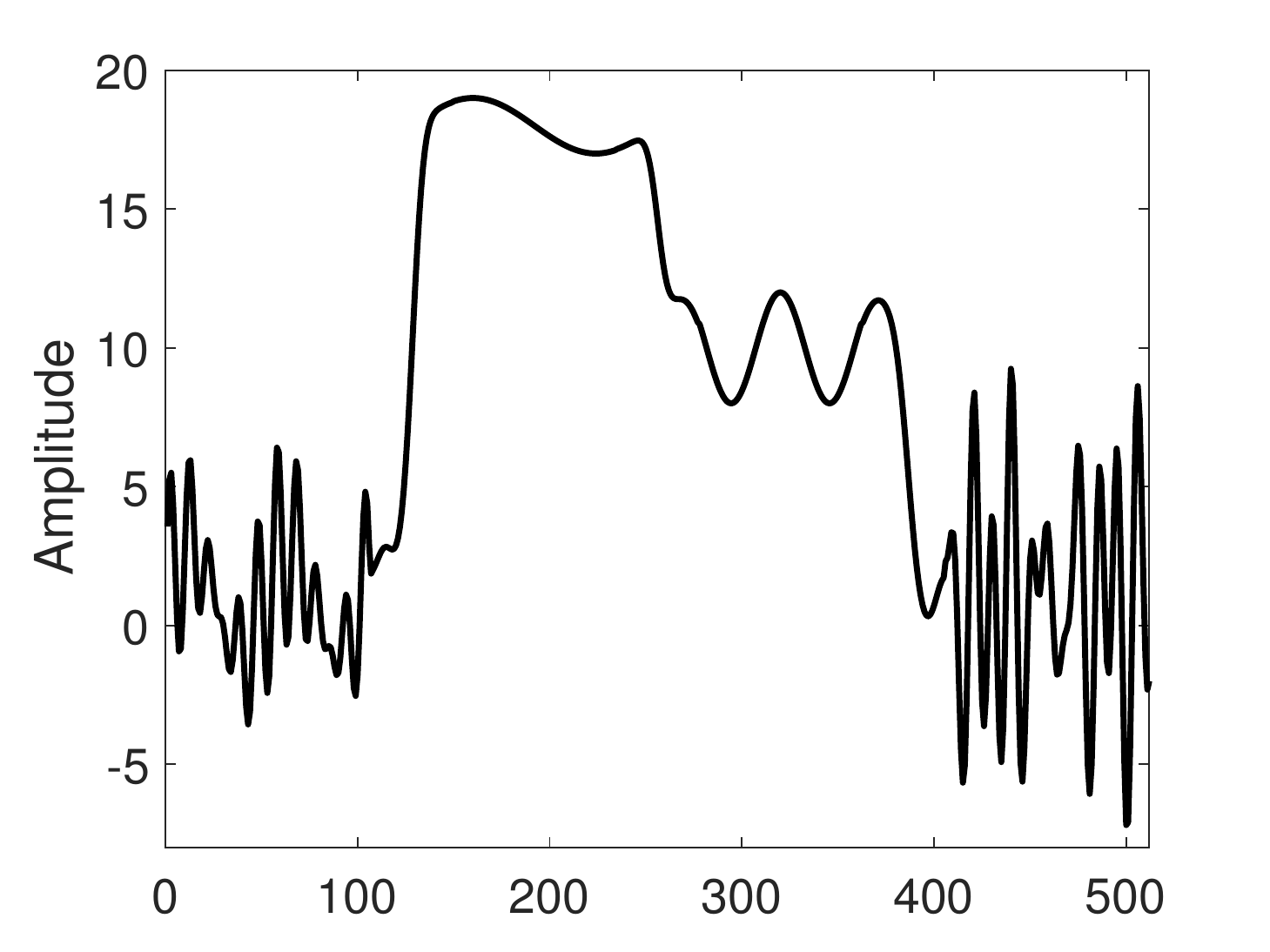}
   
  \caption{ Synthetic signal used to illustrate the localization property of the wave functions.}
\label{fig:sample}
\end{figure}

\begin{figure}[h!]
  \centering
  \includegraphics[width=0.4 \textwidth]{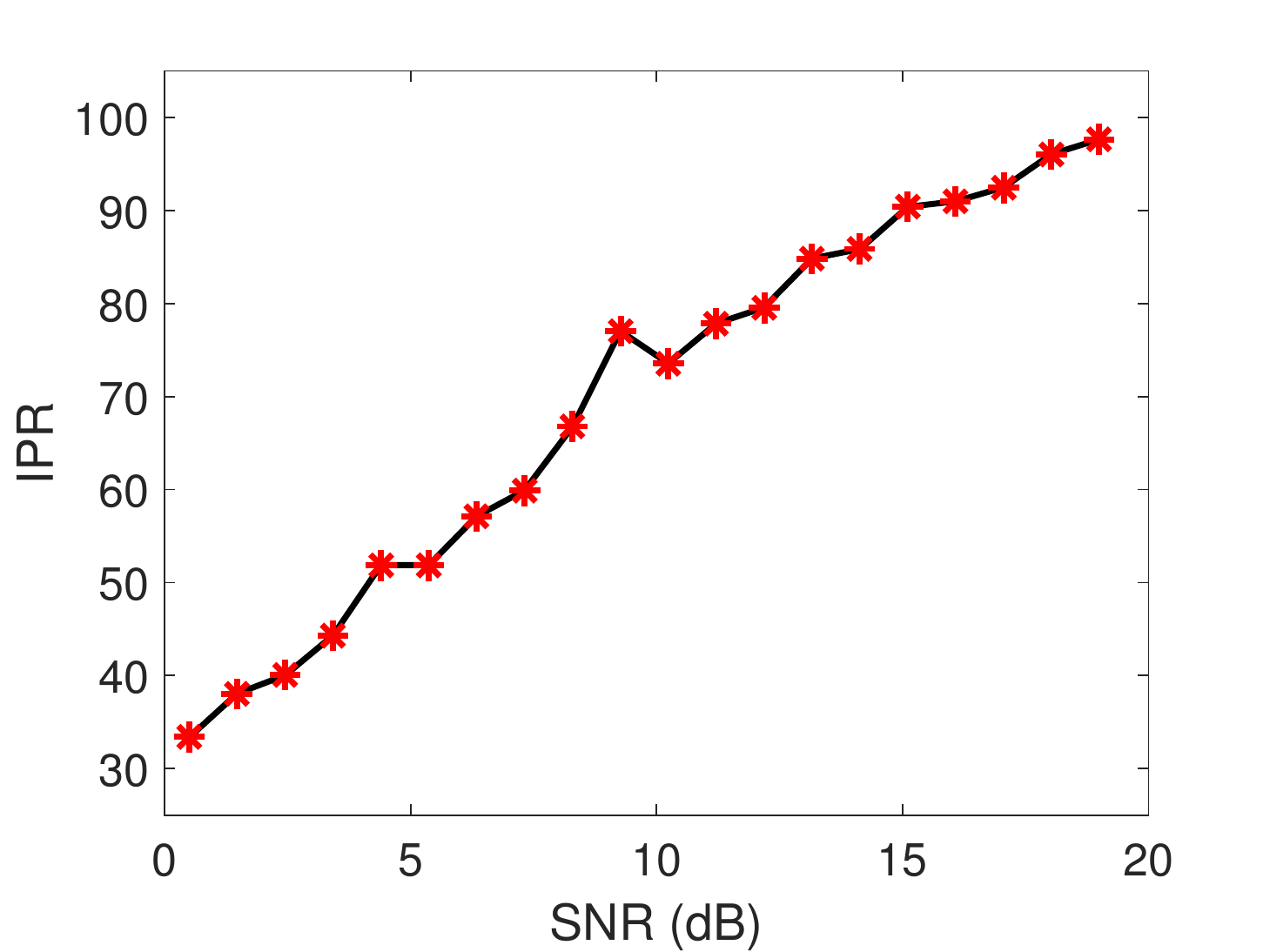}

   \caption{Quantum localization effect: IPR corresponding to the wave functions calculated from the signal in Fig.~\ref{fig:sample} degraded by an additive Gaussian noise for several SNR. The size of the signal was 512. The IPR is computed through \eqref{eq:IPR} and averaged over all 512 wave functions of the adaptive basis.}
\label{fig:IPR}
\end{figure}

\begin{figure*}[h!]
	\centering
	\subfigure[]{\includegraphics[width=0.22\textwidth]{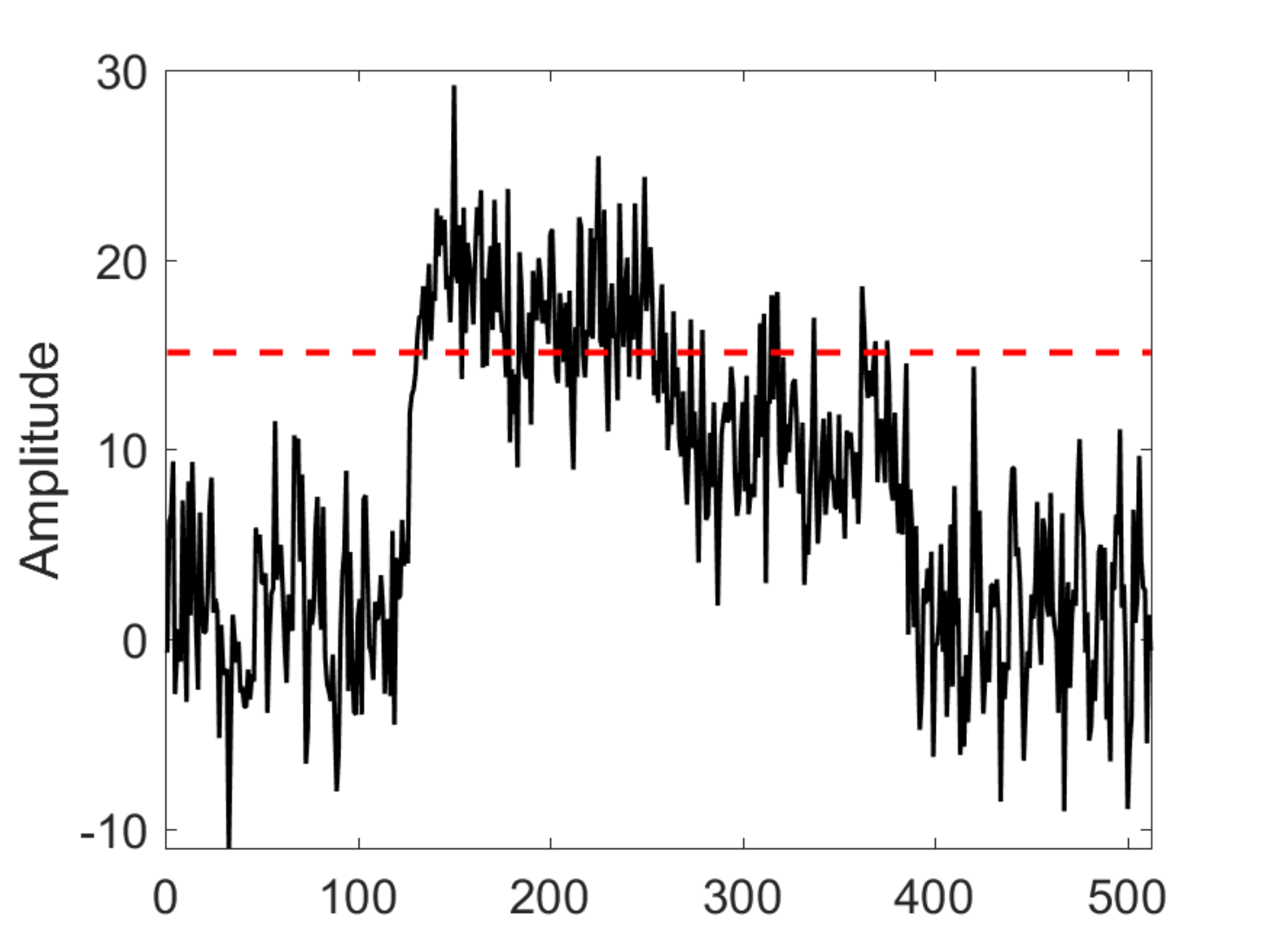}}
	\subfigure[]{\includegraphics[width=0.22\textwidth]{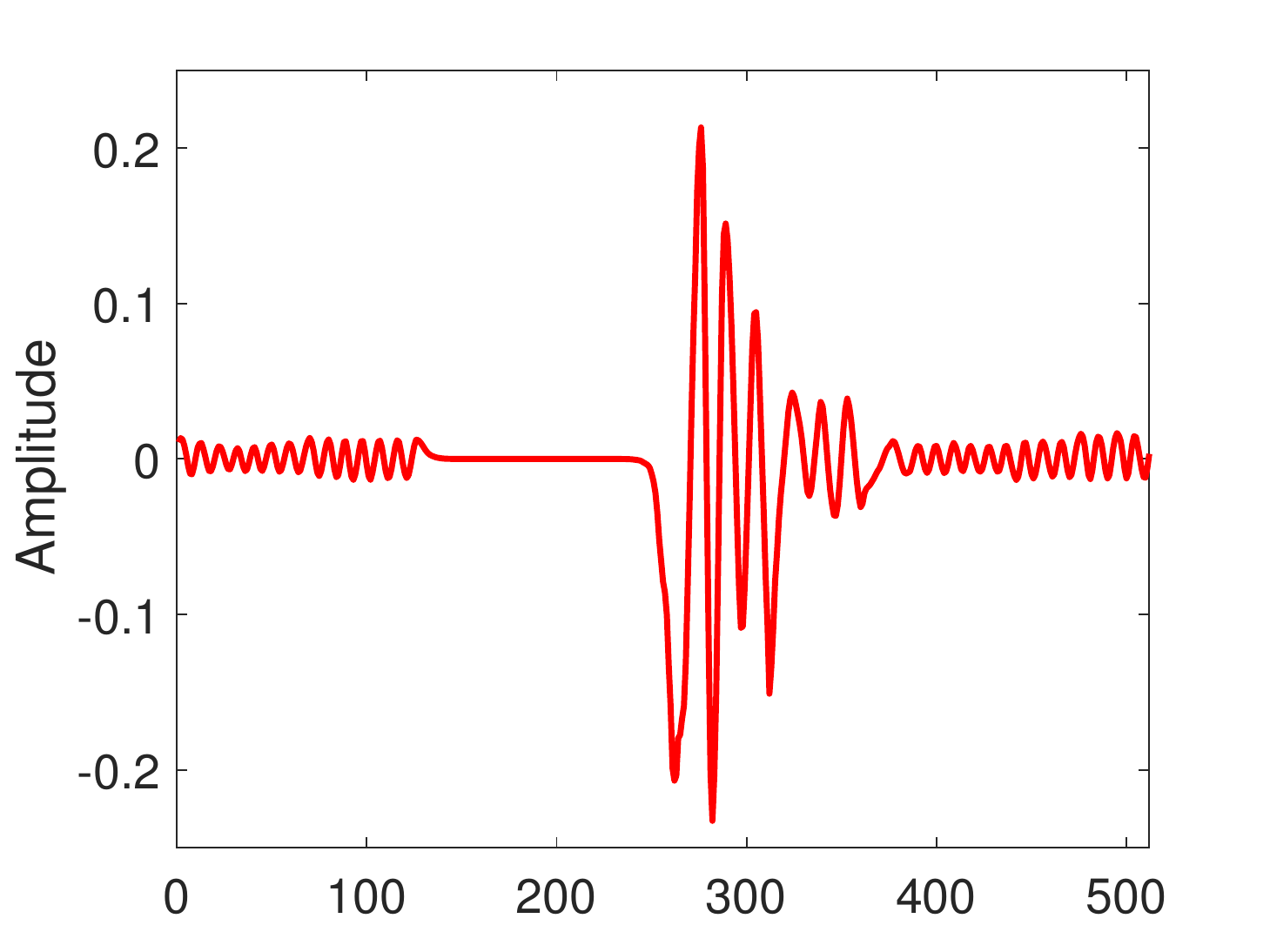}}
	\subfigure[]{\includegraphics[width=0.22\textwidth]{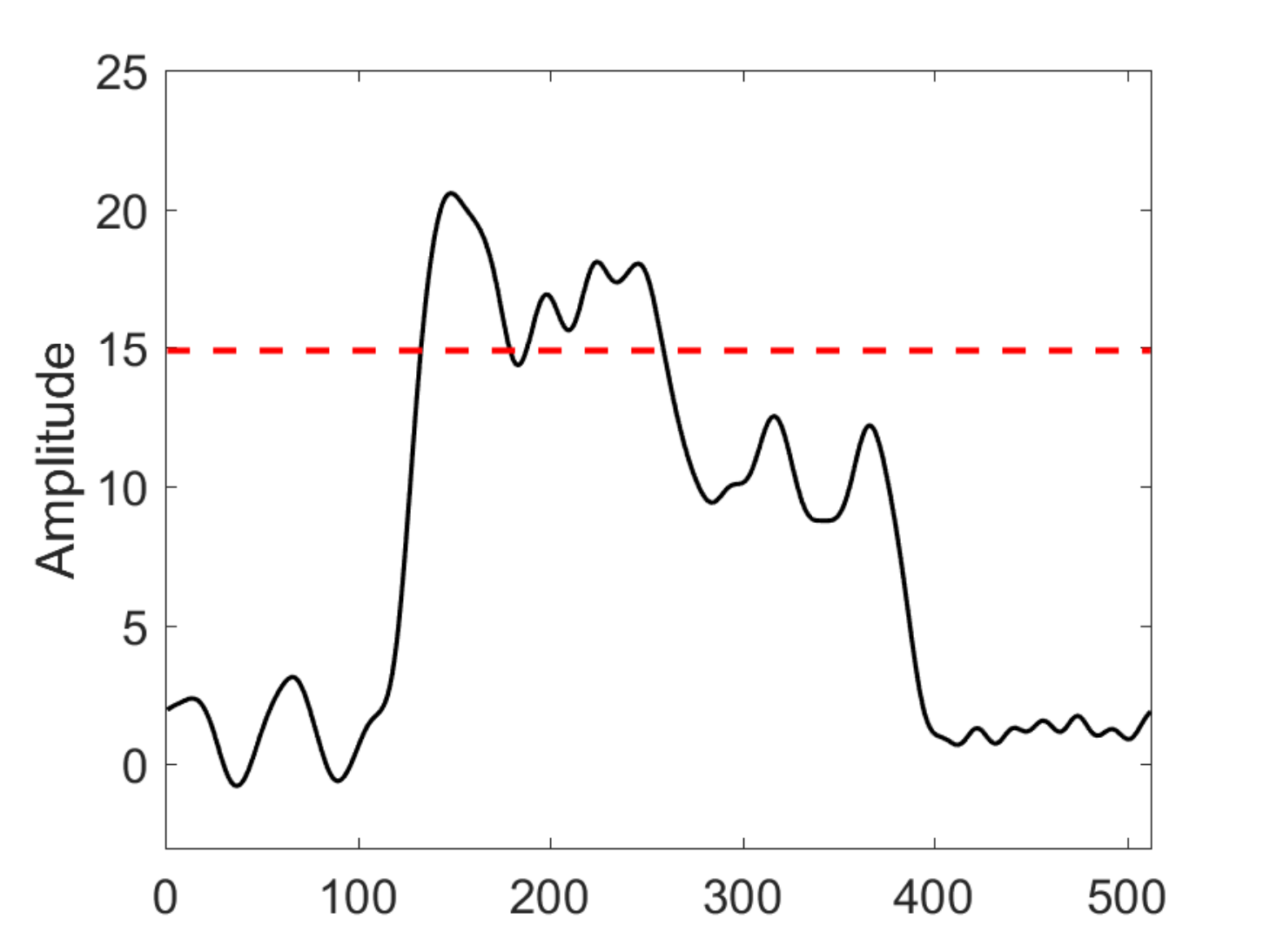}}
	\subfigure[]{\includegraphics[width=0.22\textwidth]{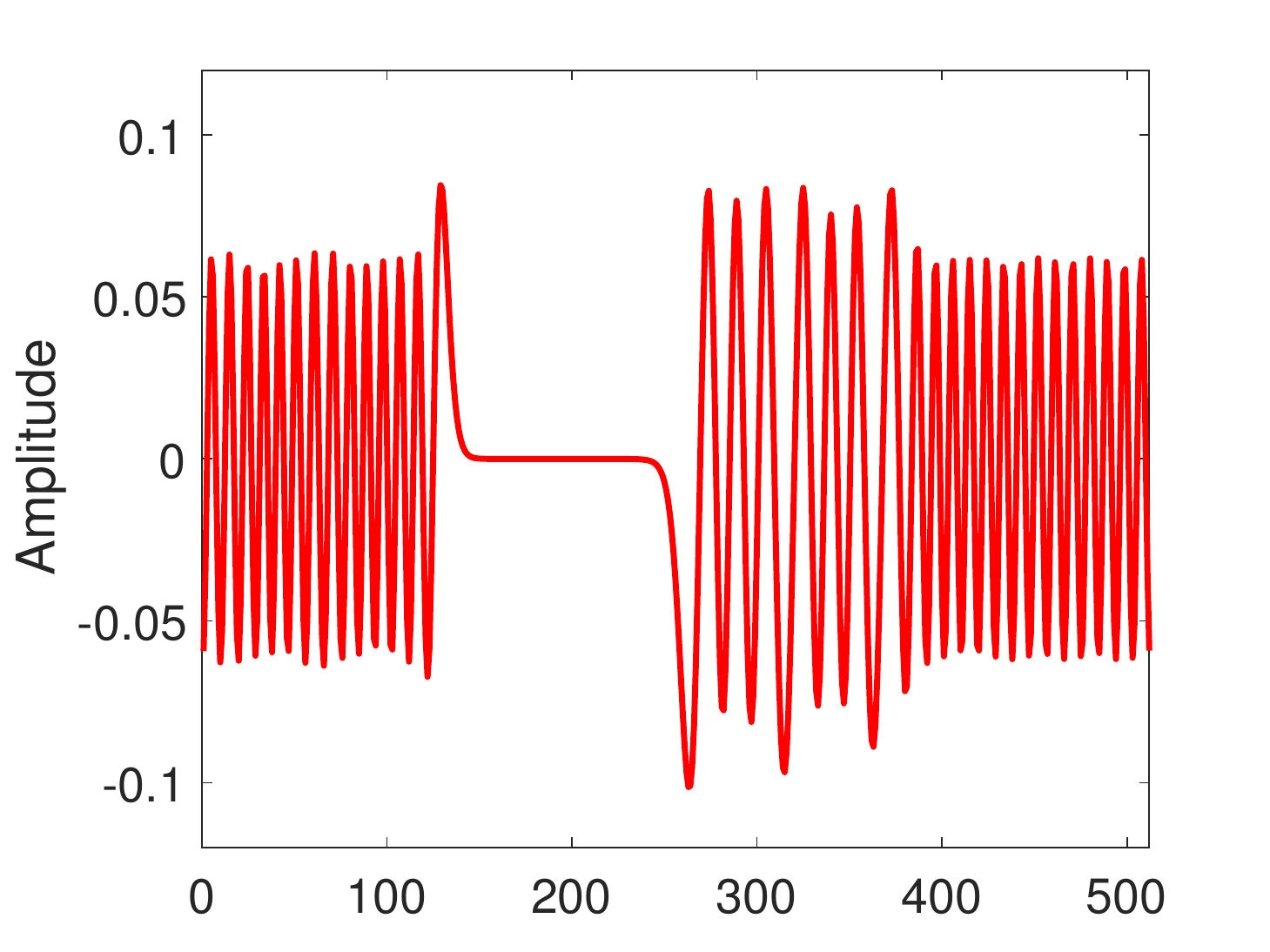}}
	
	\caption{Role of the hyperparameter $\sigma$ and localization: (a) Signal in Fig.~\ref{fig:sample} contaminated by additive Gaussian noise corresponding to a SNR of 15 dB, (b) localized wave function number $68$ calculated from the noisy signal with energy level illustrated by the dashed line in (a), (c) blurred version of the noisy signal in (a) obtained by Gaussian low-pass filter corresponding to $ \sigma^2 = 10 $, (d) delocalized wave function number $68$ calculated from the low-pass filtered signal with the same energy level illustrated by the dashed line in (c).}
	\label{fig:localization_1D}
\end{figure*}


%

\begin{figure*}[h!]
  \centering
   \subfigure[]{\includegraphics[width=0.26\textwidth]{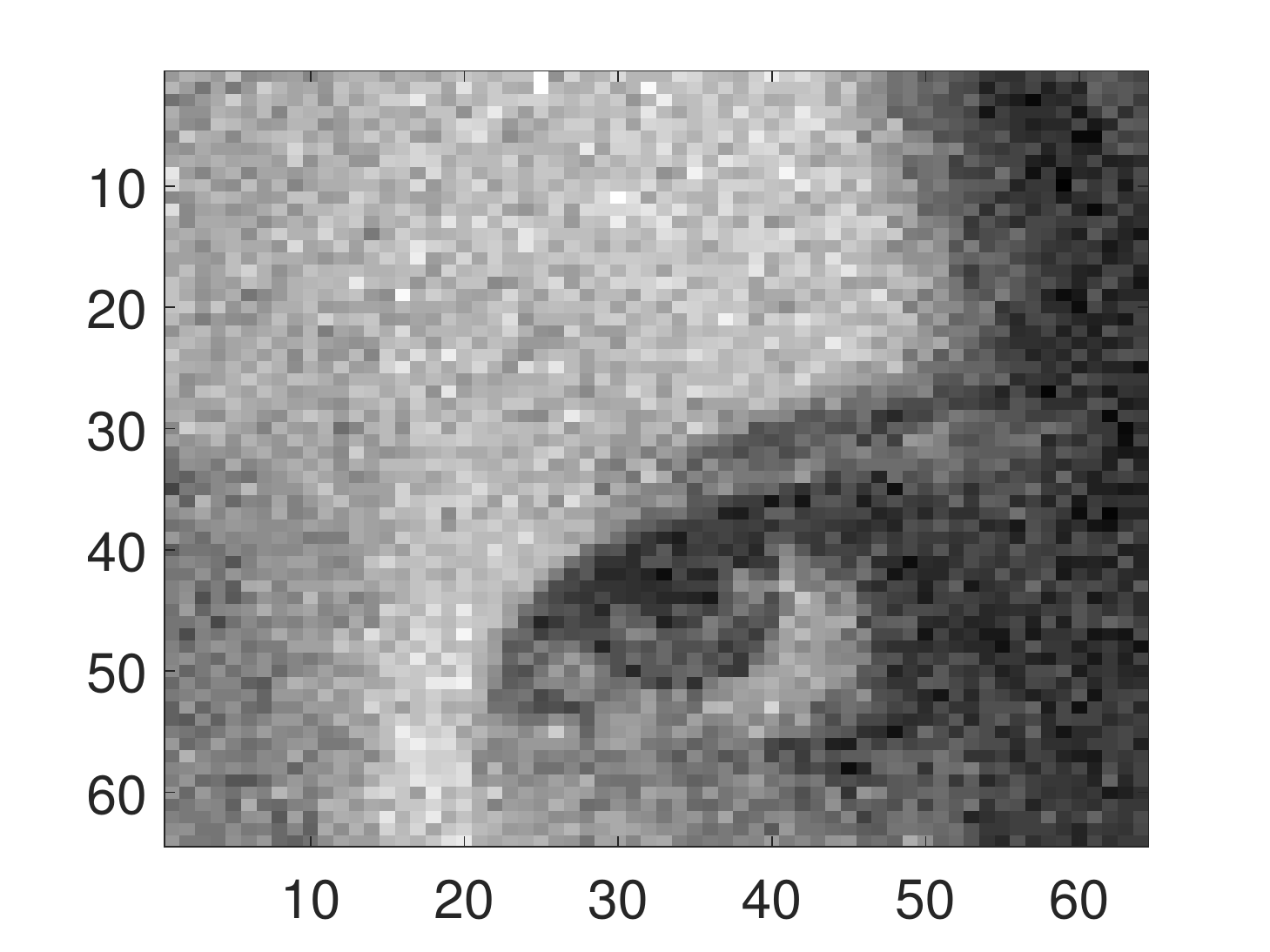}}
   \subfigure[]{\includegraphics[width=0.26\textwidth]{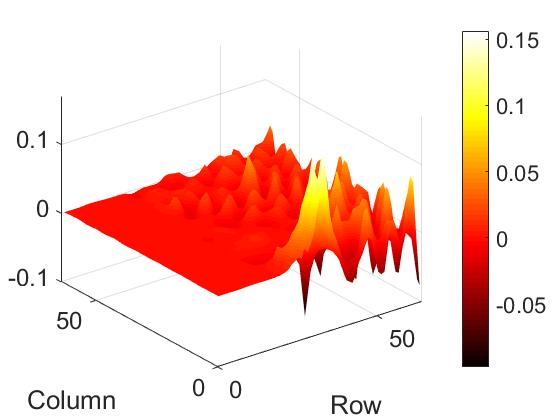}}
   \subfigure[]{\includegraphics[width=0.26\textwidth]{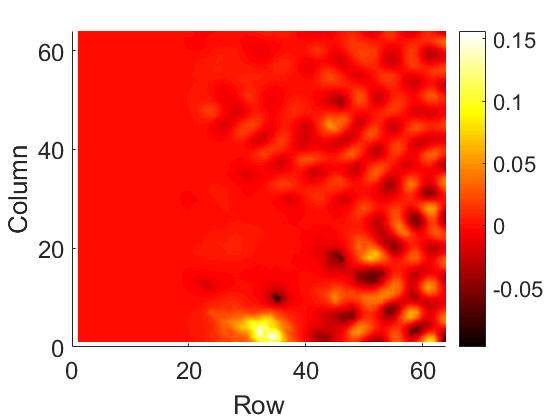}} \\
   \subfigure[]{\includegraphics[width=0.26\textwidth]{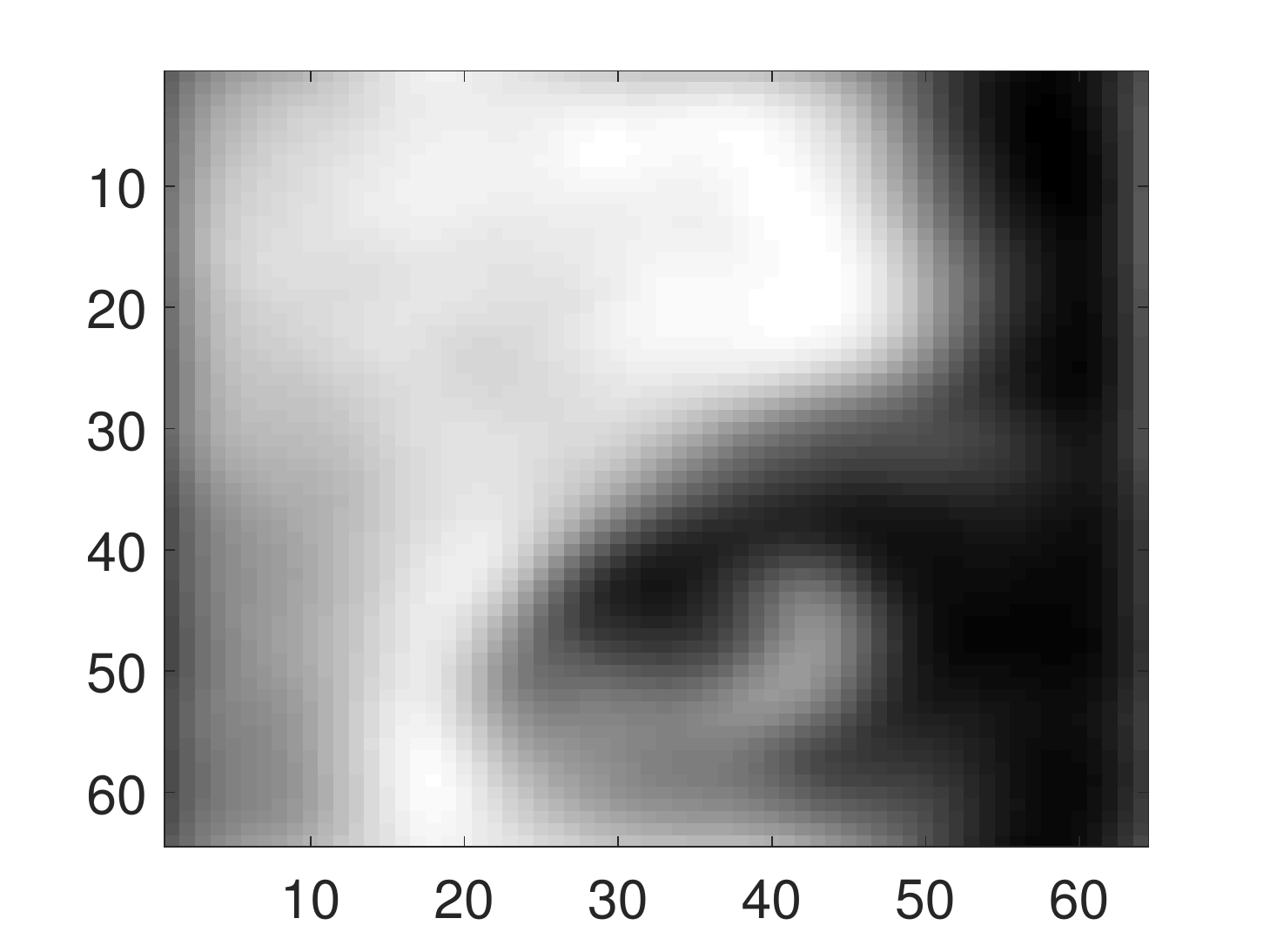}}
   \subfigure[]{\includegraphics[width=0.26\textwidth]{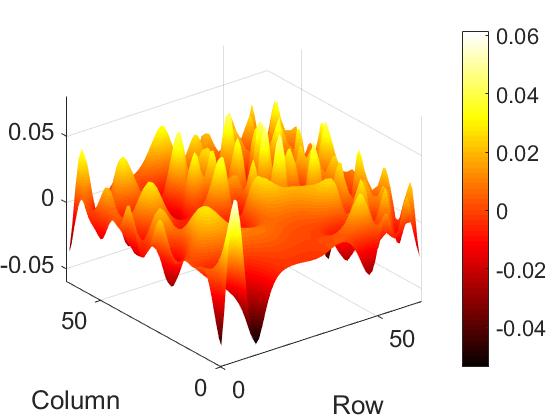}}
   \subfigure[]{\includegraphics[width=0.26\textwidth]{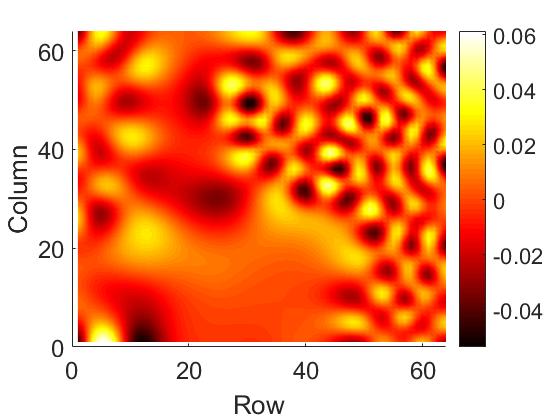}}
   
   \caption{Role of the hyperparameter $\sigma$ and localization: (a) Lena image used in Fig.~\ref{fig:relation}, contaminated by additive Gaussian noise corresponding to a SNR of 15 dB, (b,c) localized wave function number $195$ calculated from the noisy lena image (a), (d) blurred version of the noisy lena image in (a) obtained by Gaussian low-pass filter corresponding to $ \sigma^2 = 6 $, (e,f) the same wave function but delocalized due to the low pass Gaussian filter applied to the noisy image.}
\label{fig:localization_2D02}
\end{figure*}

In Fig.~\ref{fig:IPR} the averaged IPR of all functions of the adaptive basis is shown for a synthetic signal (Fig.~\ref{fig:sample}) degraded by an additive Gaussian noise with different signal to noise ratios (SNR). The localization property is clearly seen: the IPR decreases with decreasing SNR, indicating that noisy signals tend to localize the basis.


To modify this characteristic of the basis, we use a smoothed adaptation of the noisy signal or image to construct the Hamiltonian matrix, computed by a simple
convolution with a Gaussian kernel whose standard deviation is denoted by $\sigma$. This is not part of the denoising process, it is just a technical trick to delocalize the adaptive basis while keeping the main features of the signal/image. In our framework, this standard deviation $\sigma$ is an additional free parameter. If $\sigma$ is chosen too large, then the noisy signal or image becomes so smooth that many characteristics needed for the adaptive basis will be lost. On the opposite, if $\sigma$ is too small the basis vector will remain strongly localized. To balance both sides one needs to tune the parameter $\sigma$ to get the best achievable outcome.


Fig.~\ref{fig:localization_1D}(b) and Fig.~\ref{fig:localization_2D02}(b-c) show examples of wave functions calculated from a noisy signal and image.  From these examples, one observes again that the wave functions are completely localized in a specific location and present a fast decrease due to the destructive interference. On the contrary, in the case where the same wave functions are calculated from low-pass filtered versions of the noisy signal and image (i.e. a smoothed version of the potential), they are shown to delocalize and spread over the whole available space as illustrated in Fig~\ref{fig:localization_1D}(d) and Fig.~\ref{fig:localization_2D02}(e-f).

\begin{figure}[h!]
	\centering
	\includegraphics[width=0.5\textwidth]{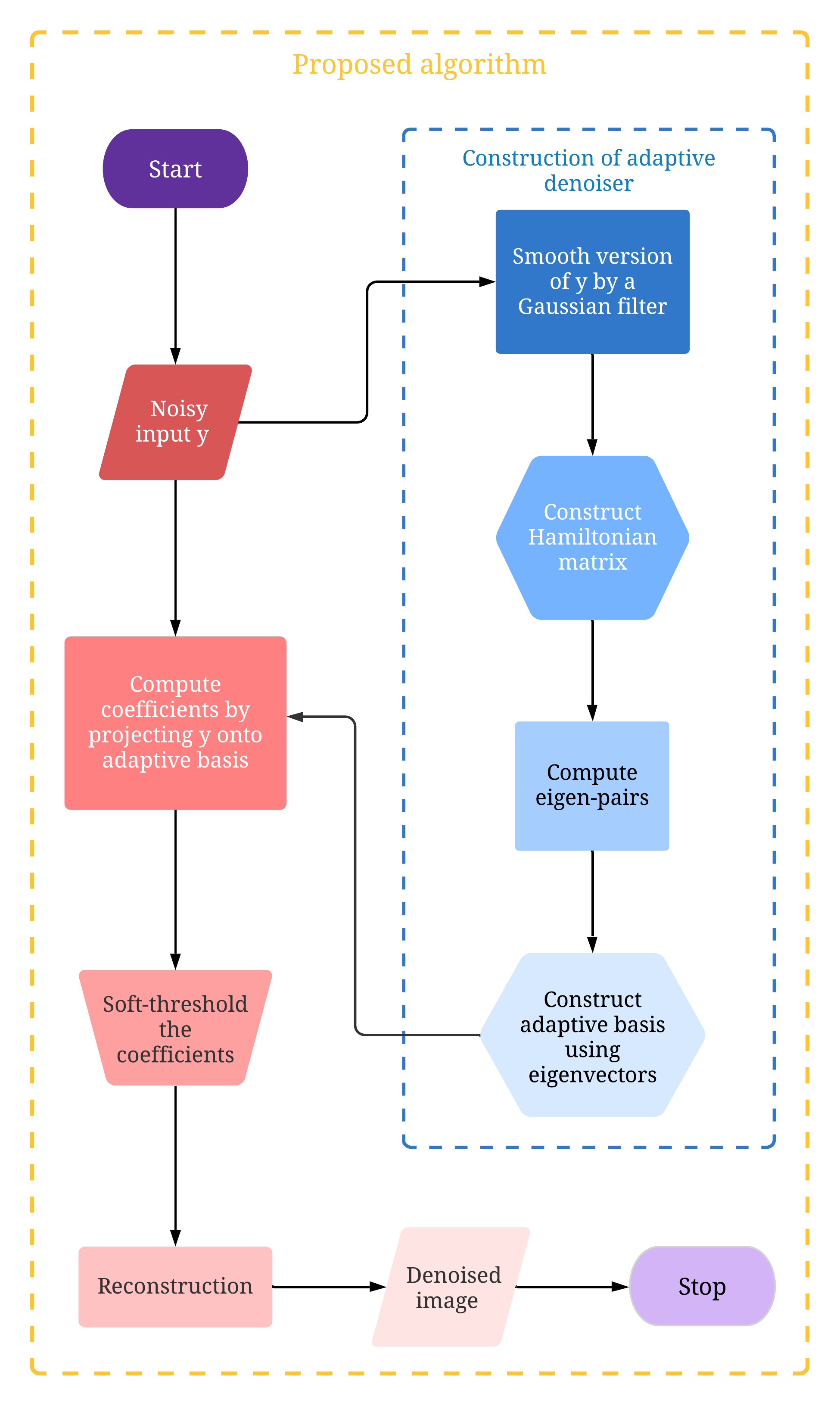}
		
	\caption{Flowchart of the proposed denoising algorithm.}
	\label{fig:sample2}
\end{figure}

\subsection{Application to the denoising problem}

This section explain in details the application of the proposed adaptive basis from quantum mechanics to the denoising problem.
The significant difficulties for signal or image denoising are to sharpen the edges without blurring and preserve the image textures without generating artifacts. The most common denoising strategies are based on three primary steps. To distinguish the useful information and the noise, the noisy signal or image is projected onto a dictionary. This is then accompanied by a hard or soft thresholding process in the transformed space. Finally, the revised coefficients are back projected to the time or space domain, so that the denoised signal or image could be retrieved. We will apply the same procedure using the adaptive basis defined by the eigenvectors $\bpsi_{i}$ obtained by solving the Schroedinger equation \eqref{eq:Schroedinger}. 

The basic assumptions is that the noise is more present in high frequency components of the signal or image, corresponding to eigenvectors associated with large energy eigenvalues. The thresholding will therefore be performed in energy, leaving out the components of the signal or image on high energy eigenvectors. The fact that our basis has frequencies which vary depending on the position should be an asset, especially for signal or image dependent noise (e.g. Poisson noise). In the following, we will show that it is indeed the case in some examples of signals and images with various types of noise.

The denoising process unfolds as follows; for a noisy signal or image denoted by $\bsx$, the denoised signal or image is rebuilt through:
\begin{eqnarray}
\label{eq:recons}
\hat{\bsx}= \sum _{i = 1} ^{N^2} \balpha_{i}\bpsi_{i}\tau_{i},
\end{eqnarray}
\noindent with
\begin{eqnarray}
\label{eq:thr}
\tau_{i}= \left \{
   \begin{array}{r c l}
      1 &  & for \; i \leq s ,\\
      1 - \frac{i-s}{\rho} & & for \; i > s \;  and \; for \; 1 - \frac{i - s}{\rho} > 0 ,\\
      0 & & otherwise.
   \end{array}
   \right .
\end{eqnarray}
where $\balpha_i = \langle \bsx,\bpsi_i \rangle$ are the coefficients representing the signal or image $\bsx$ in the proposed adaptive basis. $s$ and $\rho$ are two hyperparameters, used to define the thresholding function for the proposed denoising algorithm. 
 
 In order to use this procedure, we will need to specify which values of the parameter $\hbar^2/2m$ should be selected. As we will see, there is a relatively large range of values where the algorithm is efficient, meaning that it can be set to a specific value independent of the signal or image on which the algorithm is used. 
 


\subsection{Algorithm description}
\label{sec:algorithm}

Denoising a signal or an image using the proposed method requires the computation of eigenvalues and eigenvectors of the discretized Hamiltonian matrix \eqref{eq:H} for appropriate values of the parameters  $\hbar^2/2m$ and $\sigma$, project the signal or image on this basis, threshold the coefficients by an appropriate threshold in energy, and reconstruct from this a denoised signal or image. These steps are summarized in Algorithm~\ref{Algo:Denoising} and Fig.~\ref{fig:sample2} \footnote{The Matlab code of the proposed denoising algorithm is available at \href{https://github.com/SayantanDutta95/QAB-denoising.git}{github.com/SayantanDutta95/QAB-denoising.git}}.

\begin{algorithm}[h!]
\label{Algo:Denoising}
\KwIn{$\bsx$, $\frac{\hbar ^2}{2m}$, $s$, $\rho$, $\sigma$}
 {Compute a smooth version of $\bsx$ by Gaussian filtering}\\
 {Construct the Hamiltonian matrix $H$ based on the smoothed version of $\bsx$} using \eqref{eq:H}\\
 {Calculate the eigenvectors $\bpsi_{i}$ of $H$}\\
 {Compute the coefficients $\balpha_{i}$ by projecting $\bsx$ onto the basis formed by $\bpsi_{i}$}\\ 
 {Threshold the coefficients $\balpha_{i}$ and recover the denoised signal or image following \eqref{eq:thr} and \eqref{eq:recons}}\\
\KwOut{$\hat{\bsx}$}
\caption{Denoising algorithm using the proposed adaptive transform.}
\DecMargin{1em}
\end{algorithm}

For very large signals and images, where the size of the matrix \eqref{eq:H} becomes too large for practical simulations, we implement a modified version of the algorithm where the matrix \eqref{eq:H} is diagonalized for subparts of the signal or image independently, and then a complete signal or image is reconstructed:
\begin{itemize}
   \item The noisy signal or image is divided into sub-blocks of equal size, using in particular square sub-blocks in the case of images.
   \item Use algorithm~\ref{Algo:Denoising} for each sub-block.
   \item Reconstruct the denoised signal or image by integrating each denoised sub-block.
\end{itemize}

\begin{figure*}[h!]
	\centering
	\subfigure[Hyperparameter $\hbar^2/2m = 0.08$]{\includegraphics[width=0.325\textwidth]{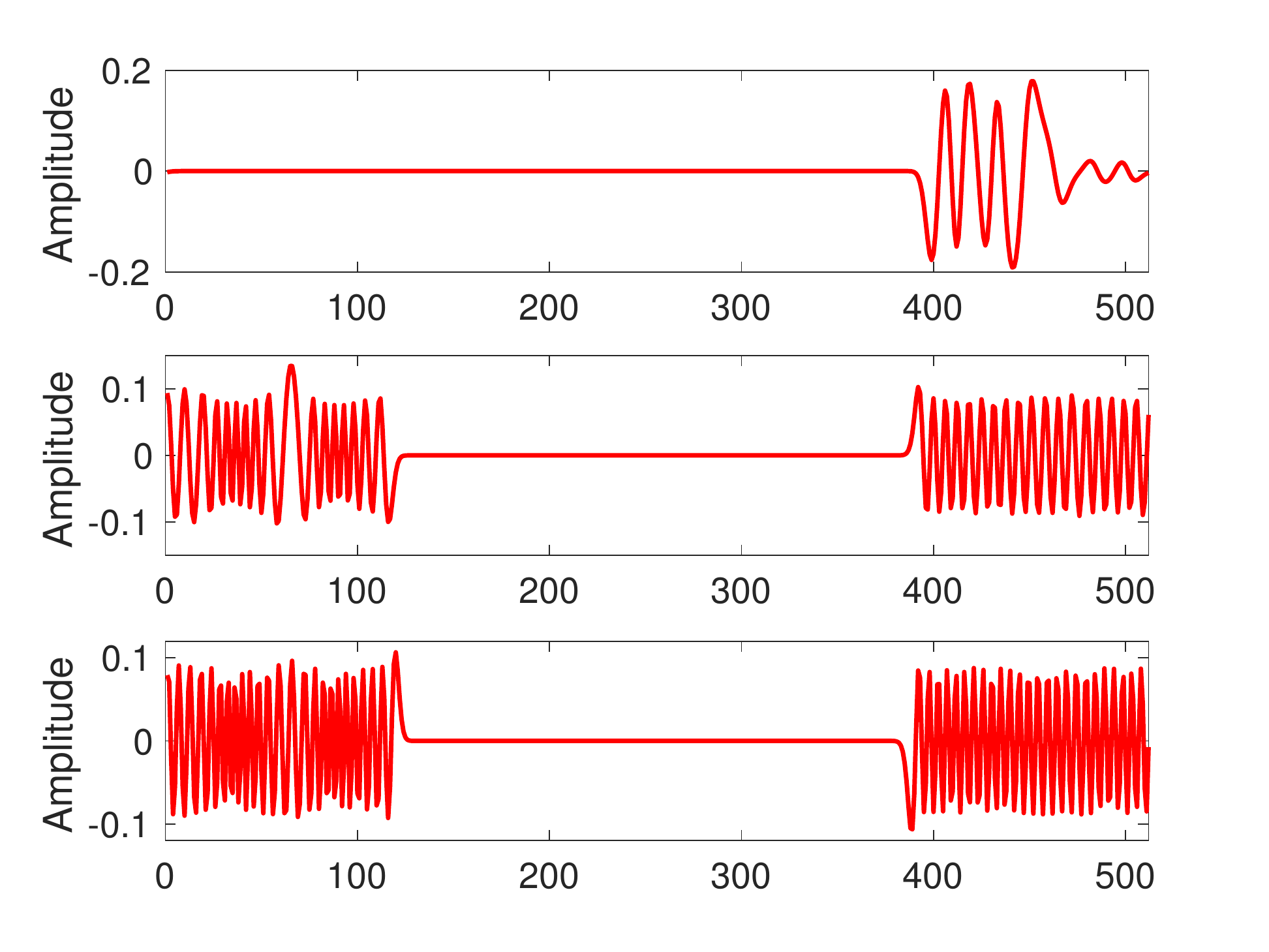}}
	\subfigure[Hyperparameter $\hbar^2/2m = 1$ ]{\includegraphics[width=0.325\textwidth]{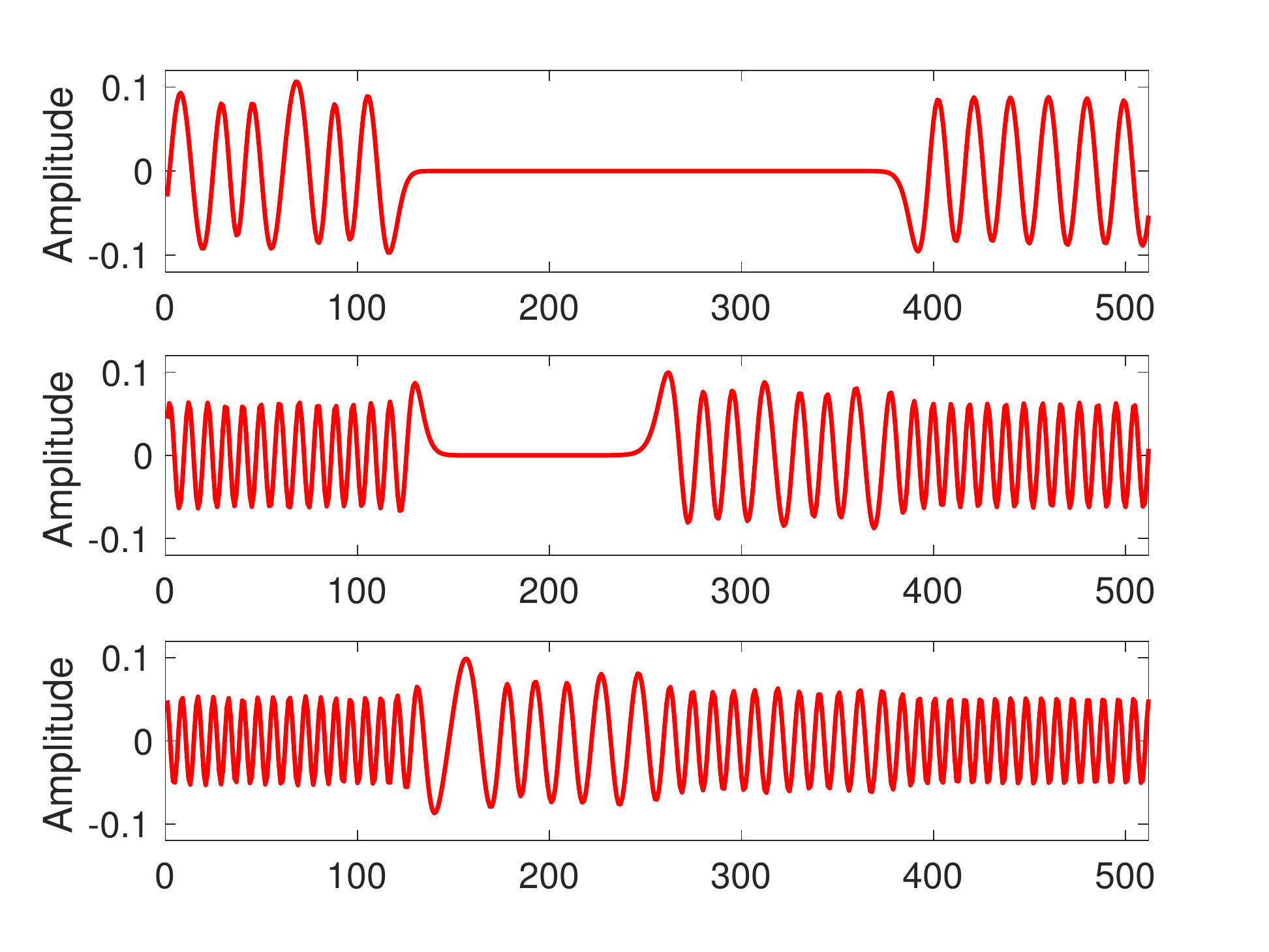}}
	\subfigure[Hyperparameter $\hbar^2/2m = 15$ ]{\includegraphics[width=0.325\textwidth]{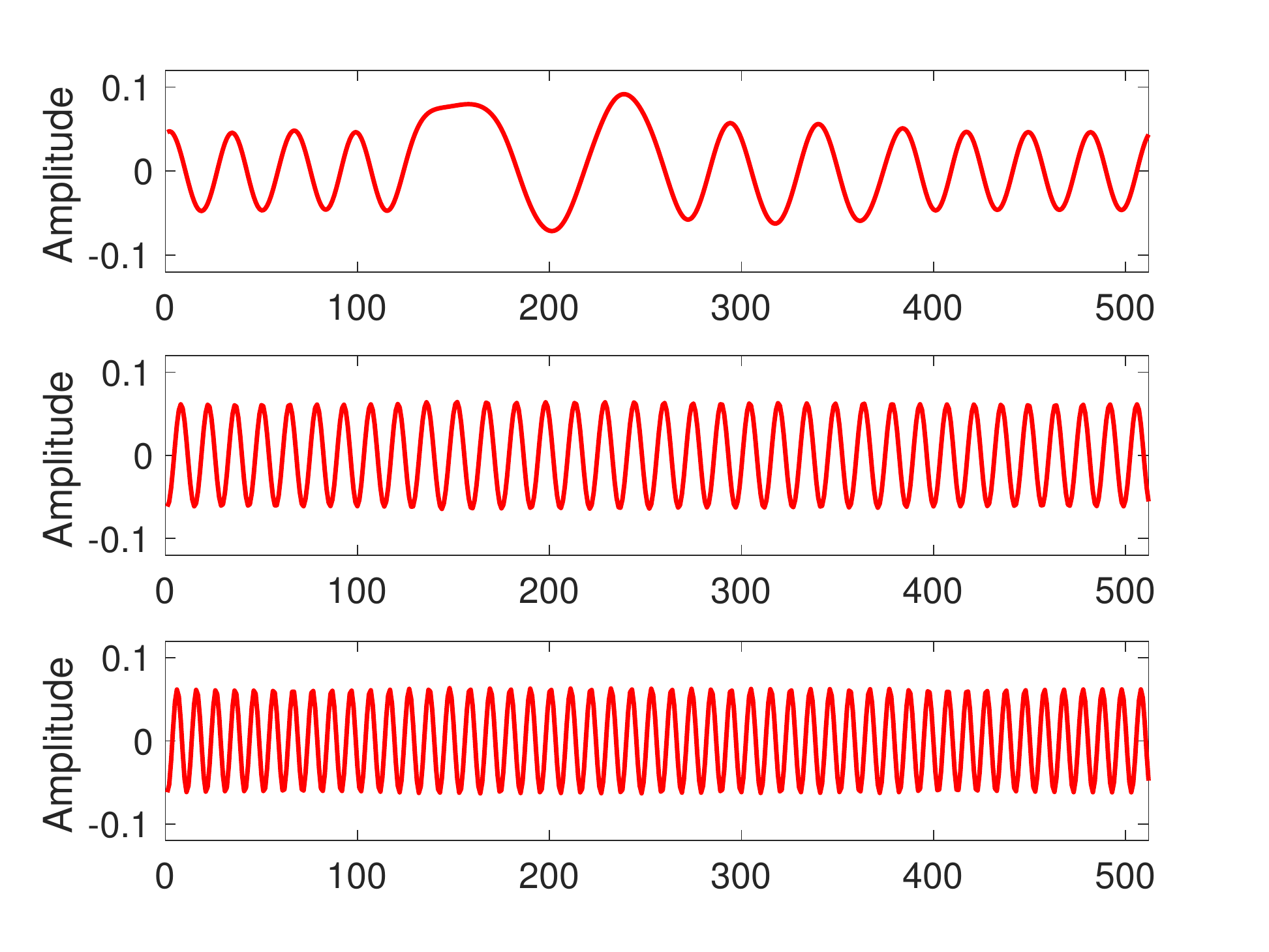}}
	
	\caption{Role of the hyperparameter $\hbar^2/2m$: adaptive basis functions (wave functions) number $25$, $70$ and $100$ calculated from the signal Fig.~\ref{fig:sample} are shown from top to bottom with different values of the hyperparameter $\hbar^2/2m$.}
	\label{fig:parameter_h}
\end{figure*}

\begin{figure*}[b!]

  \centering
   \subfigure[]{\includegraphics[width=0.245\textwidth]{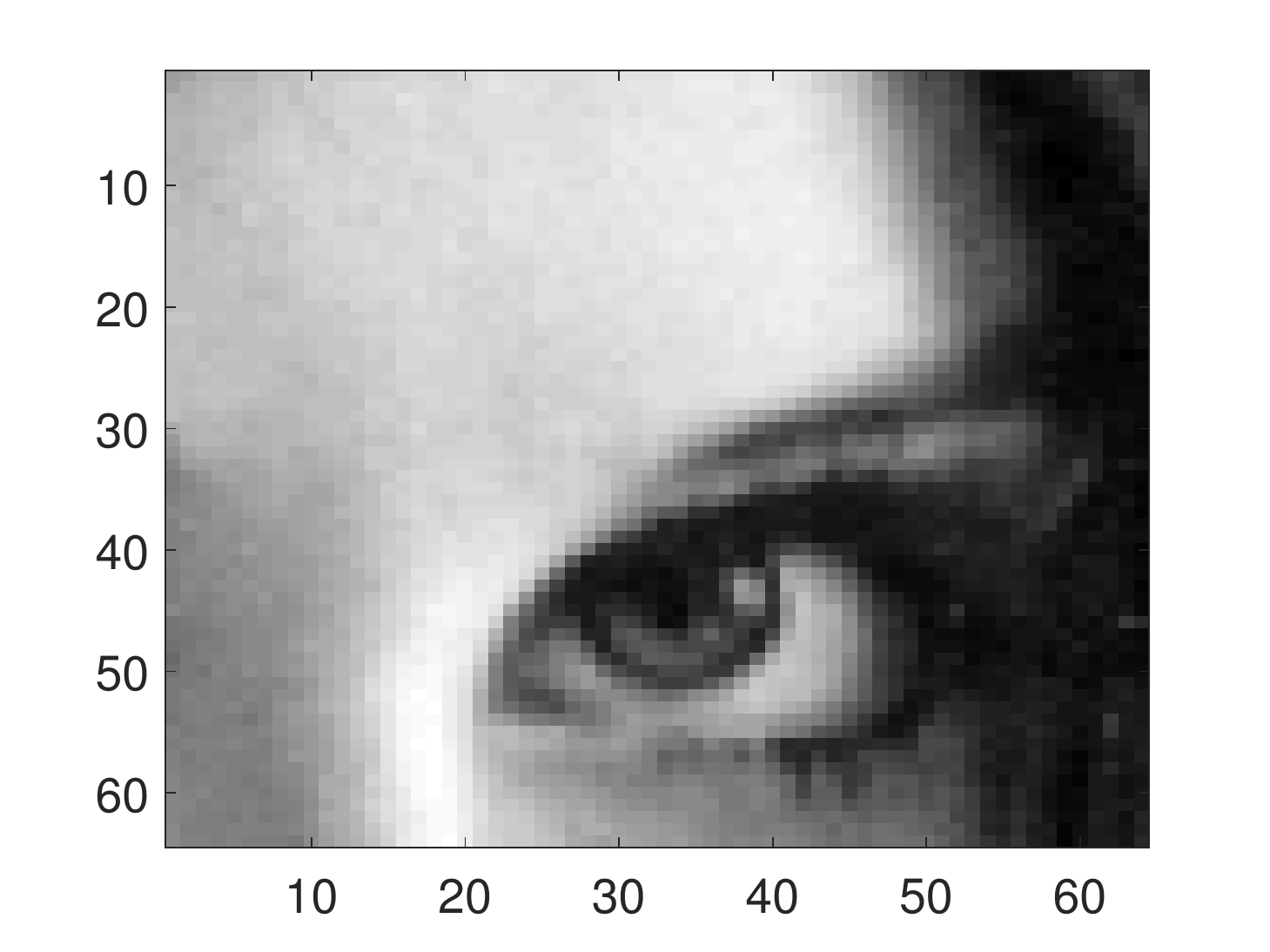}}
   \subfigure[]{\includegraphics[width=0.245\textwidth]{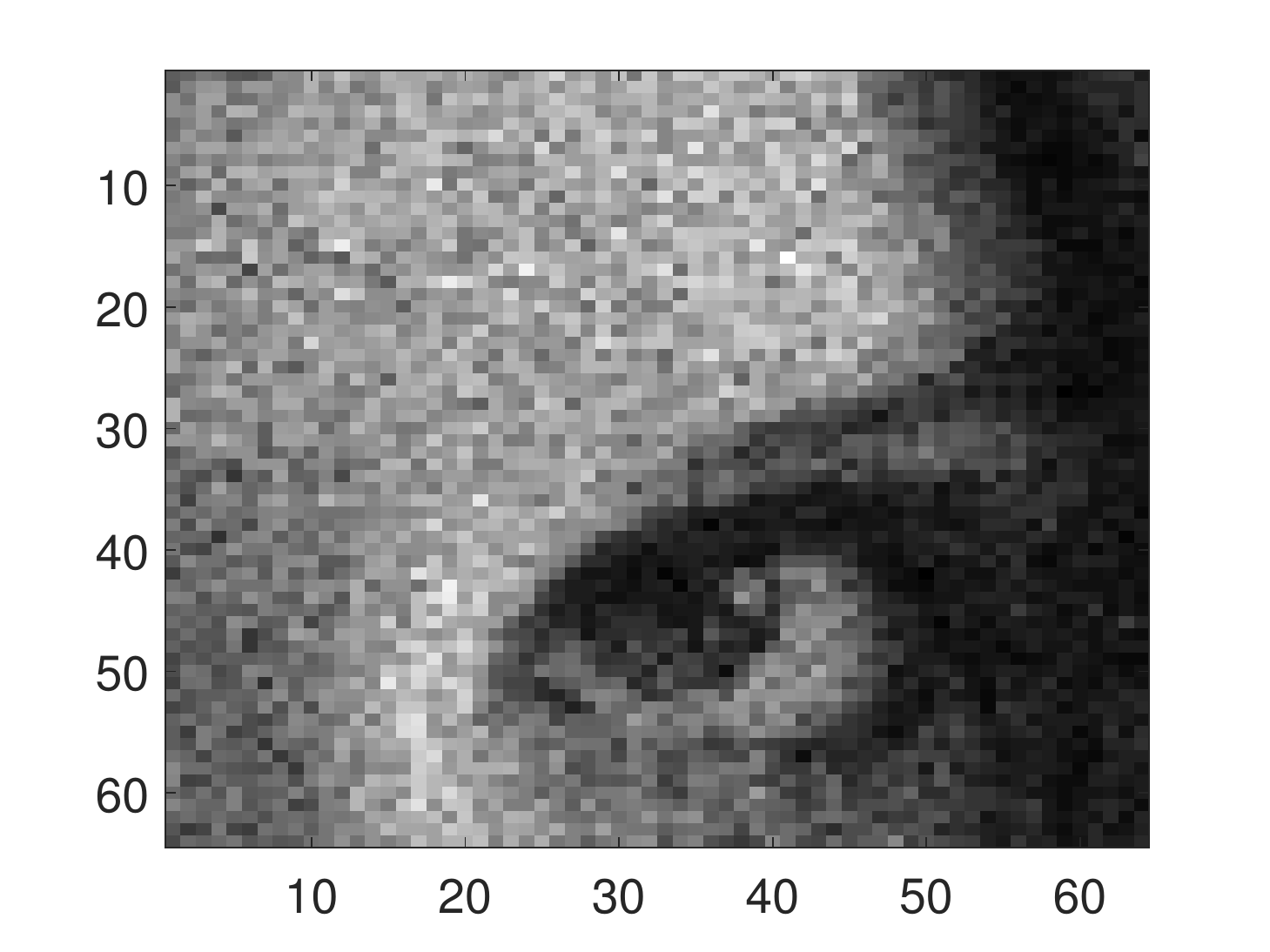}} 
   \subfigure[]{\includegraphics[width=0.245\textwidth]{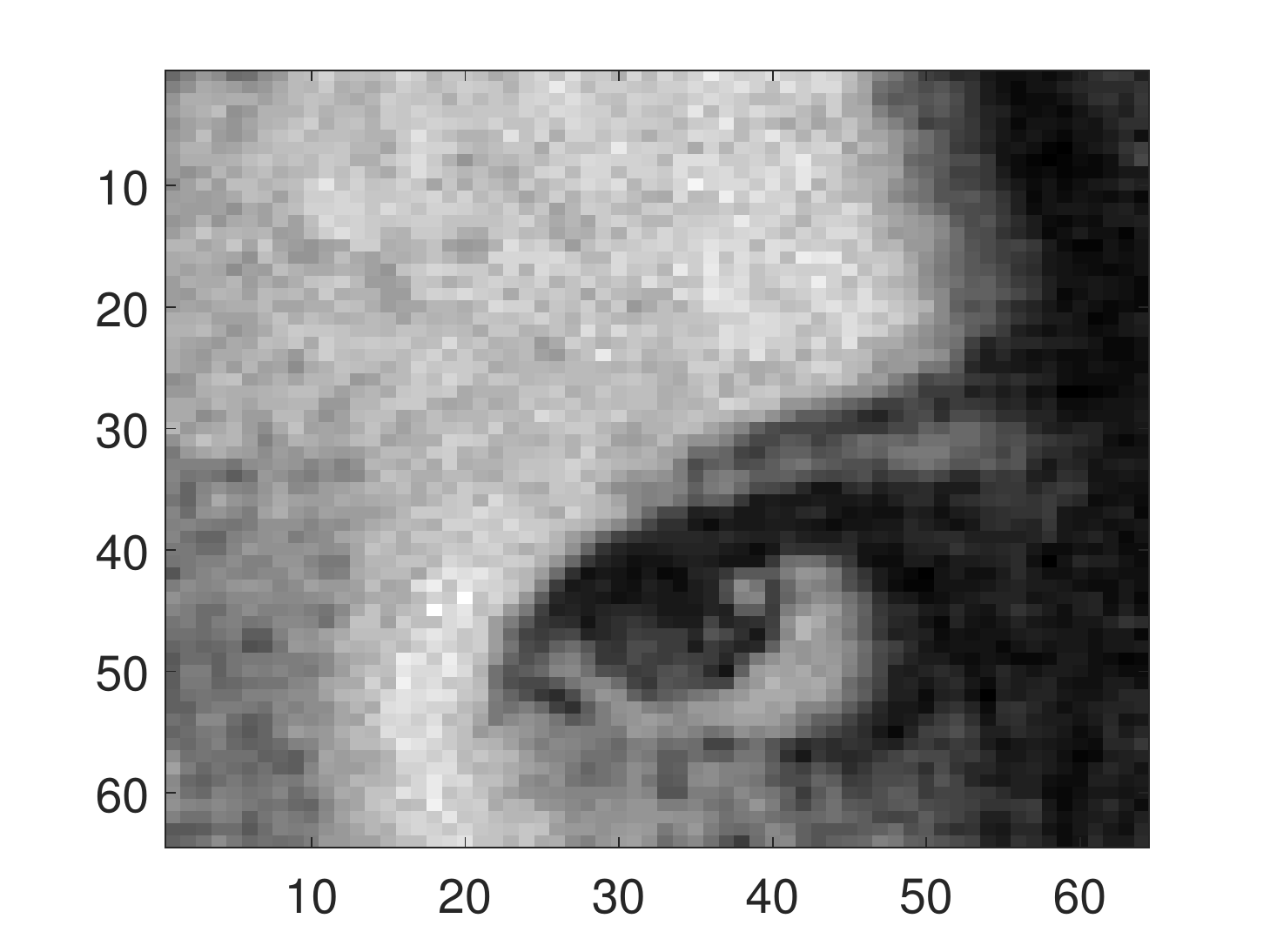}}
   \subfigure[]{\includegraphics[width=0.245\textwidth]{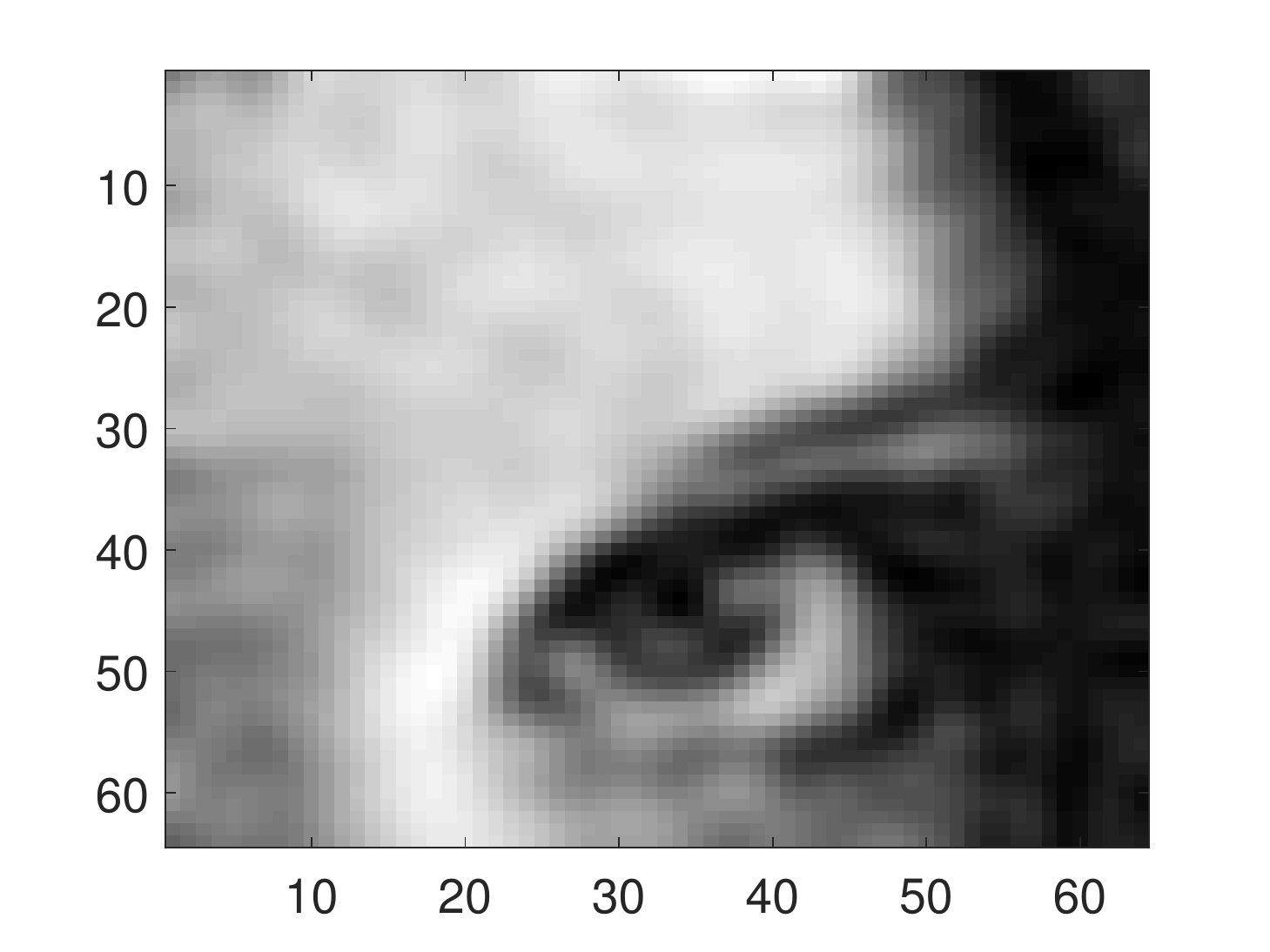}}
   
   \caption{Role of the hyperparameter $\sigma$: (a) Cropped version of clean Lena, (b) cropped version of noisy Lena contaminated by Poisson noise corresponding to a SNR of 15 dB, (c) denoised result with hyperparameter $\sigma^2=0$, giving a PSNR = 25.37 dB, of the image (b), (d) denoised result with hyperparameter $\sigma^2=4$, giving a PSNR = 28.81 dB. The hyperparameters are $\hbar^2/2m = 0.6$, $\rho = 1$, and $s = 600$ for each set of experiment.}
\label{fig:localization_lena}
\end{figure*}

\section{Results}
\label{sec:results}

This section regroups results showing the interest of the proposed approach in signal and image denoising and analyze the optimal choice of parameters. 
Subsection~\ref{sec:infparameters} elaborates the dependence of the proposed denoising method on the choice of the hyperparameters $ \hbar^2/2m $, $ \sigma $, $s$ and $\rho$. Subsection~\ref{sec:denoisingresults} compares the denoising results obtained with the proposed approach to several state of the art methods. Finally, the section ends with an example of real medical application in Subsection~\ref{sec:medicalimage}, showing the ability of the proposed method to denoise real world (dental) cone beam computed tomography (CBCT) images.

\subsection{Influence of hyperparameters $\hbar^2/2m$, $\sigma$, $s$ and $\rho$ on the efficiency of the algorithm}
\label{sec:infparameters}

\begin{figure*}[b!]
	\centering
	\subfigure[PSNR as a function of the thresholding hyperparameter $s$, for $\sigma^2 = 20$, $\rho = 1$ and $\hbar^2/2m = 0.5$.]{\includegraphics[width=0.42\textwidth]{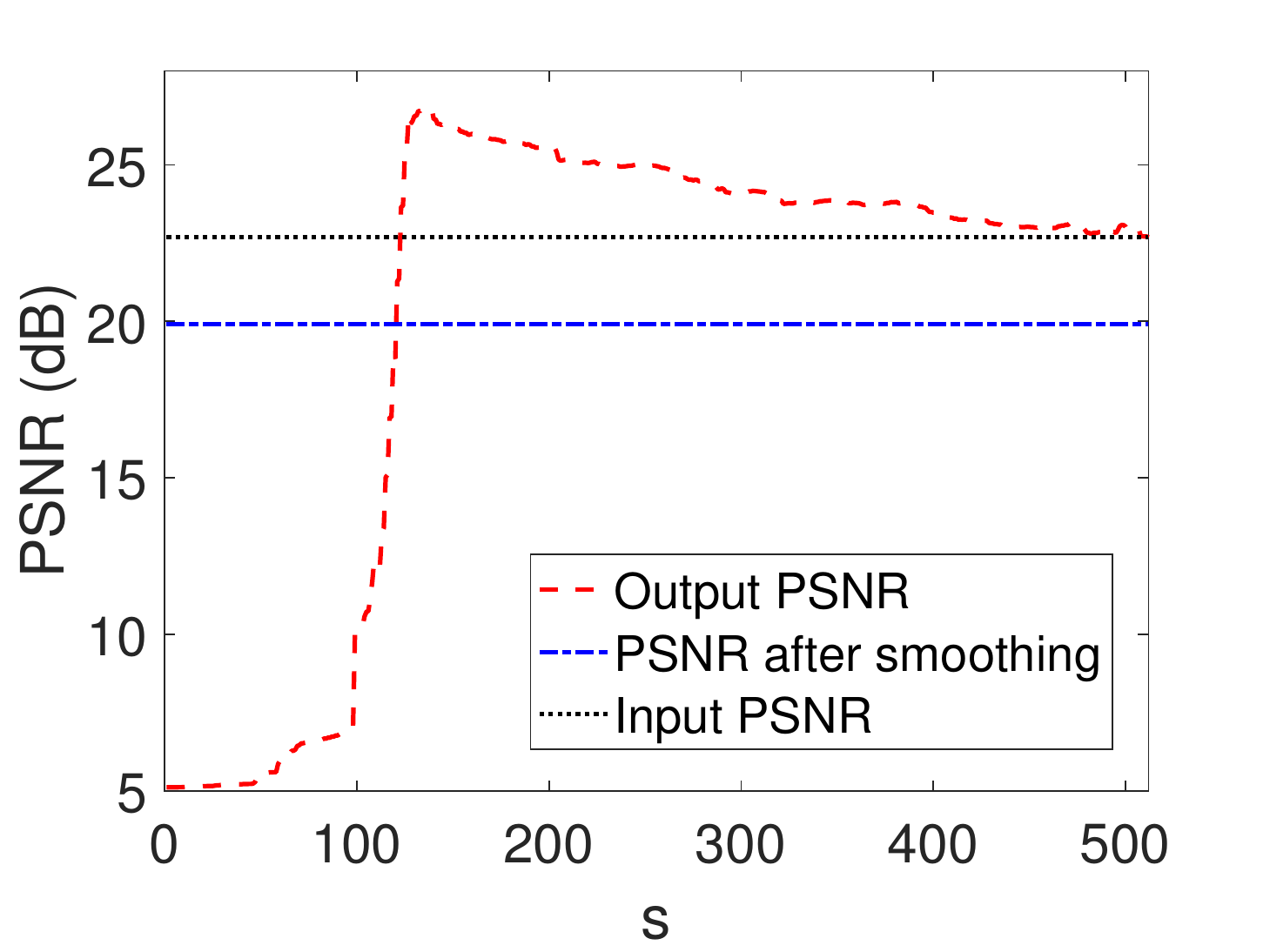}}
	\subfigure[SNR as a function of the thresholding hyperparameter $s$, for $\rho = 1$, $\hbar^2/2m = 0.4$ and four different values of $\sigma^2$ (0, 4, 8 and 40).]{\includegraphics[width=0.42\textwidth]{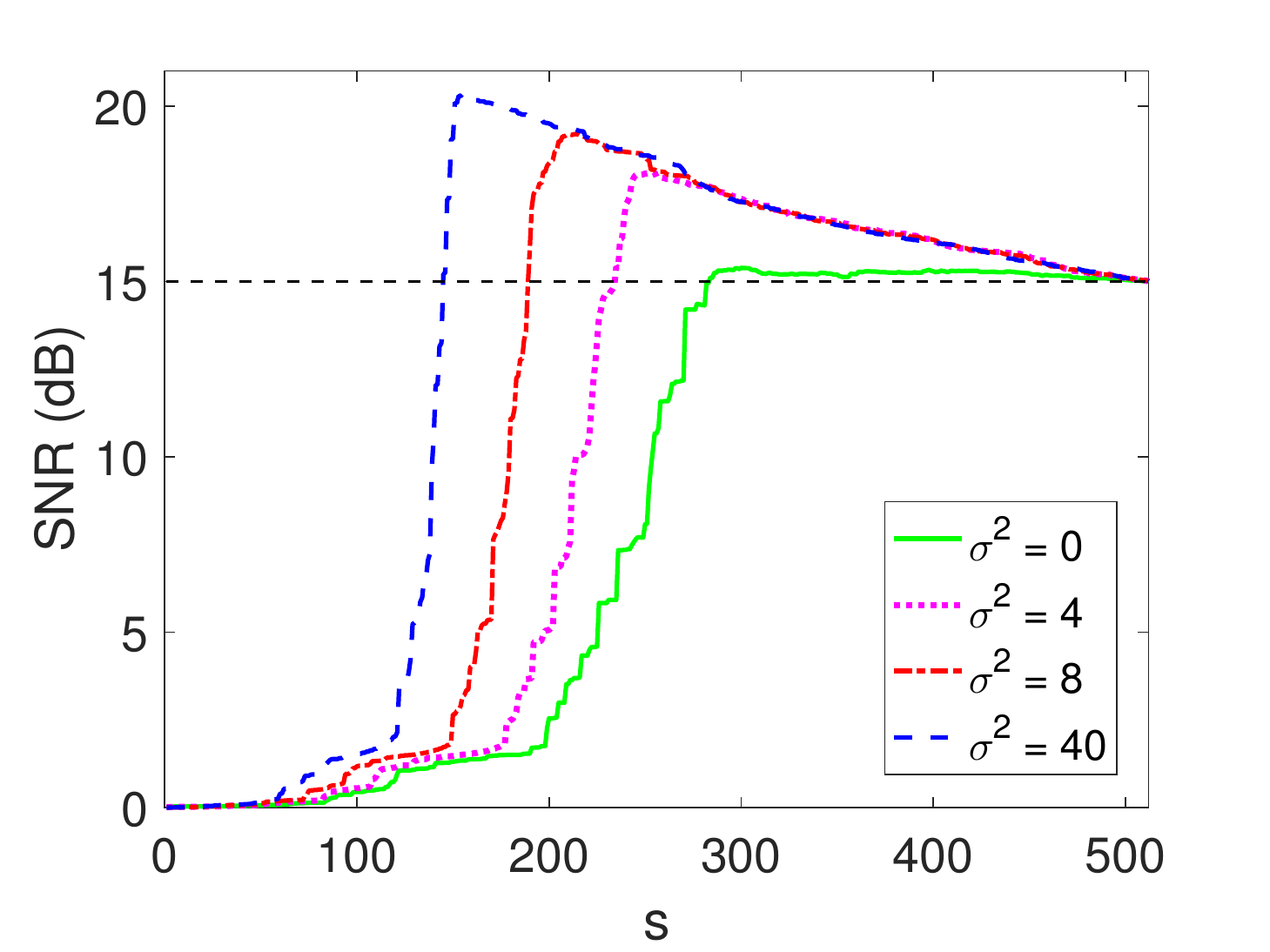}}
	
	\caption{Role of hyperparameters $s$ and $\sigma$. Simulations with  the 1D signal Fig.~\ref{fig:sample} corrupted by additive Gaussian noise corresponding to a SNR of 15 dB.}
	\label{fig:cutoff}
\end{figure*}


\begin{figure*}[h!]
	\centering
	\subfigure[]{\includegraphics[width=0.365\textwidth]{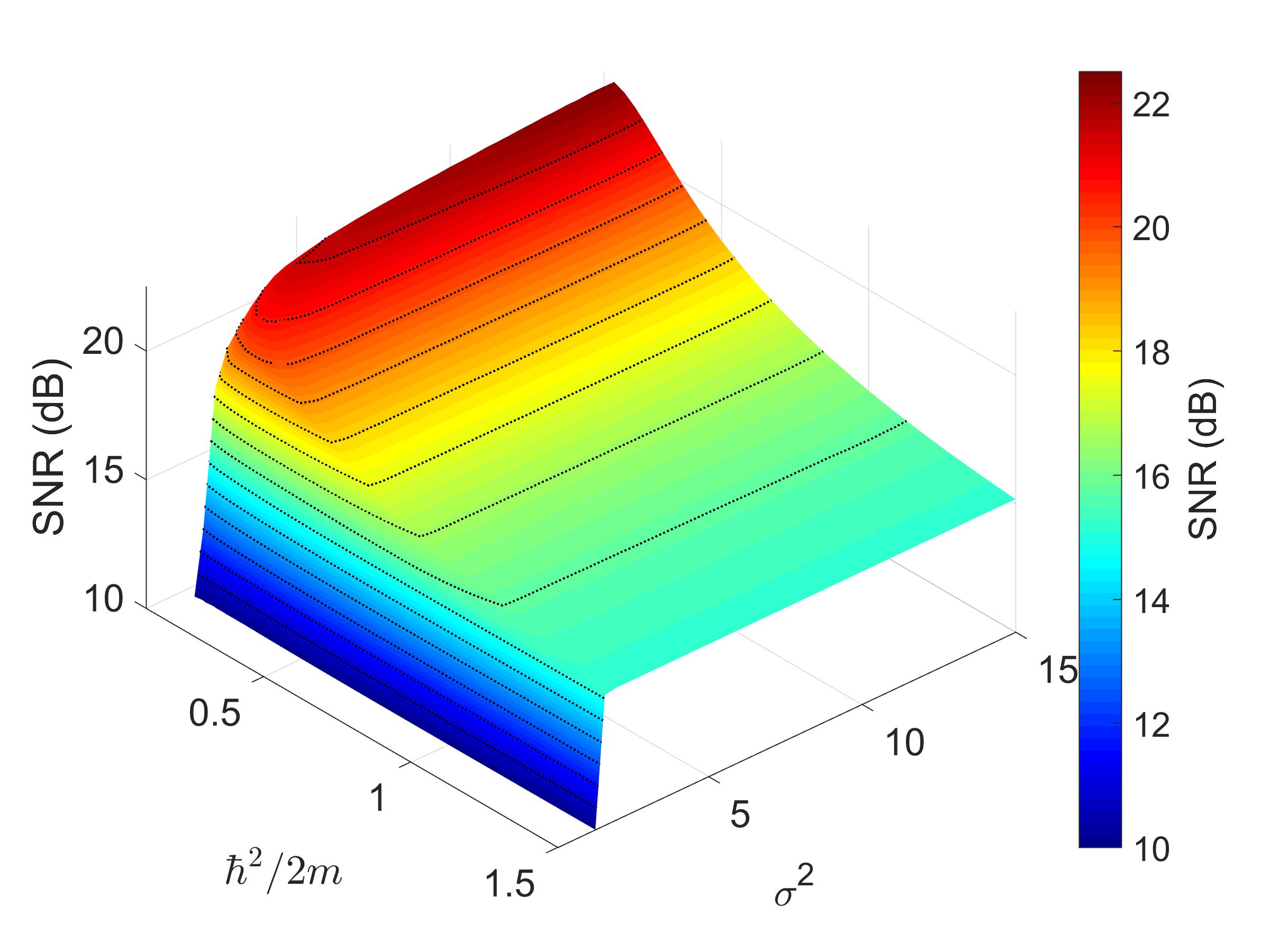}}
	\subfigure[]{\includegraphics[width=0.365\textwidth]{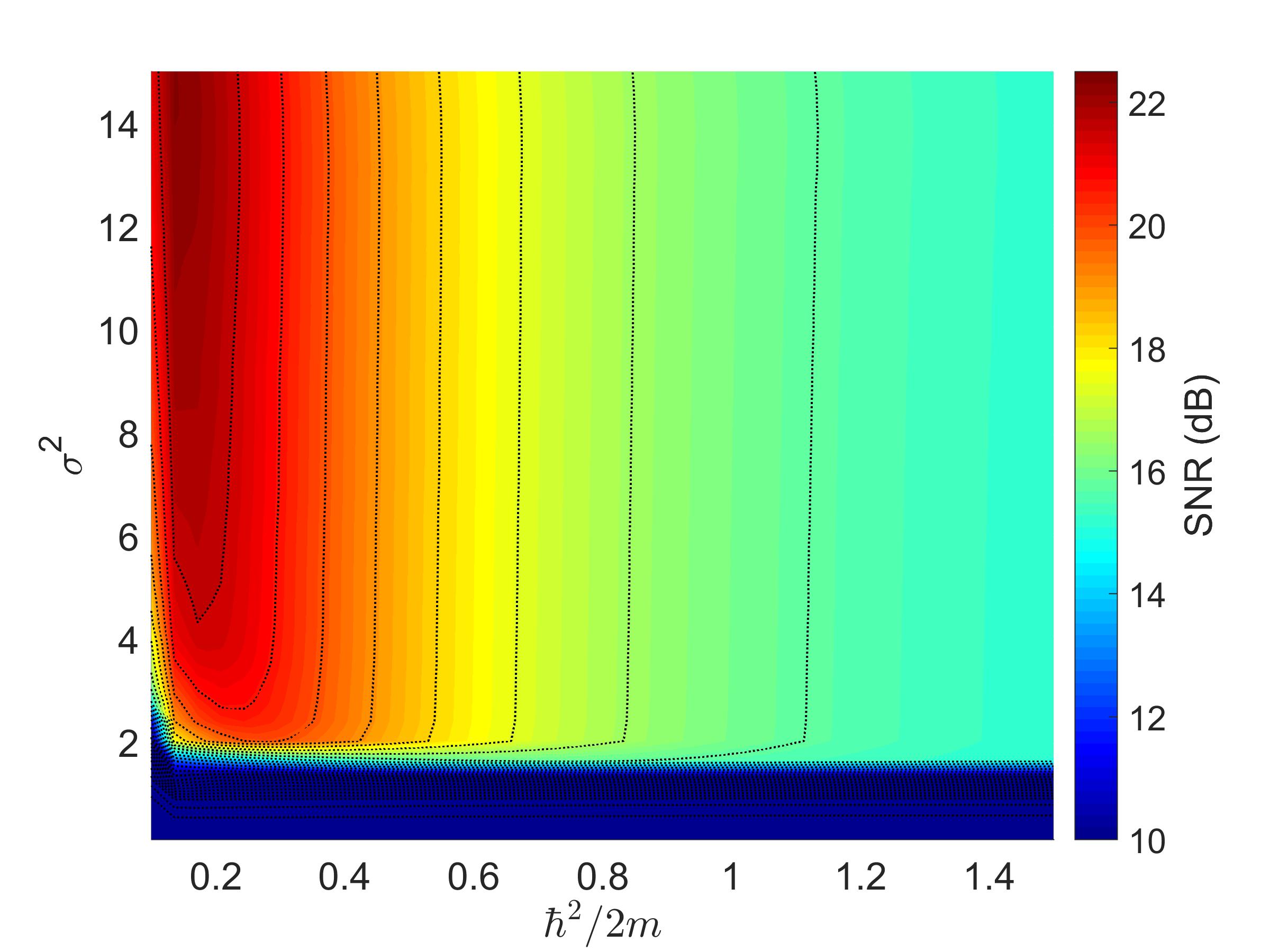}}
	\subfigure[]{\includegraphics[width=0.365\textwidth]{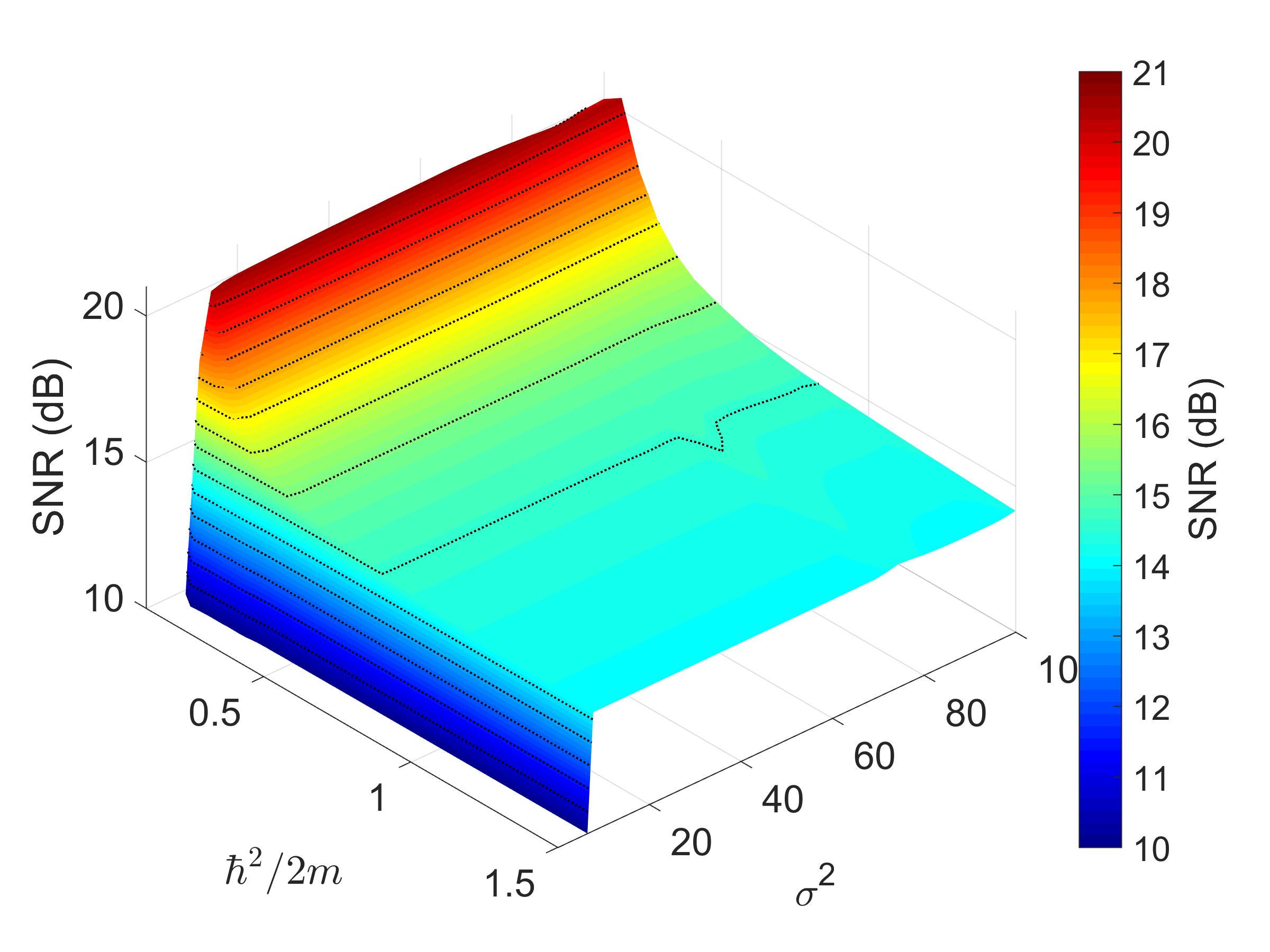}}
	\subfigure[]{\includegraphics[width=0.365\textwidth]{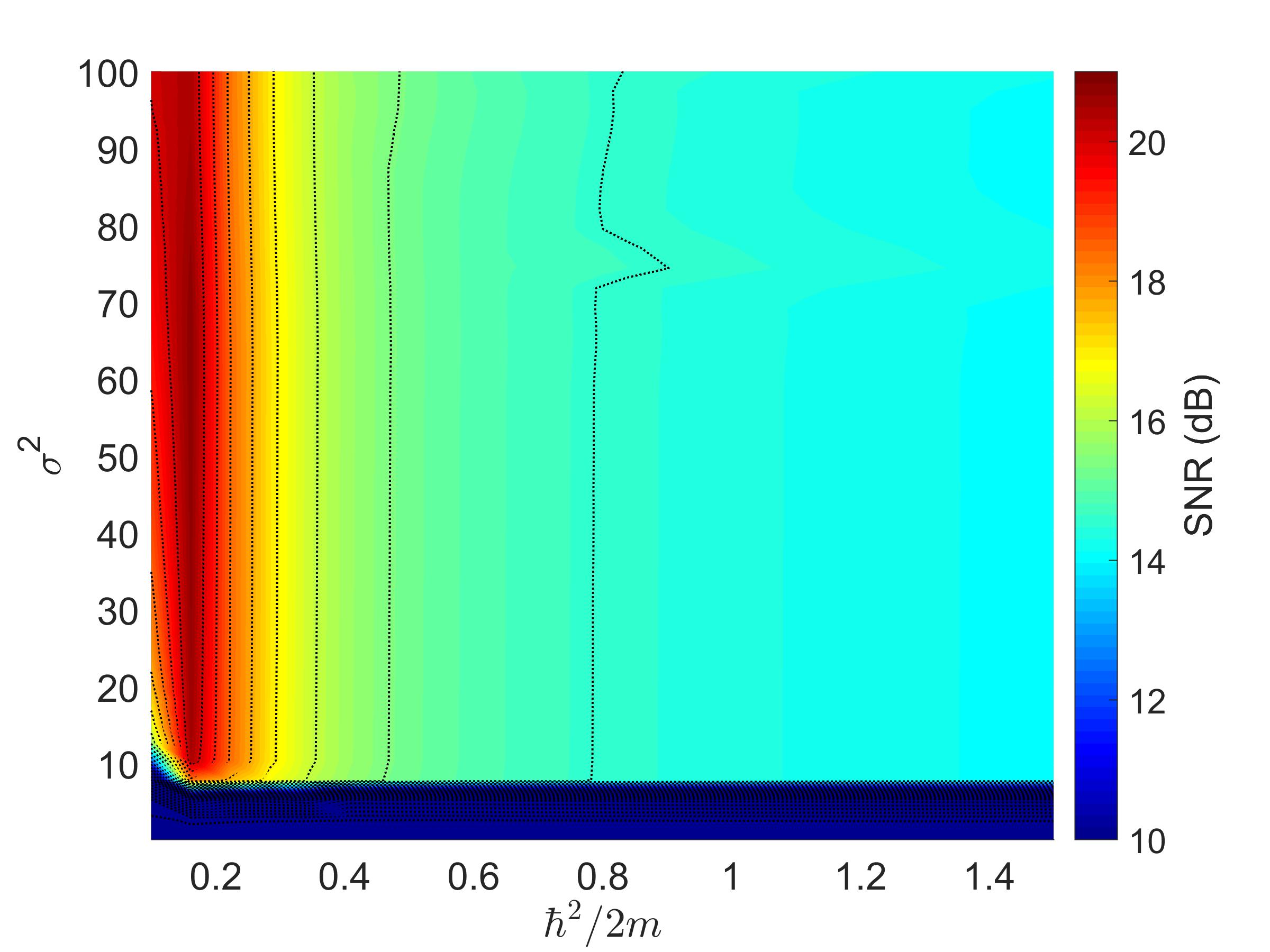}}
	\subfigure[]{\includegraphics[width=0.365\textwidth]{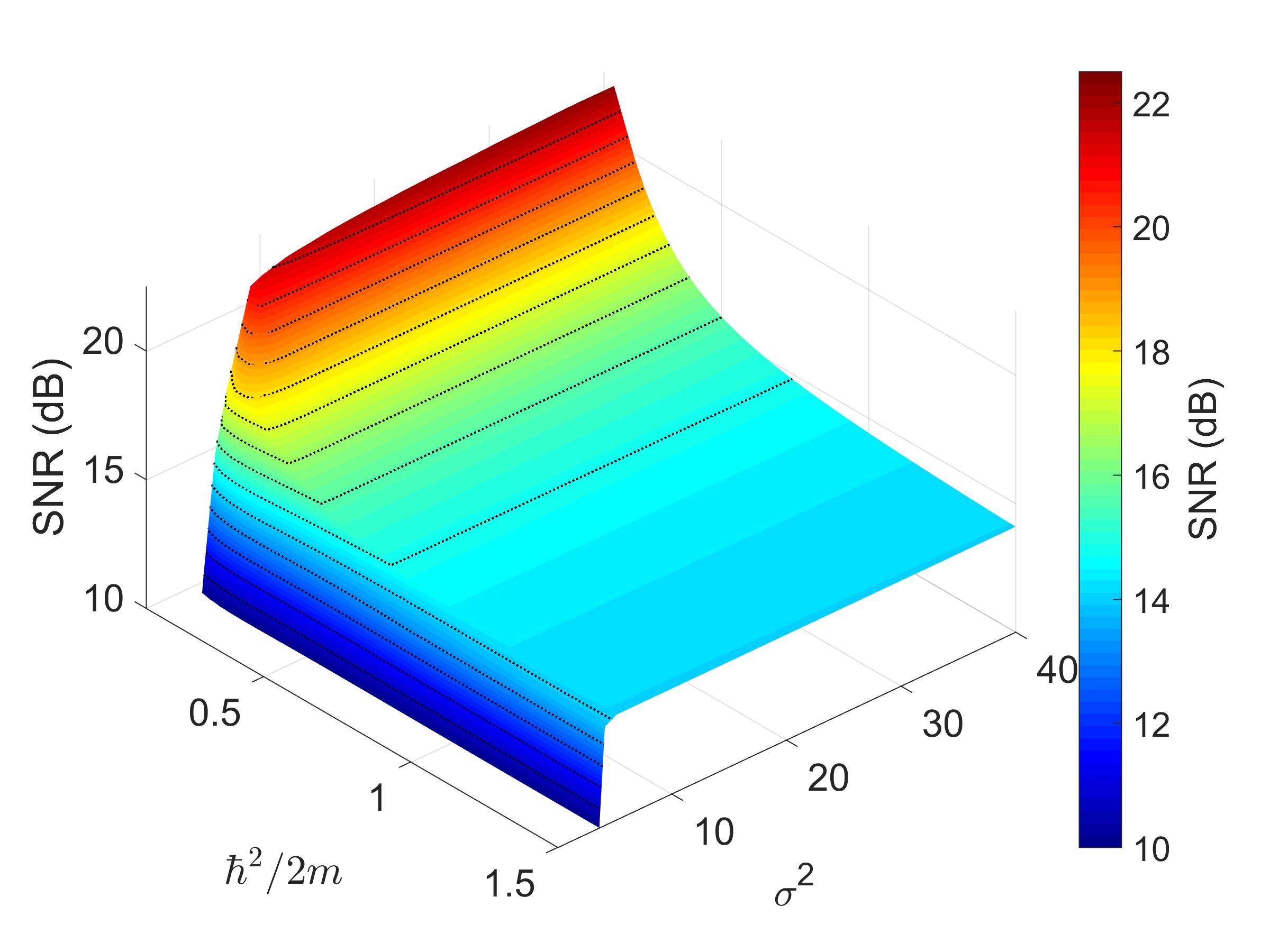}}
	\subfigure[]{\includegraphics[width=0.365\textwidth]{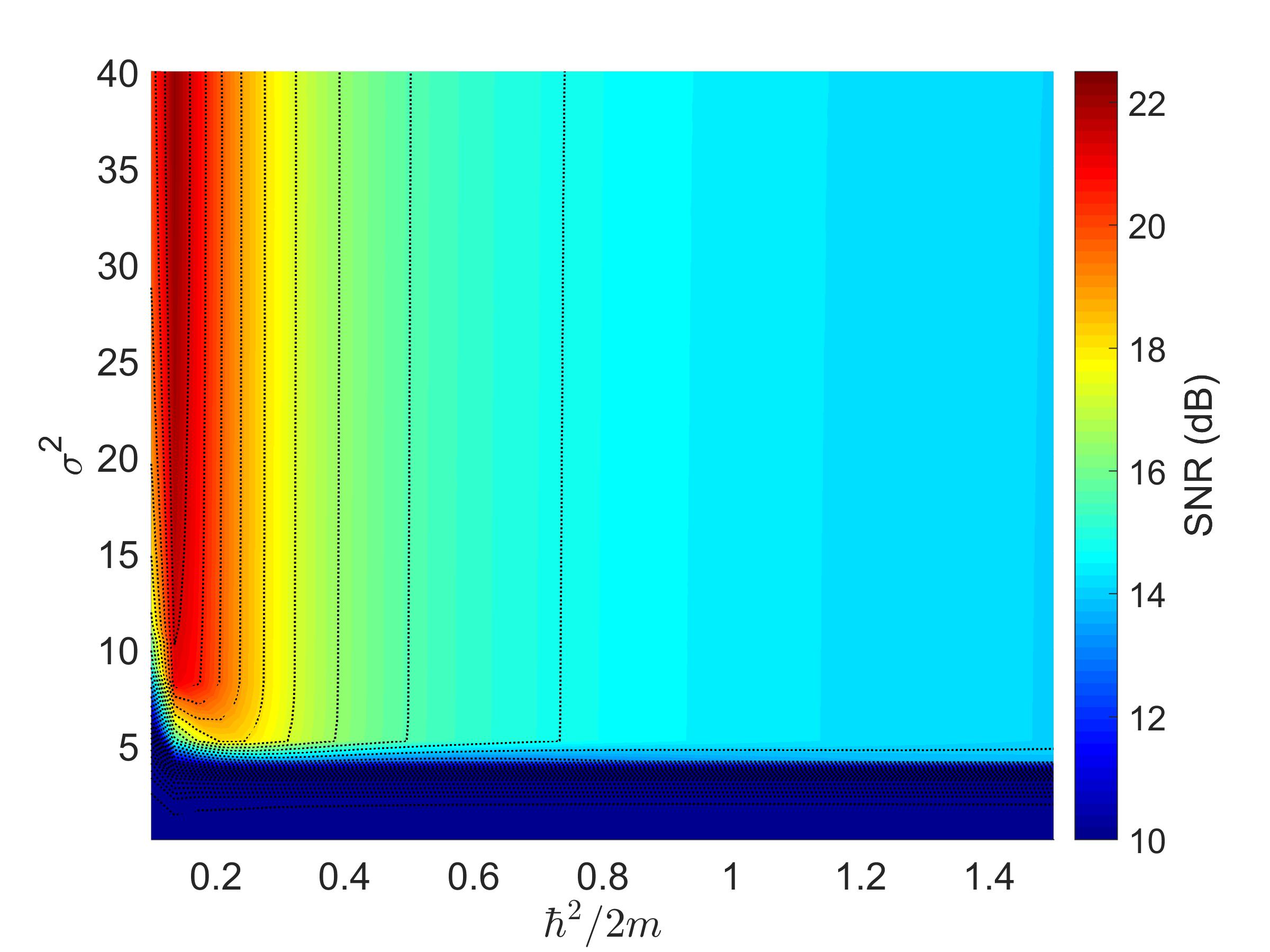}}
	
	\caption{ Influence of the hyperparameters $\hbar^2/2m$ and $\sigma$ on proposed decomposition performed on the 1D system Fig.~\ref{fig:sample} in presence of (a,b) Poisson noise, (c,d) Gaussian noise and (e,f) speckle noise corresponding to a SNR of 15 dB respectively. The hyperparameters are $\rho = 1$ and $s = 110$ for each set of experiment.}
	\label{fig:1Dparameter}
\end{figure*}

\begin{figure*}[h!]
	\centering
	\subfigure[]{\includegraphics[width=0.22\textwidth]{Figure/figures/SampleA.pdf}}
	\subfigure[]{\includegraphics[width=0.27\textwidth]{Figure/parameter/Parameter1Dpoisson.jpg}}
	\subfigure[]{\includegraphics[width=0.27\textwidth]{Figure/parameter/Parameter1Dpoisson2D.jpg}}\\   \BlankLine
	\subfigure[]{\includegraphics[width=0.22\textwidth]{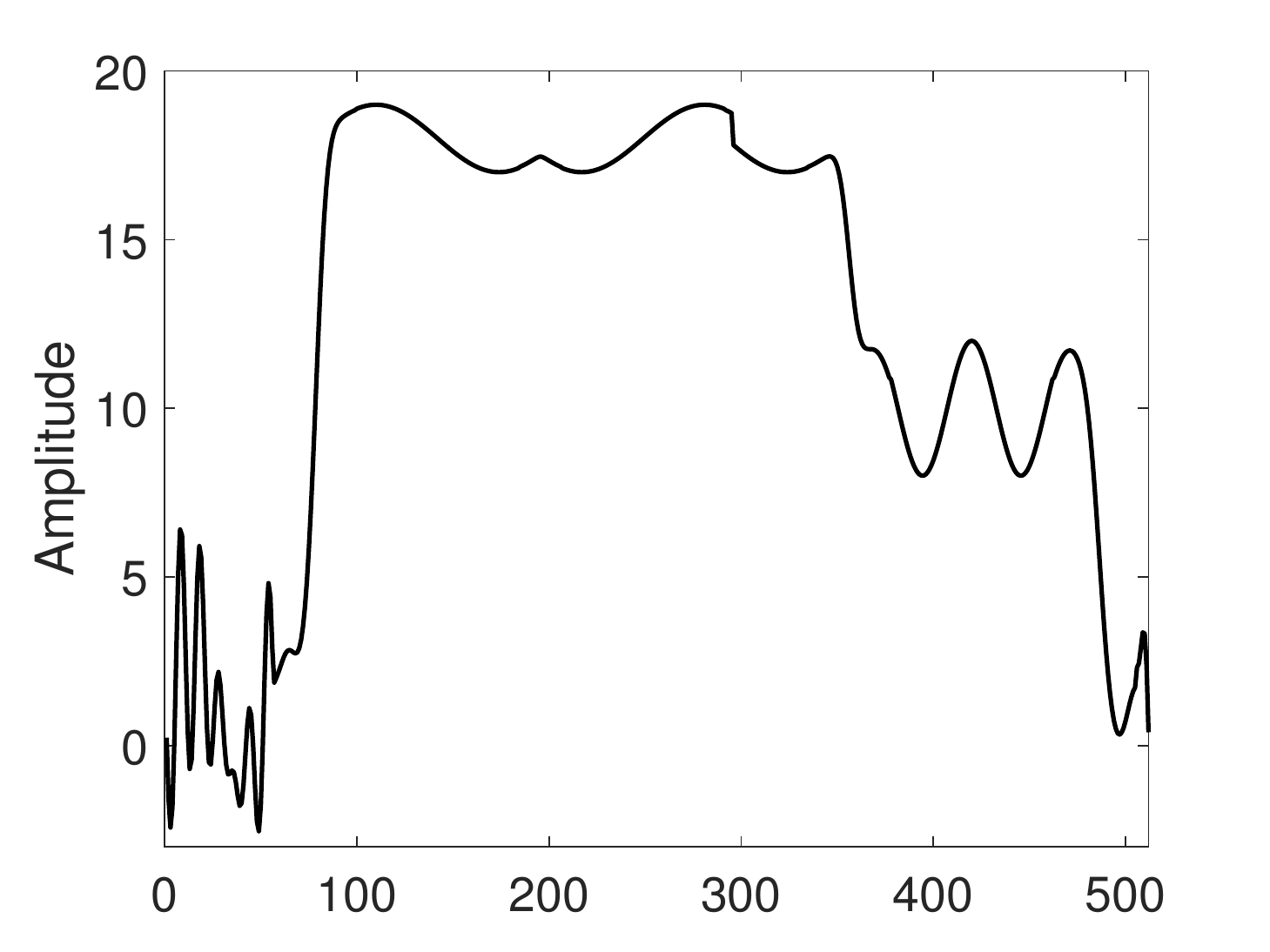}}
	\subfigure[]{\includegraphics[width=0.27\textwidth]{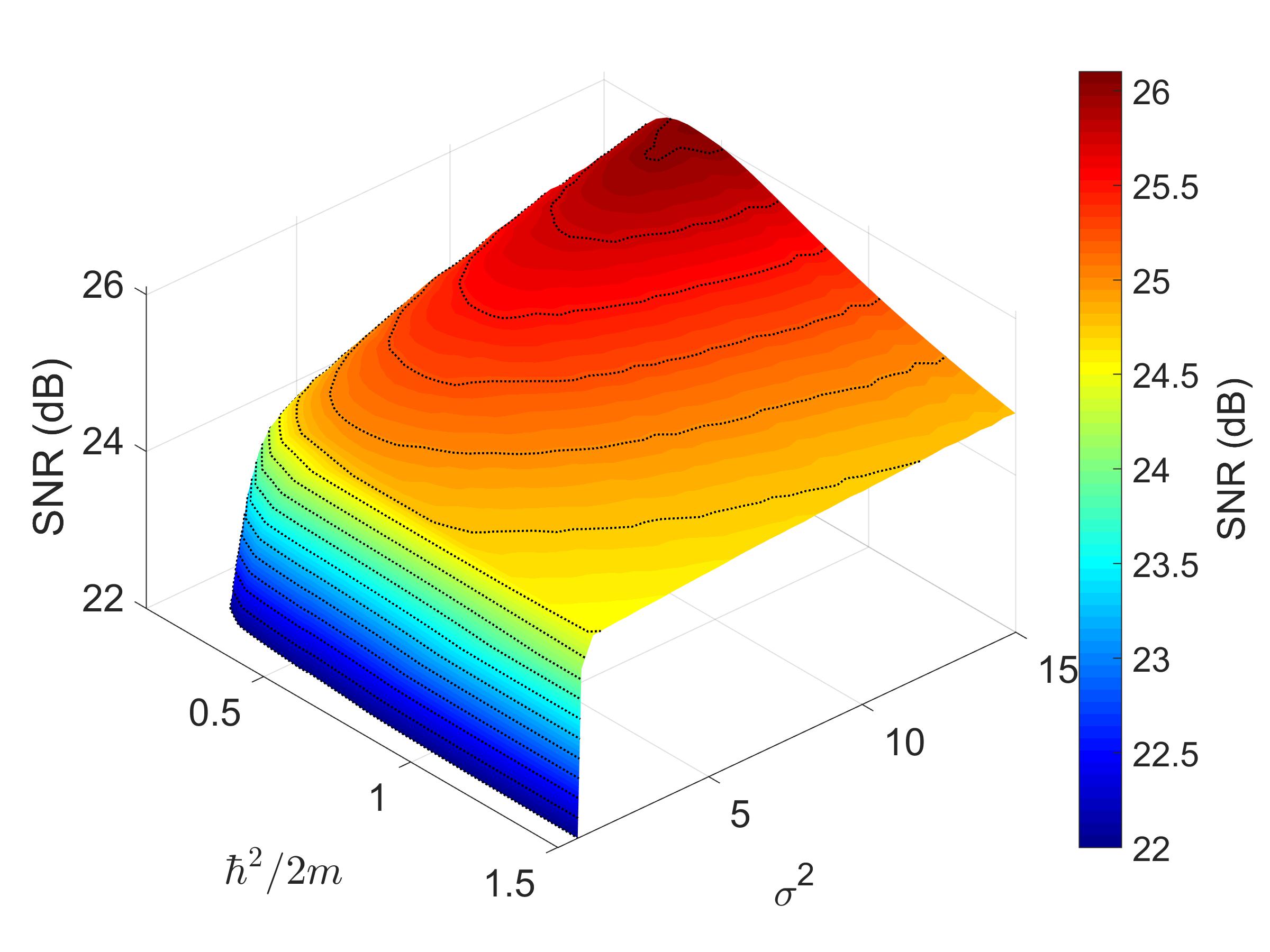}}
	\subfigure[]{\includegraphics[width=0.27\textwidth]{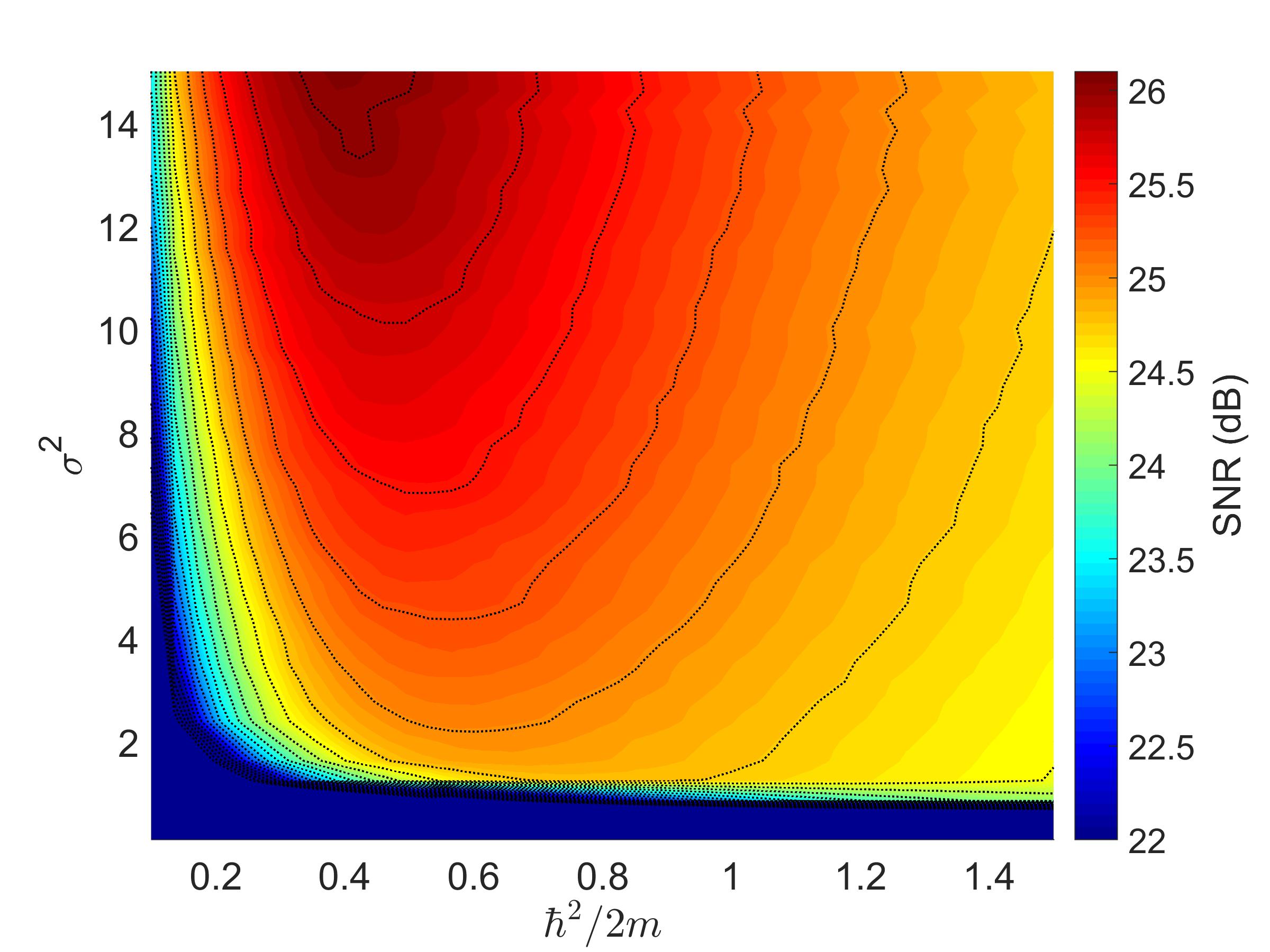}}\\   \BlankLine
	\subfigure[]{\includegraphics[width=0.22\textwidth]{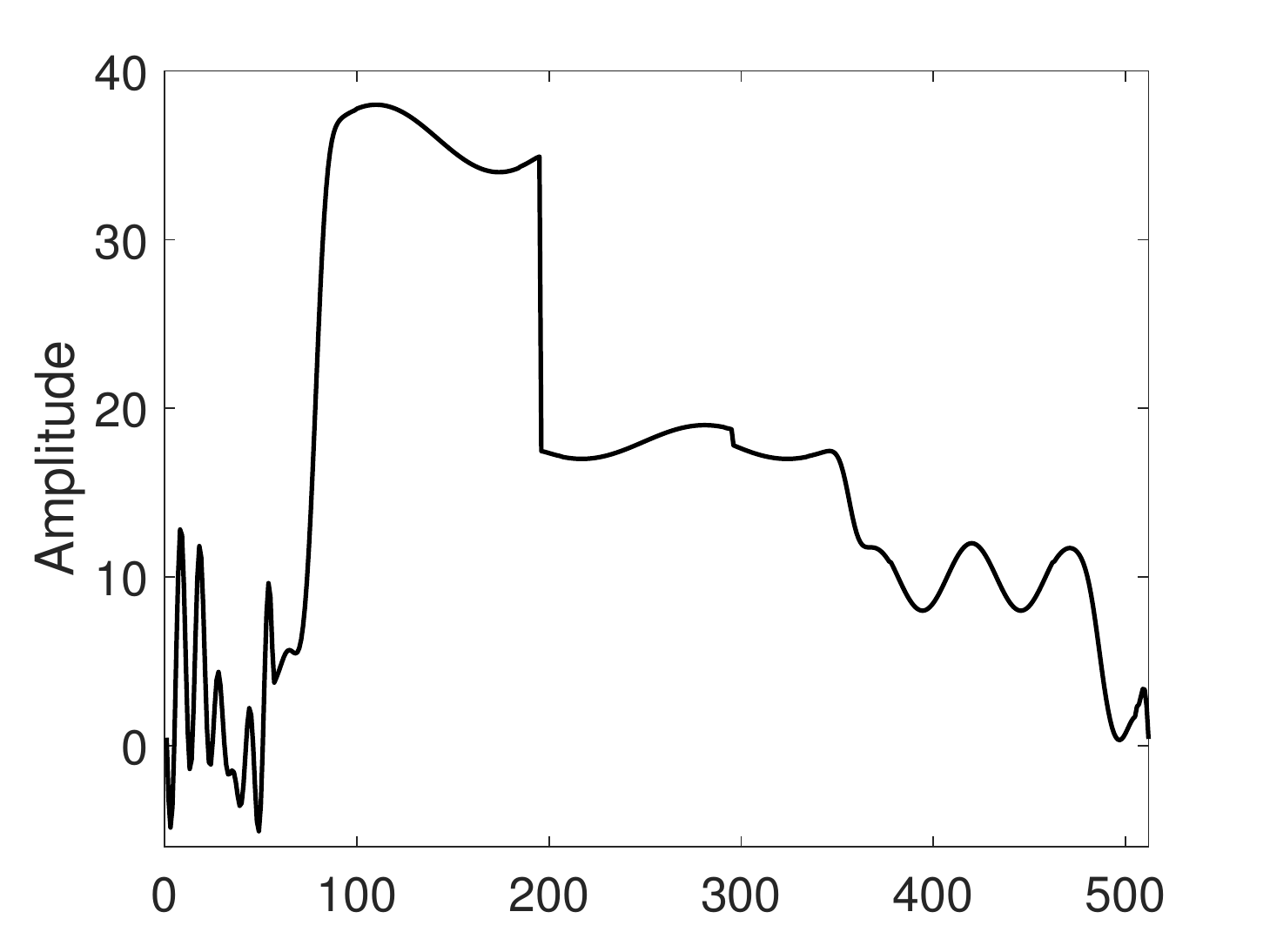}}
	\subfigure[]{\includegraphics[width=0.27\textwidth]{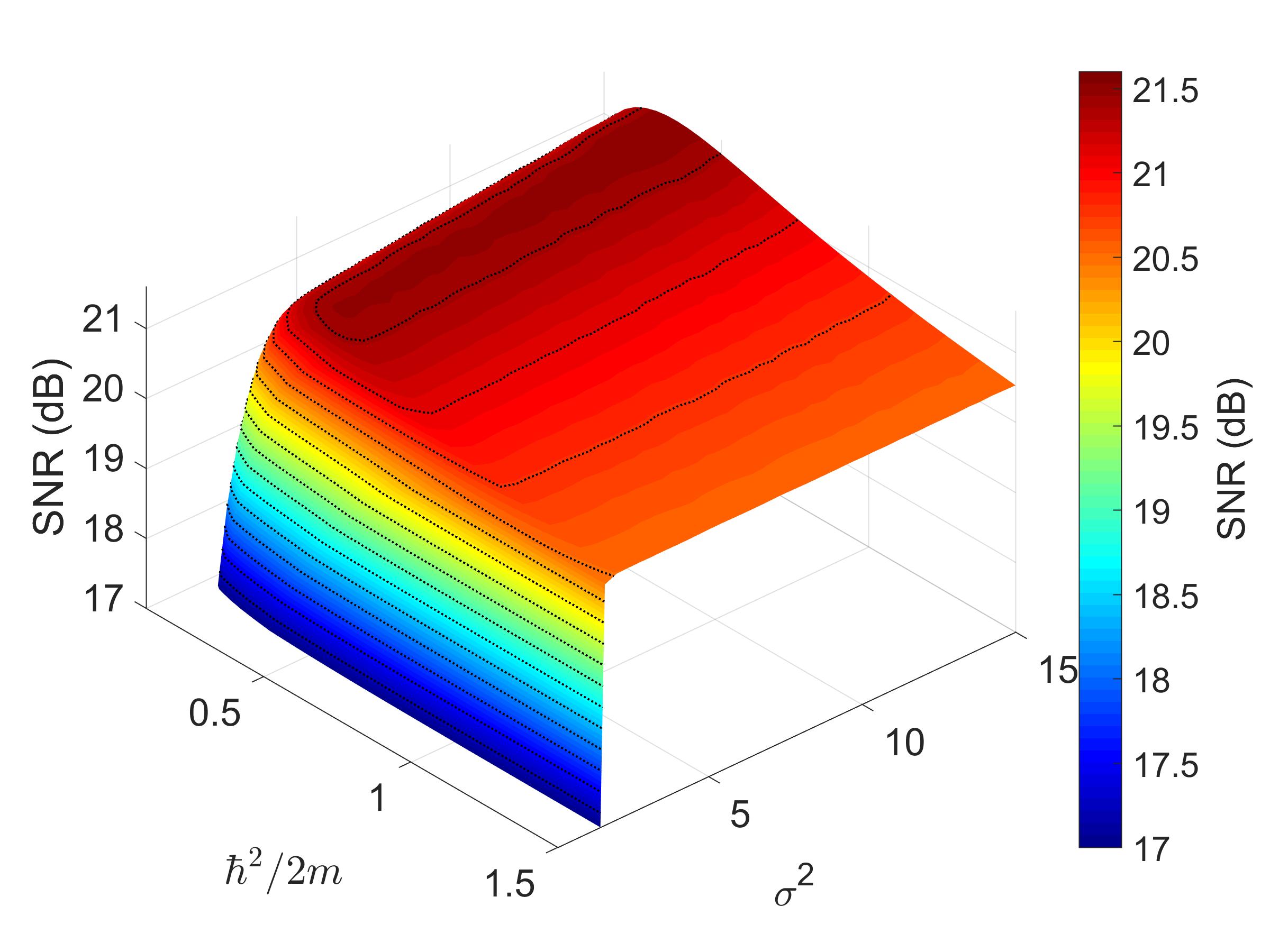}}
	\subfigure[]{\includegraphics[width=0.27\textwidth]{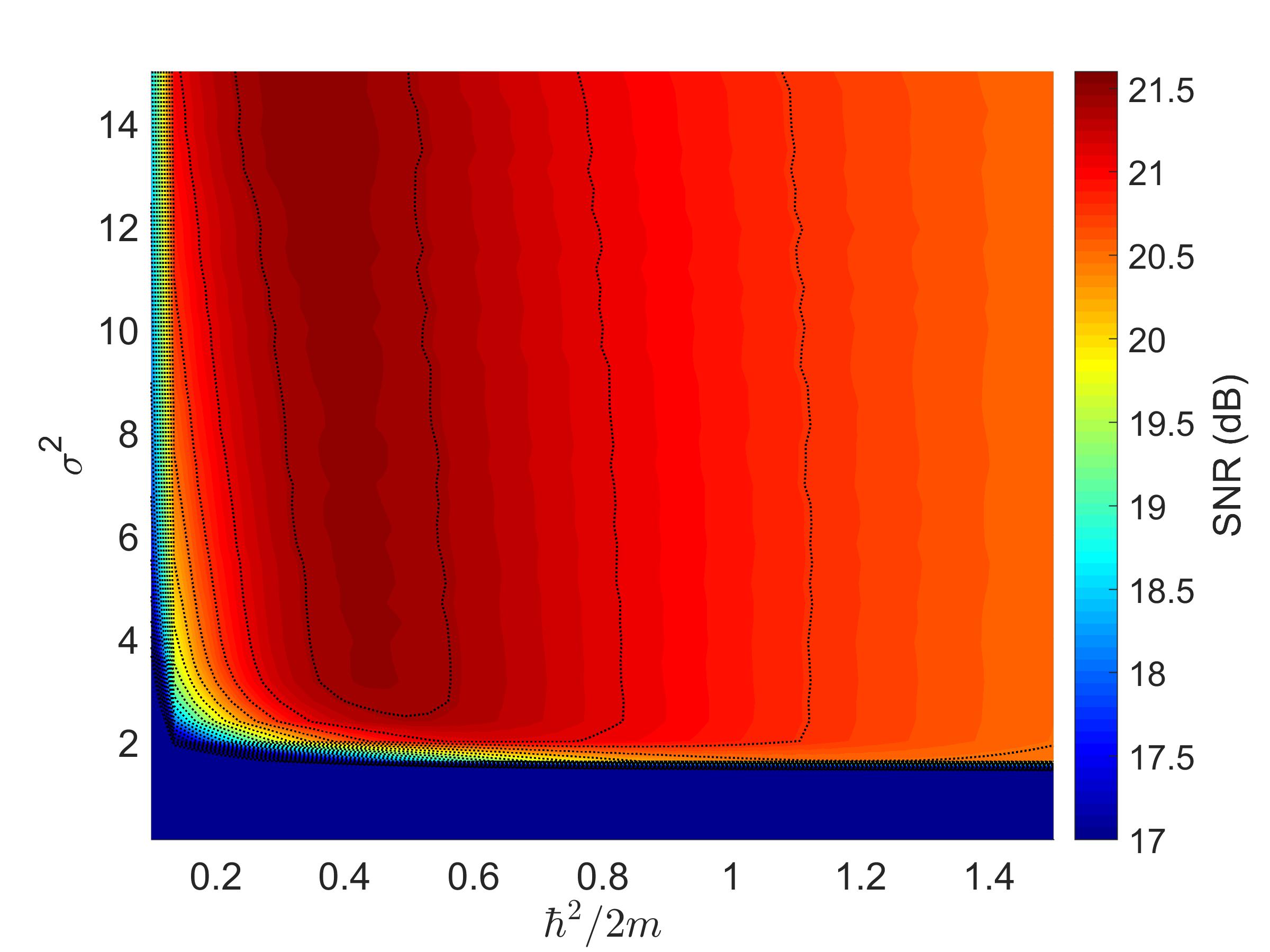}}
	
	\caption{(a) Sample A, (b,c) influence of the hyperparameters $\hbar^2/2m$ and $\sigma$ on proposed method performed on the sample A corrupted with Poisson noise corresponding to a SNR of 15 dB, (d) sample B, (e,f) influence of the hyperparameters $\hbar^2/2m$ and $\sigma$ on proposed method performed on the sample B corrupted with Poisson noise corresponding to a SNR of 15 dB, (g) sample C, (h,i) influence of the hyperparameters $\hbar^2/2m$ and $\sigma$ on proposed method performed on the sample C corrupted with Poisson noise corresponding to a SNR of 15 dB. The hyperparameters are $\rho = 1$ and $s = 110$ for each set of experiment.}
	\label{fig:1Dparameter02}
\end{figure*}

\begin{figure*}[h!]
	\centering
	\subfigure[]{\includegraphics[width=0.365\textwidth]{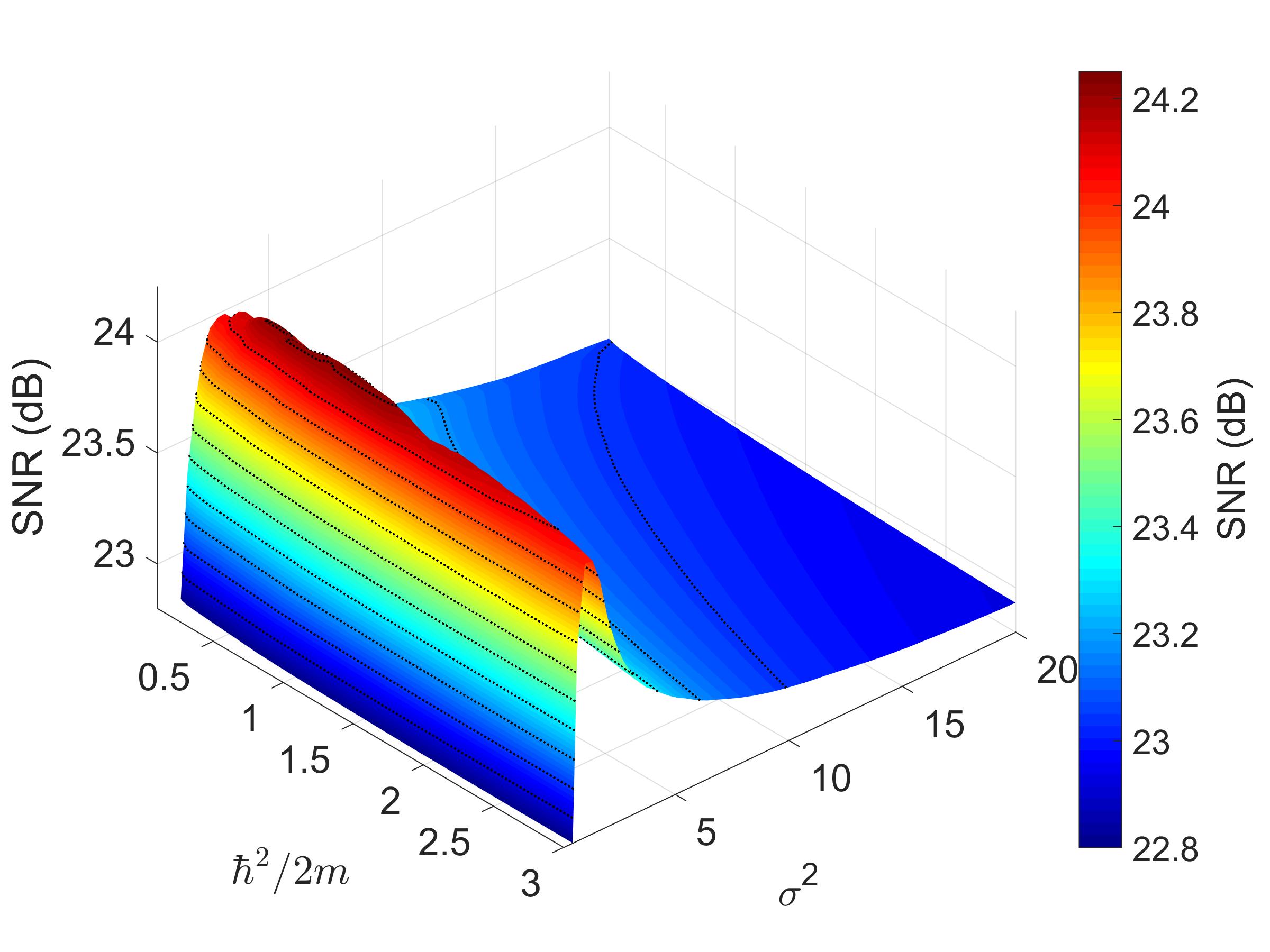}}
	\subfigure[]{\includegraphics[width=0.365\textwidth]{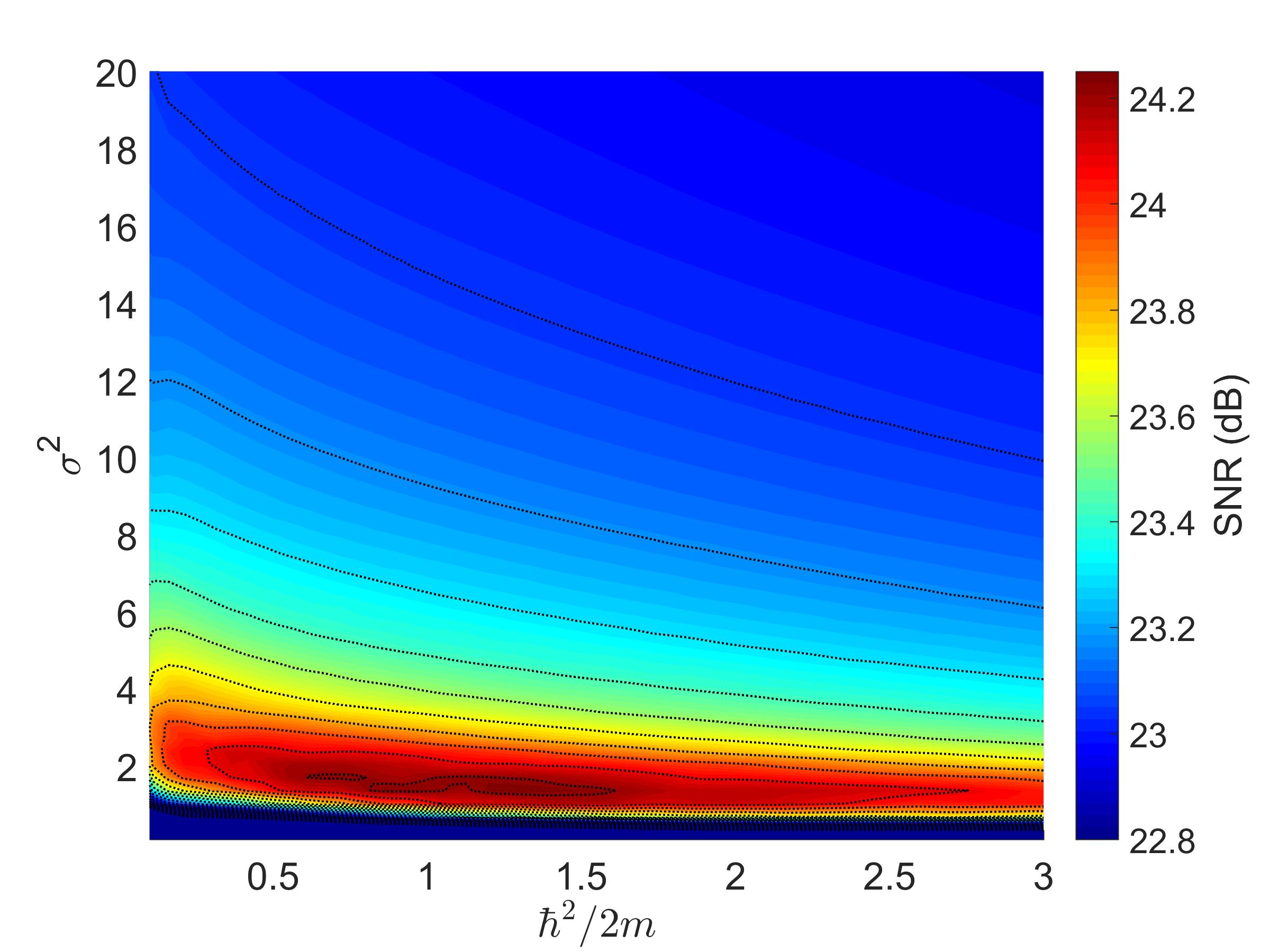}}
	\subfigure[]{\includegraphics[width=0.365\textwidth]{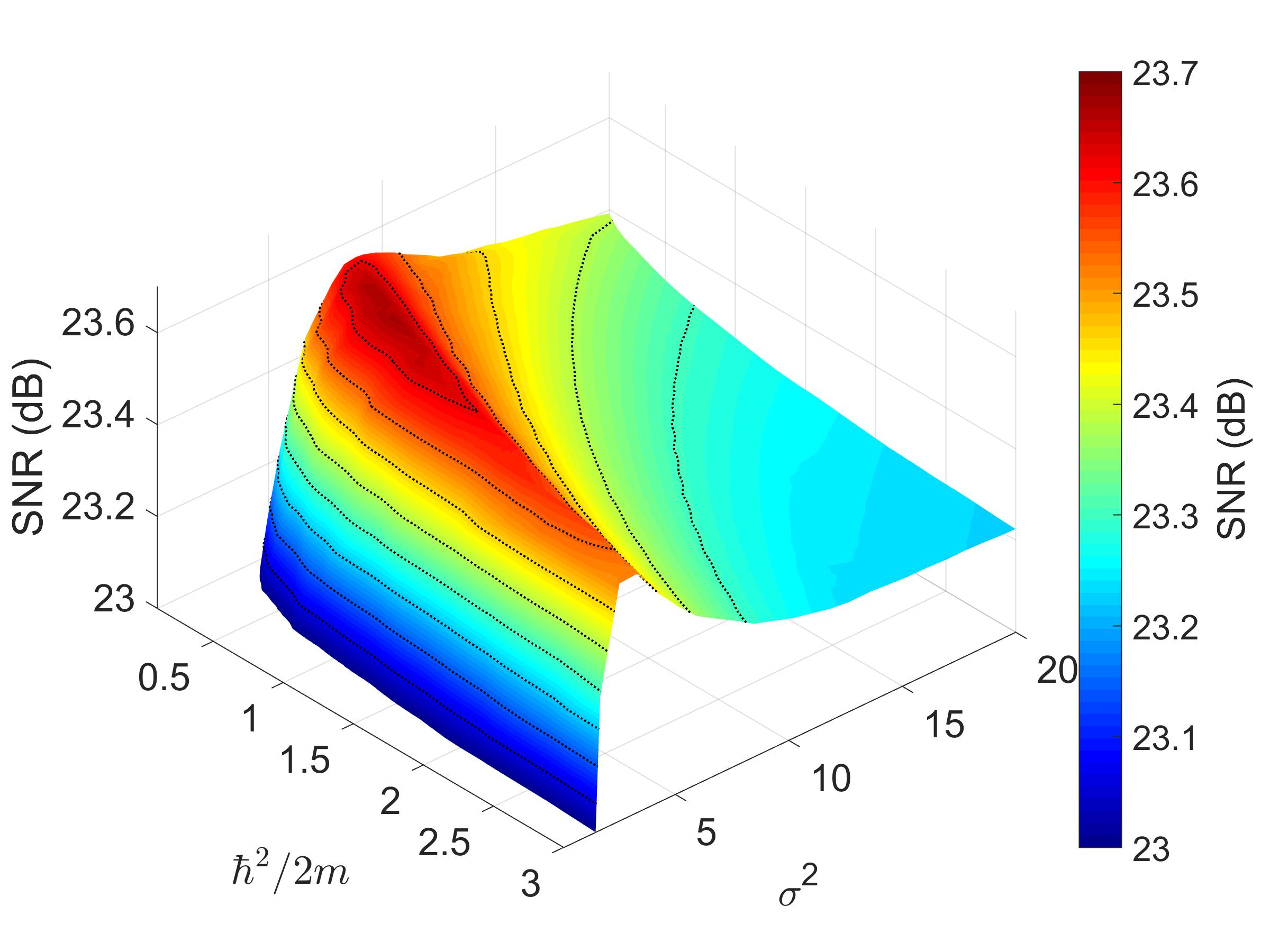}}
	\subfigure[]{\includegraphics[width=0.365\textwidth]{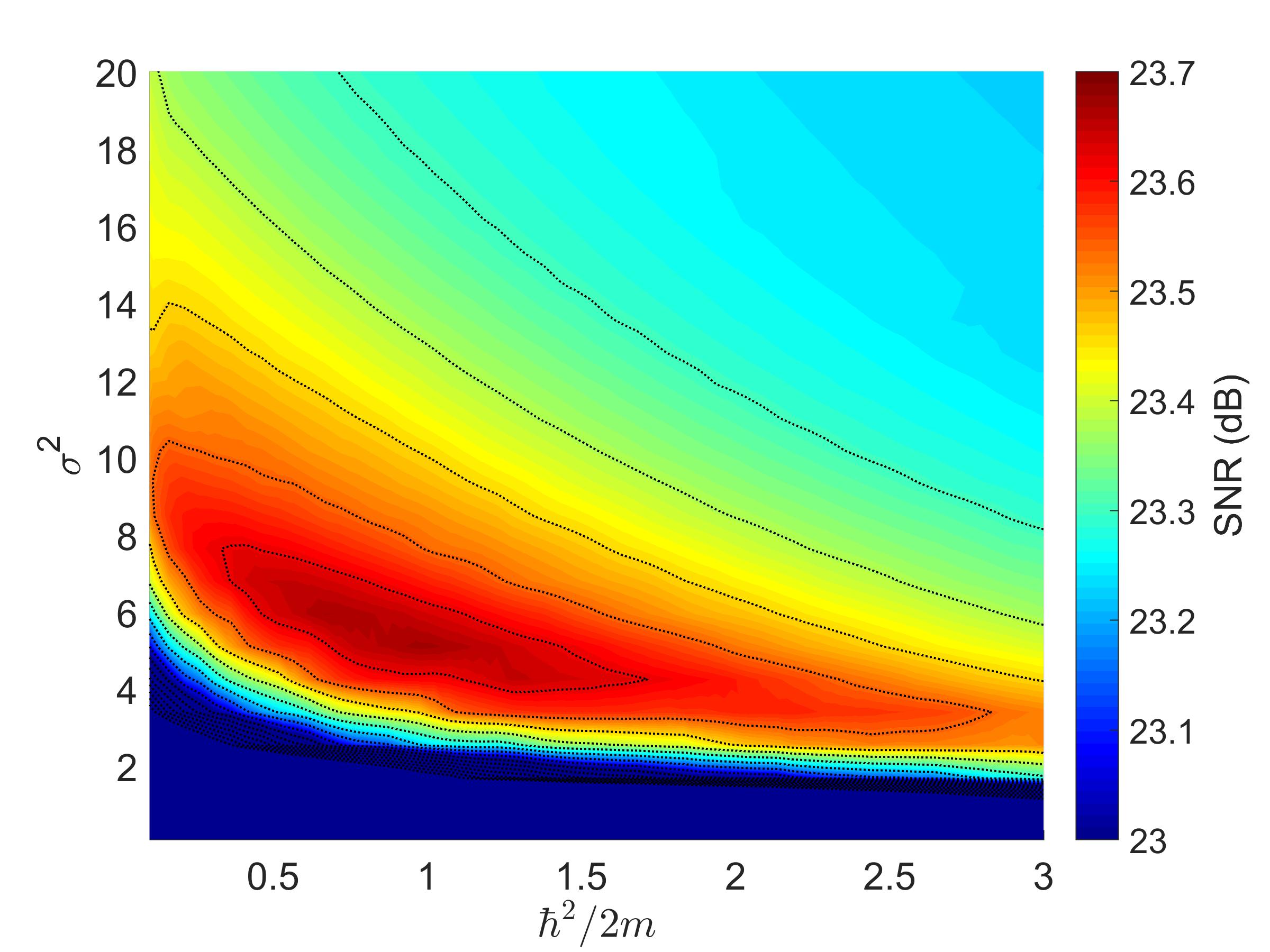}}
	\subfigure[]{\includegraphics[width=0.365\textwidth]{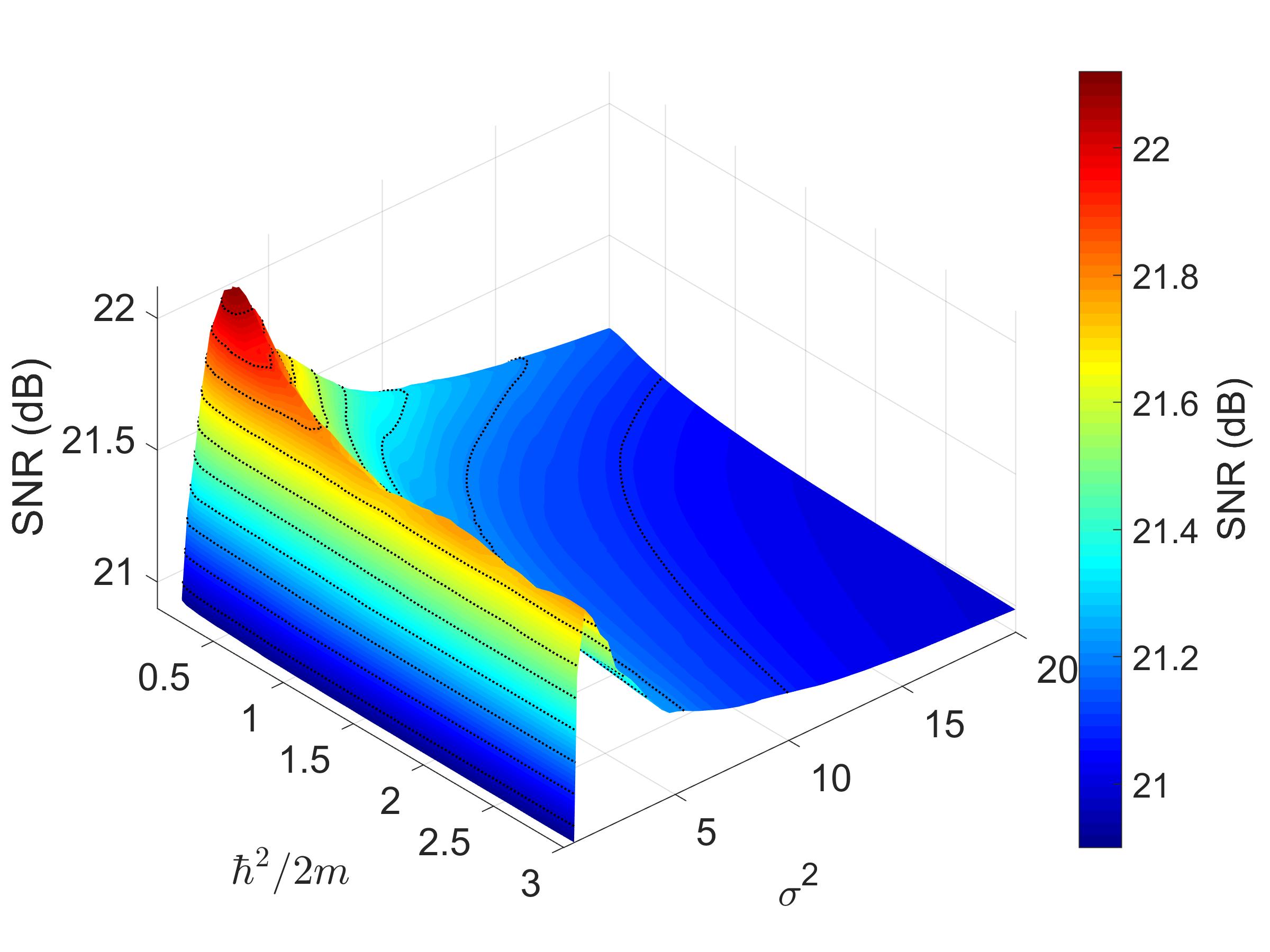}}
	\subfigure[]{\includegraphics[width=0.365\textwidth]{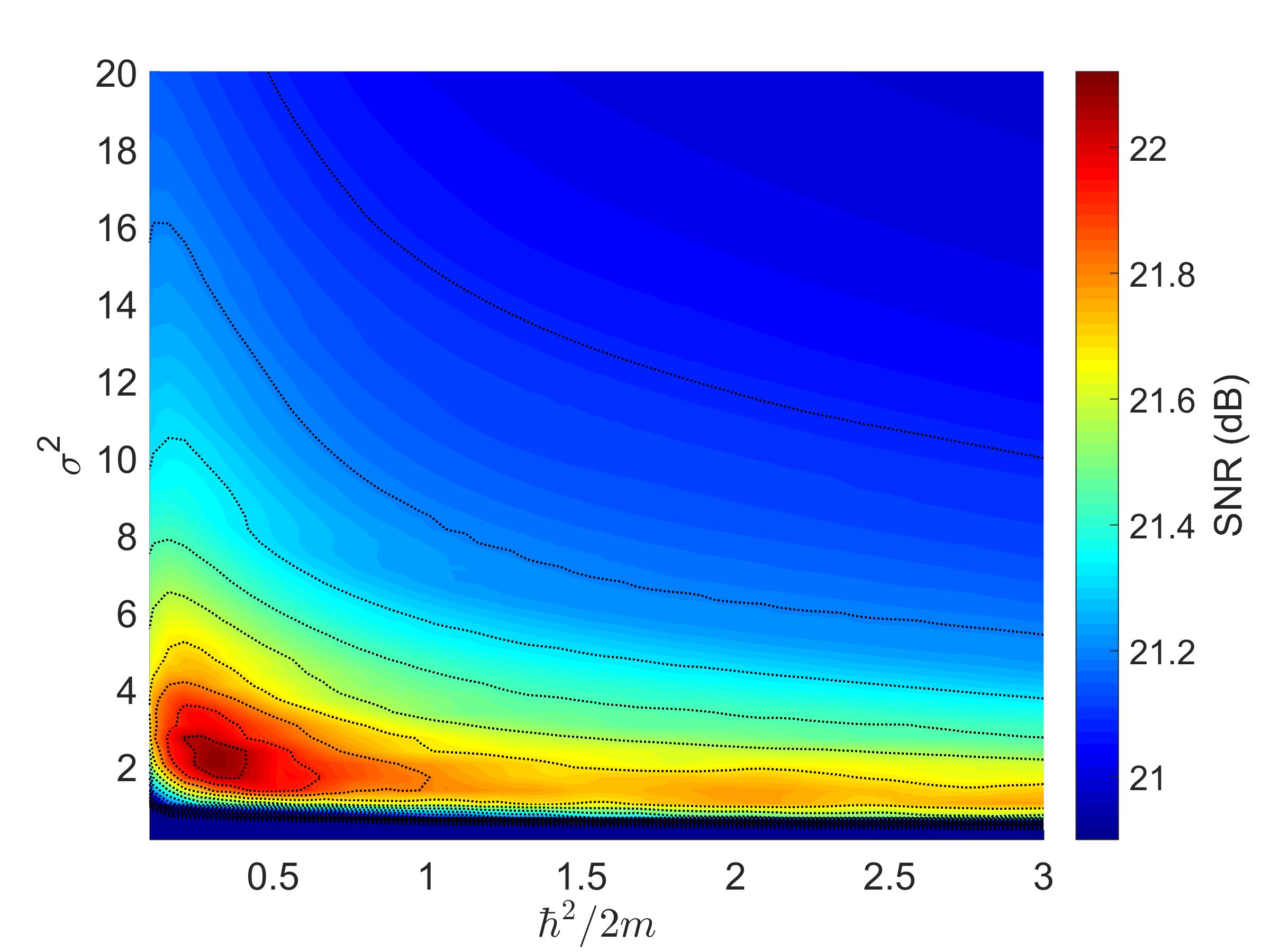}}
	
	\caption{ Influence of the hyperparameters $\hbar^2/2m$ and $\sigma$ on proposed decomposition carried out on the 2D Lena image in Fig.~\ref{fig:relation} in presence of (a,b) Poisson noise, (c,d) Gaussian noise and (e,f) speckle noise corresponding to a SNR of 15 dB respectively. The hyperparameters are $\rho = 1$ and $s = 600$ for each set of experiment.}
	\label{fig:2Dparameter}
\end{figure*}

In this subsection, we provide a detailed discussion about the influence of the hyperparameters on the proposed adapative bases.

As mentioned above, the parameter $\hbar^2/2m$ specifies  how the local frequencies of the vectors of the basis vary with the amplitude of the signal or image pixel value. To illustrate this relationship, the effect of $\hbar^2/2m$ on local frequencies is shown in Fig.~\ref{fig:parameter_h} for three distinct values of this parameter. For each case, three wave functions (number 25, 70 and 100) computed from the synthetic signal in Fig.~\ref{fig:sample} are displayed. For low values of $\hbar^2/2m$ (i.e., $0.08$ for the results in Fig.~\ref{fig:parameter_h}(a)), one may remark that the wave functions are oscillating at very high frequencies, even for higher values of the potential (i.e., of the signal). The presence of a maximal oscillation period due to the discretization of the signal implies that in this limit the high values of the signal are not taken properly into account. For very high values of $\hbar^2/2m$ ($15$ for the results in Fig.~\ref{fig:parameter_h}(c)), most of the wave functions are at an energy well above the potential values, and they discriminate less and less between the regions with different potential height. In this limit, wave functions behave very similarly to cosine functions with increasing frequencies, thus reducing the interest of the proposed bases that becomes very similar to the Fourier transform. At intermediate values of  $\hbar^2/2m$ ($1$ for the results in Fig.~\ref{fig:parameter_h}(b)), wave functions explore the different regions but with clearly different oscillation frequencies, \textit{i.e.} wave vectors have significantly larger frequencies or short wavelengths for the low potential valued regions as opposed to high potential regions.

The second hyperparameter studied in this section that has a strong impact on the proposed denoising algorithm is the 
paramater $\sigma$ which makes the adaptive basis delocalized on the system (signal or image). As explained above in Subsection~\ref{sec:techprob}, this parameter corresponds to the cut-off frequency of the Gaussian low pass filter used to smooth the noisy signal or images before computing the wave functions through \eqref{eq:Schroedinger}. This cut off frequency is fixed through the choice of the standard deviation $\sigma$ of the Gaussian filter. Again, we highlight that this parameter is not related to the denoising process itself, but to the definition of the adaptive basis to be used for denoising.

The localization of the wave functions in the presence of noise has an important impact on the proposed signal or image representation and furthermore on the efficiency of the denoising process. To illustrate this claim, 
Fig~\ref{fig:localization_lena} shows a denoising result with and without the use of the low pass Gaussian filter prior to the computation of the wave functions. In this example, the cropped version of Lena in Fig.~\ref{fig:localization_lena}(a) was degraded by a Poisson noise resulting into a SNR of $15$ dB. The denoised images in Fig.~\ref{fig:localization_lena}(c,d) were obtained using the algorithm detailed in Algo.~\ref{Algo:Denoising}. However, while the result in Fig.~\ref{fig:localization_lena}(c) exploits the image decomposition through localized wave functions computed directly from the noisy image, the result in Fig.~\ref{fig:localization_lena}(d) was obtained by filtering the noisy image by a low pass Gaussian filter before using \eqref{eq:Schroedinger}, in order to delocalize the wave functions. The interest of such delocalization can be visually appreciated in this example and allows a peak SNR (PSNR) gain of more than $3$ dB. In the following, we will always use a pre-smoothed signal or image in \eqref{eq:Schroedinger}, and the parameter $\sigma$ of the smoothing is thus an important parameter of the algorithm.

At last, in order to denoise the signal or image one has to threshold the coefficients of the signal or image on the adaptive basis; this process uses two thresholding hyperparameters $s$ and $\rho$ defined in \eqref{eq:thr}, which define respectively the threshold value and the abruptness of the cut off. In particular, the parameter $s$ corresponds to the threshold in energy of the wave functions taken into account in the expansion \eqref{eq:recons} to reconstruct the signal or image.

Fig.~\ref{fig:cutoff}(a) illustrates the PSNR as a function of the thresholding hyperparameter $s$ while reconstructing the denoised result corresponding to the signal in Fig.~\ref{fig:sample} contaminated by additive Gaussian noise of 15 dB. It is clear from that figure that initially PSNR decreases due to the low pass filtering, whereas the thresholding operation on the adaptive basis shows improvement in PSNR value. Fig.~\ref{fig:cutoff}(b) illustrates the variation of the SNR for changing values of the hyperparameter $s$. For $\sigma^2=0$, the reconstructed signal has a SNR worse or similar to the original noisy one, indicating once more the importance of the smoothing process before computing the adaptive basis through \eqref{eq:Schroedinger}.
For nonzero values of $\sigma^2$, there is a relatively small range of optimal $s$ values, where the SNR is much better than in the original noisy signal. Of course this threshold value should eventually depend on the level of noise. The adaptive transform makes the filtering of high frequencies stronger at high values of the potential, but the overall level of filtering should still depend on the noise properties.

Numerical experiments on the synthetic signal of Fig.~\ref{fig:sample} and on the Lena image in Fig.~\ref{fig:relation} were carried out to analyze the impact of $\hbar^2/2m$ and $\sigma$ on the denoising quality and subsequently to adjust these parameters to their best values for assessment of the efficiency of the algorithm. Three different types of noise were considered: Poisson, additive Gaussian and multiplicative speckle noise. In all cases, the level of noise was adjusted to correspond to a SNR of $15$ dB.

Fig.~\ref{fig:1Dparameter} show the quality of the denoising results for the synthetic signal Fig.~\ref{fig:sample}, in terms of SNR, versus the value of the hyperparameters $\hbar^2/2m$ and $\sigma^2$ for different types of noise: Poisson noise, Gaussian noise and speckle noise. Several observations can be made from these results. As expected, an optimal value arises in each case. In particular, the hyperparameter $\sigma$ should clearly be chosen to be nonzero, indicating the importance of taking into account the localization effects. However, even if an optimal value exists for the different hyperparameters, a small variation in the choice of these hyperparameters around the optimal values only slightly influences the quality of the denoising. Moreover, the optimal values are only slightly dependent on the nature of the noise. 
This means that for this type of signal the hyperparameters could be fixed beforehand at a fixed value which can be chosen independently of the type of noise present.


Next, the dependence of $\hbar^2/2m$ and $\sigma$ hyperparameters on the shape of the signals is analyzed. For this purpose, two additional synthetic signals were generated as shown in Fig.~\ref{fig:1Dparameter02}(d)(g) together with Fig.~\ref{fig:1Dparameter02}(a), which corresponds to the same synthetic signal used previously, further normalized to $1$ and corrupted by Poisson noise.
From the results in Fig.~\ref{fig:1Dparameter02}(b-c)(e-f)(h-i), it can be clearly observed that the quality of the denoising does depend on the shape of the signals, which can be expected given the nature of the adaptive basis used by the proposed approach. However, the denoising process is efficient for a fairly large interval around the optimal values. As there is a big overlap in the acceptable range of values of the hyperparameters for various signal shape, again this means that the hyperparameters could be fixed beforehand at a fixed value which can be chosen independently of the signal.

\begin{figure*}[b!]
  \centering
  \subfigure[Synthetic signal]{\includegraphics[width=0.18\textwidth]{Figure/figures/SampleA.pdf}}
  \subfigure[Synthetic image]{\includegraphics[width=0.18\textwidth]{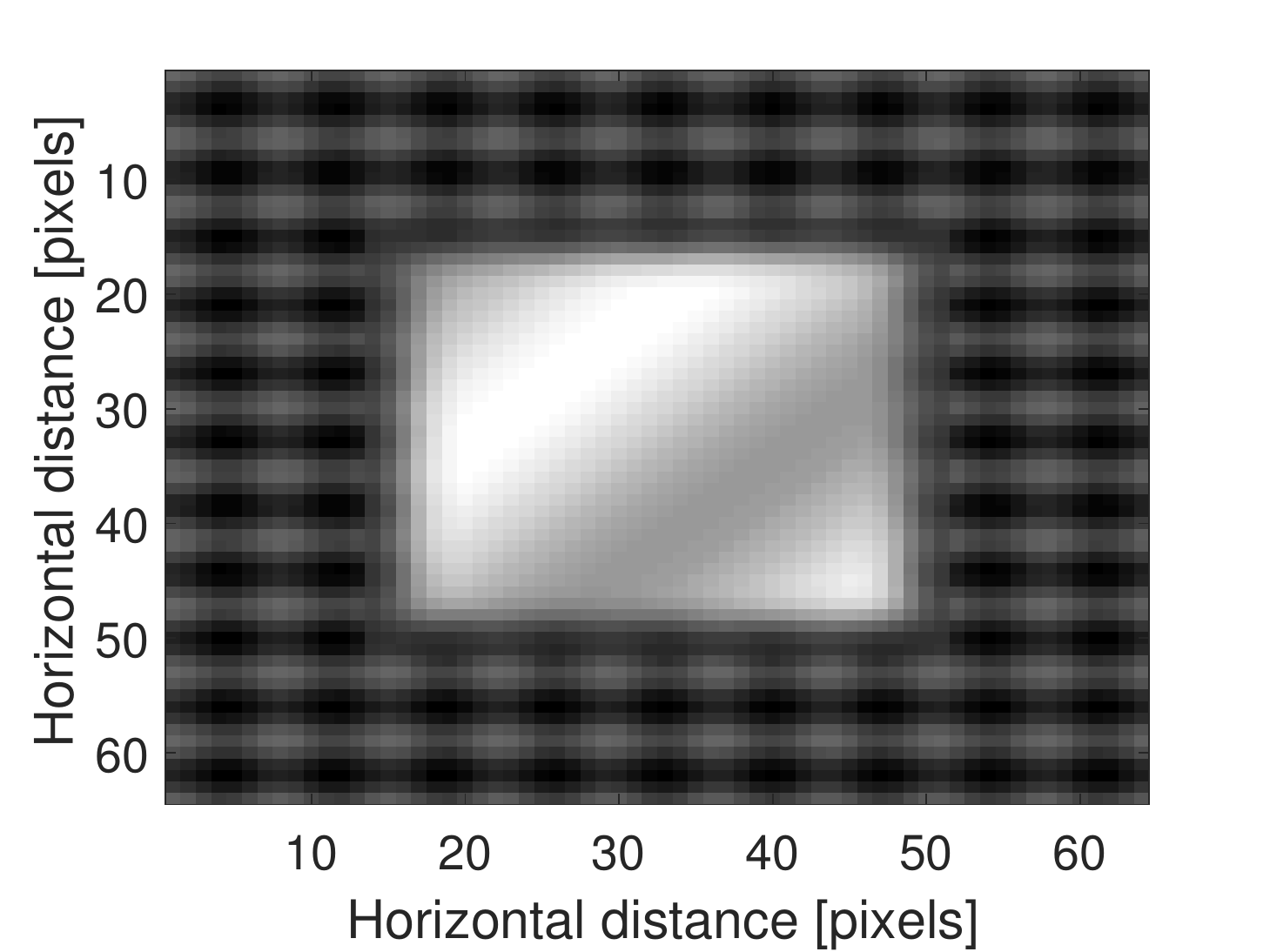}}
  \subfigure[Elaine $(512\times 512)$]{\includegraphics[width=0.145\textwidth]{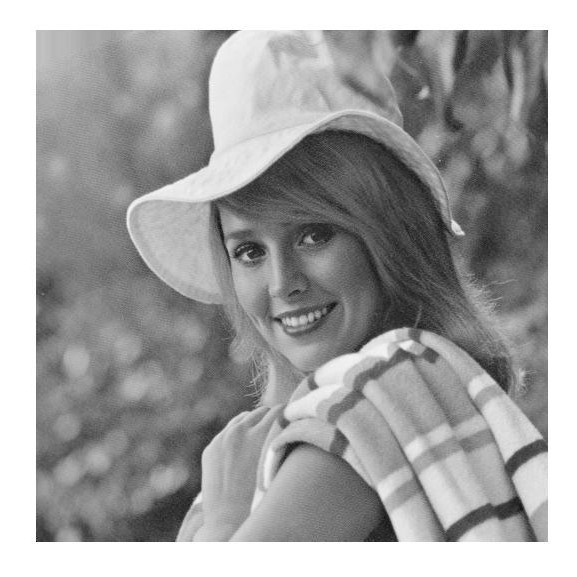}}
  \subfigure[Lena $(512\times 512)$]{\includegraphics[width=0.145\textwidth]{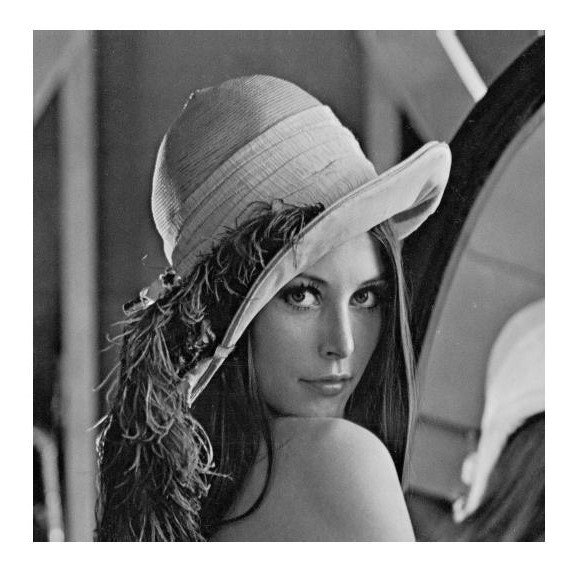}}
  \subfigure[Fruits $(512\times 512)$]{\includegraphics[width=0.145\textwidth]{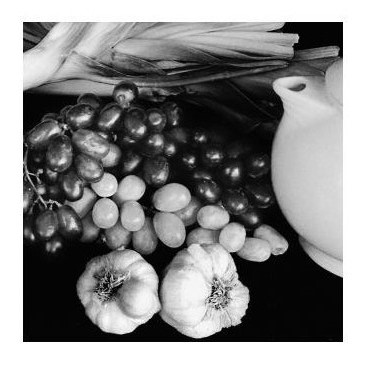}}
  \subfigure[Moon $(320\times 320)$]{\includegraphics[width=0.145\textwidth]{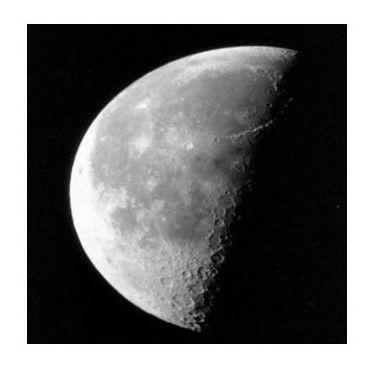}}

  \caption{Signal and images used to compare the proposed denoising method to existing algorithms. The size in number of pixels is indicated for each considered image.}
\label{fig:2Dsamples}

\end{figure*}

\begin{table*}[h!]
	\begin{center}
		\caption{Quantitative denoising results.}
		\vspace{2mm}
		\label{tab_gp}
		\begin{tabular}{c c c c c c c c c c c }
			\hline
			
			\multirow{2}{*}{Sample} & \multirow{2}{*}{Method}
			 & \multicolumn{3}{c}{Gaussian Noise (15dB)} & \multicolumn{3}{c}{Poisson Noise (15 dB)} & \multicolumn{3}{c}{Speckle Noise (15dB)}\\
			\cline{3-11}
			& & SNR (dB) & PSNR (dB) & SSIM & SNR (dB) & PSNR (dB) & SSIM & SNR (dB) & PSNR (dB) & SSIM\\

			\hline\hline
			\multirow{8}{*}{Synthetic Signal}
			& Wavelet hard		& {\color{blue}\textbf{18.84}} & {\color{blue}\textbf{25.21}} & NA & 17.21 & {\color{blue}\textbf{24.64}} & NA & {\color{blue}\textbf{17.36}} & {\color{blue}\textbf{24.13}} & NA \\
			& Wavelet soft		& 18.53 & 24.21 & NA & {\color{blue}\textbf{17.79}} & 23.62 & NA & 17.02 & 22.70 & NA \\
			& VST				& NA & NA & NA & NA & NA & NA & NA & NA & NA \\
			& TV				& 16.20 & 23.01 & NA & 15.94 & 23.45 & NA & 15.92 & 22.92 & NA \\
			& GSP				& NA & NA & NA & NA & NA & NA & NA & NA & NA \\
			& NLM				& NA & NA & NA & NA & NA & NA & NA & NA & NA \\
			& DL				& NA & NA & NA & NA & NA & NA & NA & NA & NA \\
			& Proposed			& {\color{red}\textbf{22.21}} & {\color{red}\textbf{27.50}} & NA & {\color{red}\textbf{22.51}} & {\color{red}\textbf{27.63}} & NA & {\color{red}\textbf{20.75}} & {\color{red}\textbf{26.86}} & NA \\
			
			\hline
			\multirow{8}{*}{Synthetic Image}
			& Wavelet hard		& 15.01 & 24.46 & 0.61 & 15.01 & 25.68 & 0.69 & 15.01 & 25.34 & 0.76 \\
			& Wavelet soft		& 15.71 & 25.05 & 0.64 & 15.61 & 26.20 & 0.70 & 15.49 & 25.80 & 0.77 \\
			& VST				& NA & NA & NA & 15.09 & 25.83 & 0.69 & 15.06 & 25.58 & 0.76 \\
			& TV				& 15.74 & 25.07 & 0.64 & 15.62 & 26.23 & 0.71 & 15.53 & 25.78 & 0.77 \\
			& GSP				& {\color{blue}\textbf{20.28}} & {\color{blue}\textbf{28.78}} & {\color{blue}\textbf{0.79}} & NA & NA & NA & NA & NA & NA \\
			& NLM				& 18.70 & 26.88 & 0.71 & NA & NA & NA & NA & NA & NA \\
			& DL				& 17.35 & 26.15 & 0.71 & {\color{blue}\textbf{17.14}} & {\color{blue}\textbf{27.22}} & {\color{blue}\textbf{0.75}} & {\color{blue}\textbf{17.21}} & {\color{blue}\textbf{27.48}} & {\color{blue}\textbf{0.80}} \\
			& Proposed			& {\color{red}\textbf{23.42}} & {\color{red}\textbf{31.78}} & {\color{red}\textbf{0.89}} & {\color{red}\textbf{23.92}} & {\color{red}\textbf{32.78}} & {\color{red}\textbf{0.92}} & {\color{red}\textbf{25.32}} & {\color{red}\textbf{33.50}} & {\color{red}\textbf{0.95}} \\
			
			
			\hline
			\multirow{8}{*}{Elaine}
			& Wavelet hard		& 20.52 & 27.02 & 0.52 & 20.08 & 28.17 & 0.49 & 19.75 & 25.99 & 0.48 \\
			& Wavelet soft		& 21.99 & 27.69 & 0.53 & 21.67 & 28.59 & 0.51 & 21.31 & 26.61 & 0.50 \\
			& VST				& NA & NA & NA & 21.71 & 28.64 & 0.53 & 22.51 & {\color{blue}\textbf{27.81}} & 0.56 \\
			& TV				& 23.67 & 29.63 & 0.62 & 22.03 & 28.84 & 0.55 & {\color{blue}\textbf{23.06}} & 27.61 & {\color{blue}\textbf{0.59}} \\
			& GSP				& {\color{red}\textbf{25.89}} & {\color{red}\textbf{30.73}} & {\color{red}\textbf{0.72}} & NA & NA & NA & NA & NA & NA \\
			& NLM				& 24.67 & {\color{blue}\textbf{30.70}} & 0.67 & NA & NA & NA & NA & NA & NA \\
			& DL				& {\color{blue}\textbf{24.97}} & 29.92 & {\color{blue}\textbf{0.68}} & {\color{red}\textbf{23.96}} & {\color{red}\textbf{29.84}} & {\color{blue}\textbf{0.62}} & 22.99 & 27.58 & 0.58 \\
			& Proposed			& 24.70 & 29.87 & {\color{blue}\textbf{0.68}} & {\color{blue}\textbf{23.89}} & {\color{blue}\textbf{29.03}} & {\color{red}\textbf{0.65}} & {\color{red}\textbf{23.52}} & {\color{red}\textbf{28.32}} & {\color{red}\textbf{0.64}} \\
			
			\hline
			\multirow{8}{*}{Lena}
			& Wavelet hard		& 20.84 & 28.17 & 0.72 & 20.01 & 28.89 & 0.68 & 19.22 & 27.49 & 0.66 \\
			& Wavelet soft		& 21.23 & 28.12 & 0.71 & 20.75 & 28.54 & 0.67 & 20.29 & 27.31 & 0.66 \\
			& VST				& NA & NA & NA & 20.82 & 29.50 & {\color{blue}\textbf{0.73}} & 21.24 & 28.55 & 0.69 \\
			& TV				& 21.95 & 29.32 & 0.70 & 21.34 & 29.58 & 0.68 & {\color{blue}\textbf{21.83}} & {\color{blue}\textbf{28.71}} & {\color{blue}\textbf{0.72}} \\
			& GSP				& 22.43 & 29.32 & {\color{red}\textbf{0.78}} & NA & NA & NA & NA & NA & NA \\
			& \SD{NLM}				& 22.92 & {\color{red}\textbf{30.58}} & {\color{blue}\textbf{0.77}} & NA & NA & NA & NA & NA & NA \\
			& DL				& {\color{red}\textbf{23.14}} & {\color{blue}\textbf{30.02}} & {\color{blue}\textbf{0.77}} & {\color{blue}\textbf{21.89}} & {\color{blue}\textbf{29.61}} & 0.71 & 20.35 & 27.24 & 0.71 \\
			& Proposed			& {\color{blue}\textbf{23.01}} & 29.89 & {\color{red}\textbf{0.78}} & {\color{red}\textbf{22.86}} & {\color{red}\textbf{29.95}} & {\color{red}\textbf{0.77}} & {\color{red}\textbf{23.21}} & {\color{red}\textbf{30.10}} & {\color{red}\textbf{0.78}} \\
			

			\hline
			\multirow{8}{*}{Fruits}
			& Wavelet hard		& 18.60 & 25.07 & 0.65 & 18.59 & 25.53 & 0.65 & 18.38 & 24.86 & 0.64 \\
			& Wavelet soft		& 18.84 & 25.08 & 0.71 & 18.81 & 25.29 & 0.72 & 18.51 & 24.50 & 0.71 \\
			& VST				& NA & NA & NA & 19.37 & 25.96 & {\color{blue}\textbf{0.76}} & 19.01 & 25.61 & {\color{blue}\textbf{0.76}} \\
			& TV				& 20.69  & 26.86 & {\color{blue}\textbf{0.79}} & 20.60 & 26.71 & 0.75 & 20.18 & 26.34 & 0.74 \\
			& GSP				& {\color{blue}\textbf{21.44}} & 27.43 & {\color{red}\textbf{0.81}} & NA & NA & NA & NA & NA & NA \\
			& NLM				& {\color{red}\textbf{21.48}} & {\color{blue}\textbf{28.02}} & 0.77 & NA & NA & NA & NA & NA & NA \\
			& DL				& 21.30 & 27.37 & {\color{blue}\textbf{0.79}} & {\color{blue}\textbf{20.87}} & {\color{blue}\textbf{27.16}} & 0.71 & {\color{blue}\textbf{20.39}} & {\color{blue}\textbf{27.08}} & 0.72 \\
			& Proposed			& 21.39 & {\color{red}\textbf{28.07}} & 0.77 & {\color{red}\textbf{21.93}} & {\color{red}\textbf{28.31}} & {\color{red}\textbf{0.79}} & {\color{red}\textbf{21.83}} & {\color{red}\textbf{28.29}} & {\color{red}\textbf{0.82}} \\

			\hline
			\multirow{8}{*}{Moon}
			& Wavelet hard		& 22.91 & 30.02 & 0.70 & 21.45 & 29.90 & 0.72 & 21.19 & 29.07 & 0.71 \\
			& Wavelet soft		& 23.09 & 30.98 & 0.74 & 22.14 & 30.51 & 0.80 & 21.90 & 29.79 & 0.79 \\
			& VST				& NA & NA & NA & 22.58 & 31.17 & {\color{blue}\textbf{0.85}} & 22.11 & 30.01 & 0.84 \\
			& TV				& 23.35  & 32.19 & 0.80 & {\color{blue}\textbf{23.51}} & {\color{blue}\textbf{32.21}} & {\color{red}\textbf{0.86}} & {\color{blue}\textbf{22.91}} & {\color{blue}\textbf{30.84}} & {\color{blue}\textbf{0.86}} \\
			& GSP				& 23.33 & 31.22 & {\color{blue}\textbf{0.85}} & NA & NA & NA & NA & NA & NA \\
			& NLM				& {\color{red}\textbf{25.79}} & {\color{red}\textbf{33.94}} & {\color{red}\textbf{0.86}} & NA & NA & NA & NA & NA & NA \\
			& DL				& 23.82 & 32.71 & 0.81 & 22.95 & 31.65 & {\color{blue}\textbf{0.85}} & 22.32 & 30.07 & 0.84 \\
			& Proposed			& {\color{blue}\textbf{24.81}} & {\color{blue}\textbf{33.11}} & 0.83 & {\color{red}\textbf{24.65}} & {\color{red}\textbf{33.34}} & {\color{red}\textbf{0.86}} & {\color{red}\textbf{23.48}} & {\color{red}\textbf{31.55}} & {\color{red}\textbf{0.89}} \\
			
			\hline
		\end{tabular}
		\vspace{3mm}
	\end{center}
\end{table*}

Finally, Fig.~\ref{fig:2Dparameter} regroups the results for the cropped Lena image for the three types of noise. The same conclusions can be drawn as from the results on 1D signals in Fig.~\ref{fig:1Dparameter}: as expected and similar to any other denoising method, the choice of the hyperparameters does have an impact on the results, and the optimal range of parameters depend on the noise. However, even though the acceptable range of parameters seems smaller than for the 1D signal, there is still a relatively large parameter region where the denoising is very effective. This again makes realistic the possibility to set these parameters beforehand in the algorithm independently from the signal or image. Additionnally, there is a large overlap between the optimal parameter ranges for Poisson and speckle noise, with a marked difference for Gaussian noise. This seems to indicate that the choice of the parameters may differ according to the broad class to which the noise of interest belongs, an information that is usually known beforehand in many cases.
\subsection{Efficiency of the denoising process}
\label{sec:denoisingresults}

This section presents denoising results on a synthetic signal, a synthetic image and four standard testing images of size $512 \times 512$ and $320 \times 320$ pixels, shown in Fig.~\ref{fig:2Dsamples}.

\begin{figure*}[h!]
	
	\centering
	\subfigure[\scriptsize{Clean}]{\includegraphics[width=0.15\textwidth]{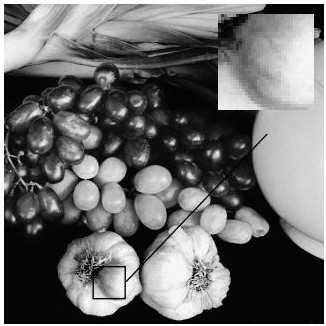}}
	\subfigure[\scriptsize{Noisy}]{\includegraphics[width=0.15\textwidth]{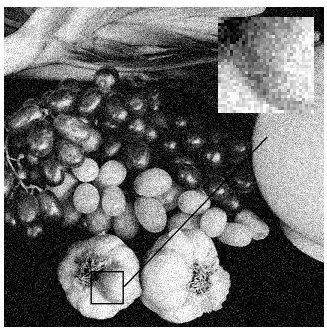}}
	\subfigure[\scriptsize{PSNR=25.07 dB}]{\includegraphics[width=0.15\textwidth]{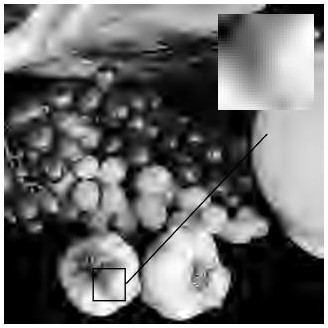}}
	\subfigure[\scriptsize{PSNR=25.08 dB}]{\includegraphics[width=0.15\textwidth]{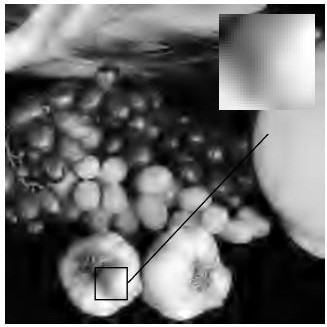}}
	\subfigure[\scriptsize{PSNR=26.86 dB}]{\includegraphics[width=0.15\textwidth]{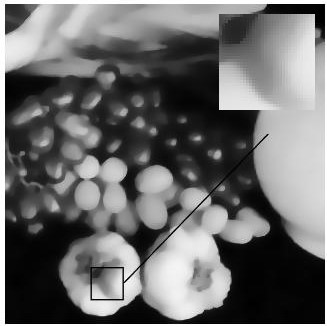}}
	
	\subfigure[\scriptsize{PSNR=27.43 dB}]{\includegraphics[width=0.15\textwidth]{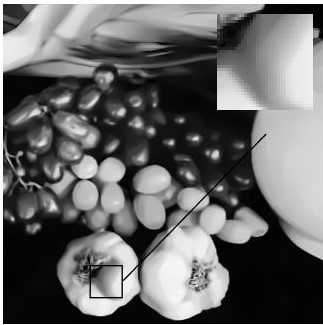}}
	\subfigure[\scriptsize{PSNR=28.02 dB}]{\includegraphics[width=0.15\textwidth]{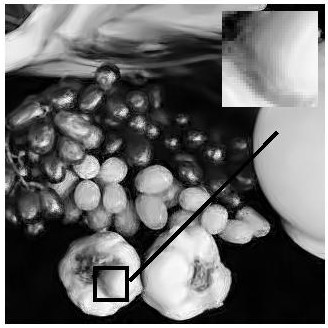}}
	\subfigure[\scriptsize{PSNR=27.37 dB}]{\includegraphics[width=0.15\textwidth]{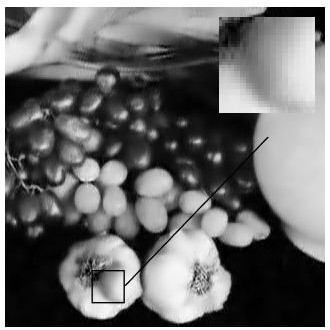}}
	\subfigure[\scriptsize{PSNR=28.07 dB}]{\includegraphics[width=0.15\textwidth]{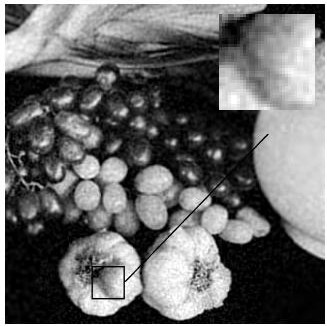}}
	
	\caption{Result of the denoising algorithm compared with other methods: (a) Clean Fruits image, (b) Image corrupted with Gaussian noise corresponding to a SNR of 15 dB. Denoising results obtained using, (c) wavelet hard thresholding, (d) wavelet soft thresholding, (e) total variation regularization, (f) graph signal processing, (g) non-local means, (h) dictionary learning and (i) proposed method. The proposed adaptive transform was computed with the hyperparameters $\hbar^2/2m = 0.23$, $\sigma^2 = 7.5$, $\rho = 2$ and $s = 560$.}
	\vspace{2mm}
	\label{fig:stillResults}
\end{figure*}

\begin{figure*}[h!]
  \centering
  \subfigure[\scriptsize{Clean}]{\includegraphics[width=0.15\textwidth]{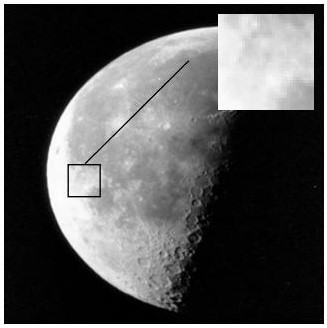}}
  \subfigure[\scriptsize{Noisy}]{\includegraphics[width=0.15\textwidth]{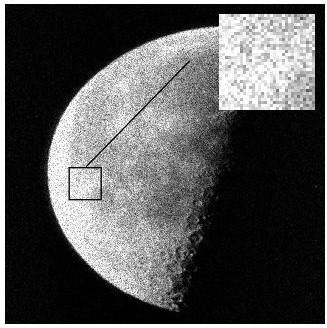}}
  \subfigure[\scriptsize{PSNR=29.90 dB}]{\includegraphics[width=0.15\textwidth]{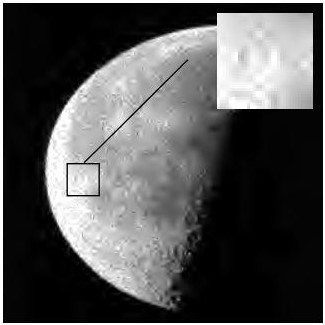}}
  \subfigure[\scriptsize{PSNR=30.51 dB}]{\includegraphics[width=0.15\textwidth]{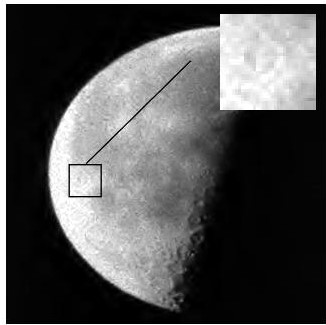}}
  
  \subfigure[\scriptsize{PSNR=31.17 dB}]{\includegraphics[width=0.15\textwidth]{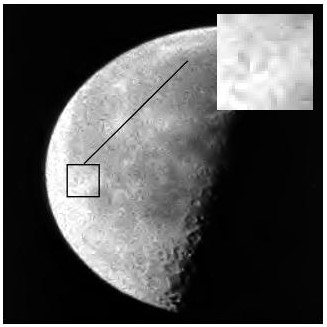}}
  \subfigure[\scriptsize{PSNR=32.21 dB}]{\includegraphics[width=0.15\textwidth]{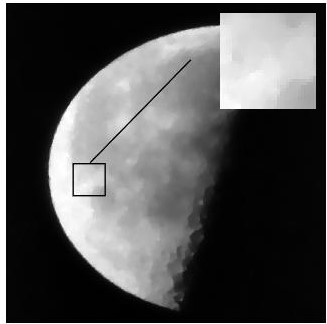}}
  \subfigure[\scriptsize{PSNR=31.65 dB}]{\includegraphics[width=0.15\textwidth]{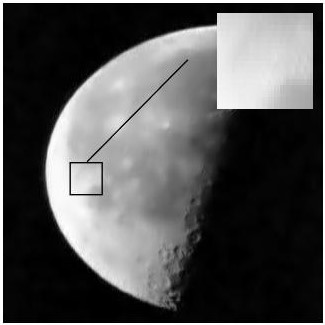}}
  \subfigure[\scriptsize{PSNR=33.34 dB}]{\includegraphics[width=0.15\textwidth]{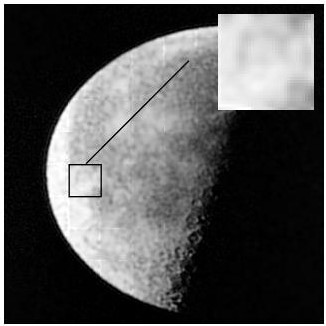}}

  \caption{Result of the denoising algorithm compared with other methods: (a) Clean moon image, (b) Image corrupted with Poisson noise corresponding to a SNR of 15 dB. Denoising results obtained using, (c) wavelet hard thresholding, (d) wavelet soft thresholding, (e) variance stabilization transform, (f) total variation regularization, (g) dictionary learning and (h) proposed method. The proposed adaptive transform was computed with the hyperparameters $\hbar^2/2m = 0.32$, $\sigma^2 = 2.5$, $\rho = 1$ and $s = 520$.}
\label{fig:moonResults}
\end{figure*}

\begin{figure*}[h!]
  \centering
  \subfigure[\scriptsize{Clean}]{\includegraphics[width=0.15\textwidth]{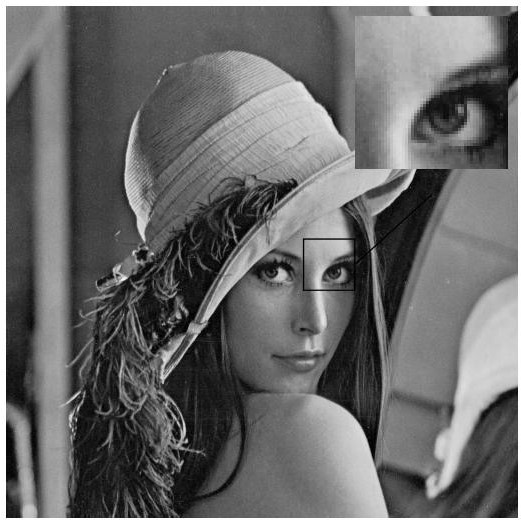}}
  \subfigure[\scriptsize{Noisy}]{\includegraphics[width=0.15\textwidth]{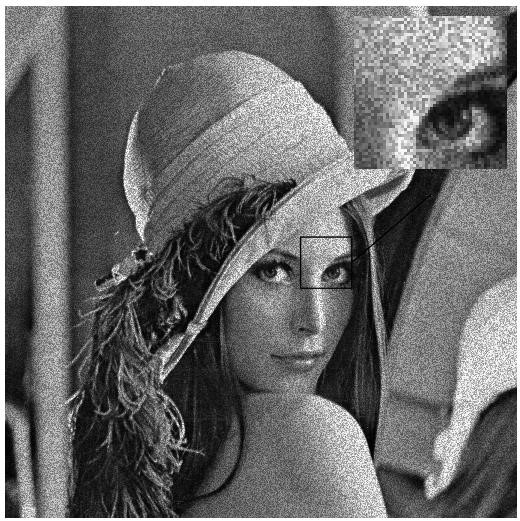}}
  \subfigure[\scriptsize{PSNR=27.49 dB}]{\includegraphics[width=0.15\textwidth]{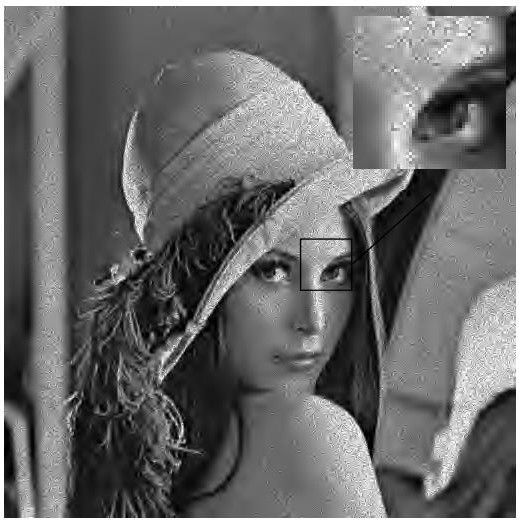}}
  \subfigure[\scriptsize{PSNR=27.31 dB}]{\includegraphics[width=0.15\textwidth]{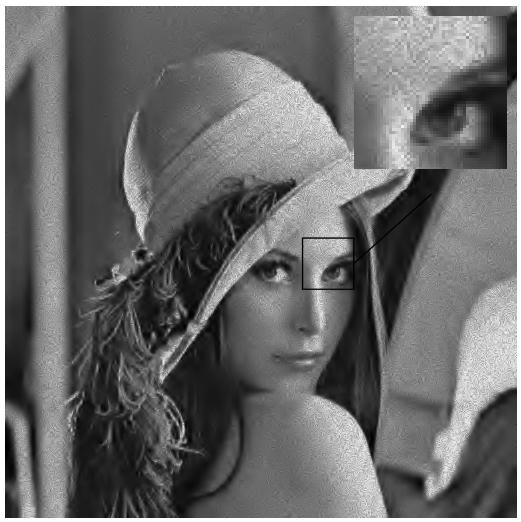}}
  
  \subfigure[\scriptsize{PSNR=28.55 dB}]{\includegraphics[width=0.15\textwidth]{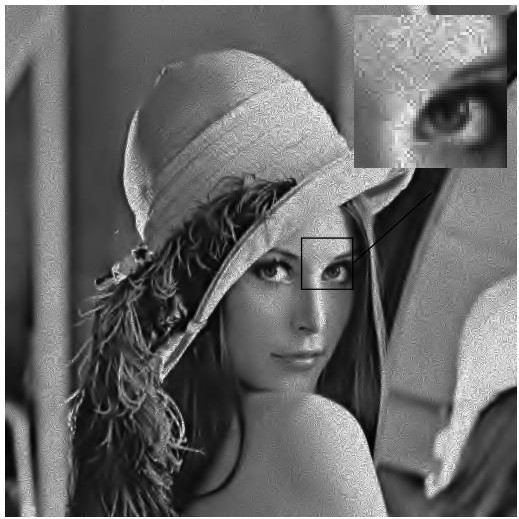}}
  \subfigure[\scriptsize{PSNR=28.71 dB}]{\includegraphics[width=0.15\textwidth]{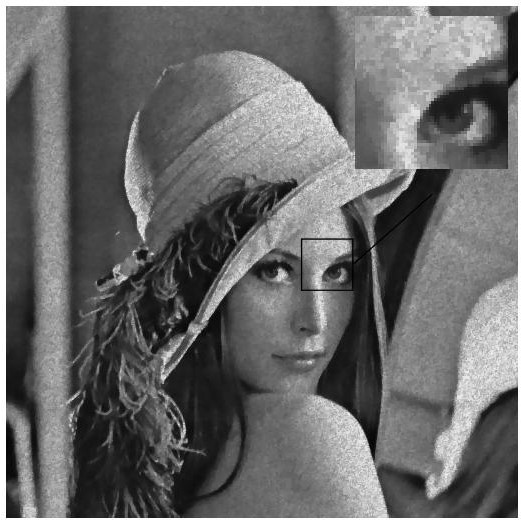}}
  \subfigure[\scriptsize{PSNR=27.24 dB}]{\includegraphics[width=0.15\textwidth]{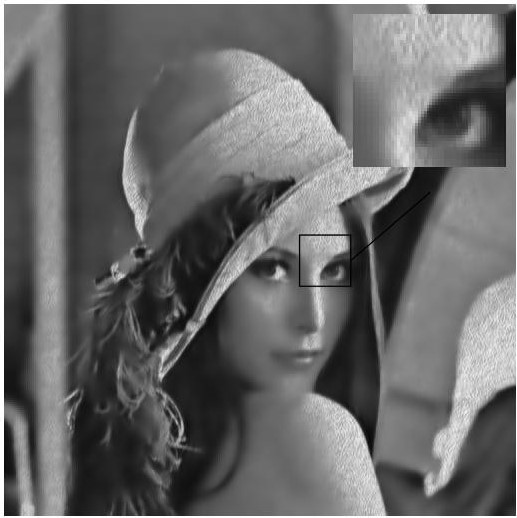}}
  \subfigure[\scriptsize{PSNR=30.10 dB}]{\includegraphics[width=0.15\textwidth]{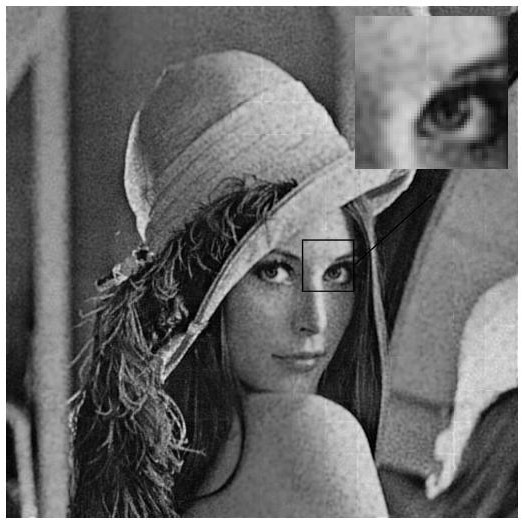}}

  \caption{Result of the denoising algorithm compared with other methods: (a) Clean Lena image, (b) Image corrupted with speckle noise corresponding to a SNR of 15 dB. Denoising results obtained using, (c) wavelet hard thresholding, (d) wavelet soft thresholding, (e) variance stabilization transform, (f) total variation regularization, (g) dictionary learning and (h) proposed method. The proposed adaptive transform was computed with the hyperparameters $\hbar^2/2m = 0.36$, $\sigma^2 = 1.35$, $\rho = 2$ and $s = 600$.}
\label{fig:lenaResults}
\end{figure*}

Denoising is an extensively explored research field that prevents an exhaustive comparison of the proposed approach to all the existing methods. Moreover, we remind that the most important contribution herein is to investigate a novel way of decomposing signals or images, which is not meant to outperform all the denoising algorithms in any scenario. Five algorithms from the literature were used for comparison purpose: i) wavelet denoising based on hard and soft thresholding of detail coefficients \cite{donoho1994ideal,donoho1995wavelet}, ii) the variance stabilization transform (VST) relevant for data dependent noise models \cite{makitalo2011closed}, iii) an optimization-based approach using the total variation (TV) semi-norm to regularize the solution \cite{Figueiredo10,rudin1994total}, iv) a graph signal processing (GSP) method by constructing an optimal graph and corresponding graph Laplacian regularizer \cite{Pang2017Graph}, v) a non-local means (NLM) image denoising method that uses principal component analysis approach \cite{tasdizen2009principal}, and vi) a dictionary learning (DL) method exploiting sparse and redundant representations over learned patch-based dictionaries \cite{elad2006image}. Note that for all the methods and for all the simulation scenarios, their hyperparameters were manually tuned to obtain optimal denoising results in the sense of the quantitative measurements employed. We used the Matlab implementations available in the Numerical tours website \cite{NumericalTours}.

\begin{figure*}[h!]
  \centering
    \subfigure[Clean CBCT image]{\includegraphics[width=0.213\textwidth]{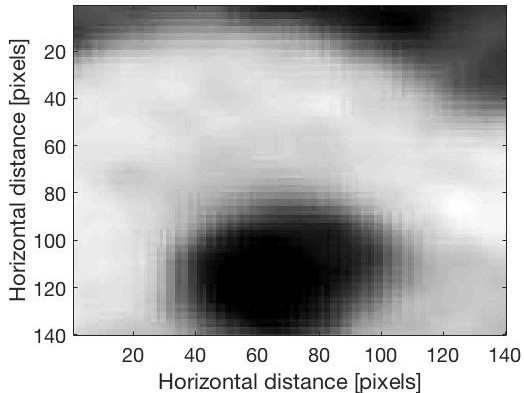}}
  \subfigure[Noisy CBCT image]{\includegraphics[width=0.213\textwidth]{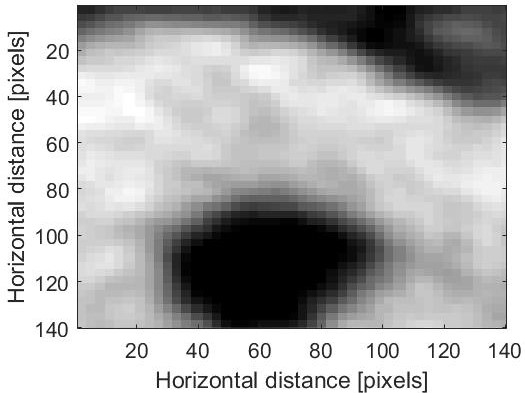}}
  \subfigure[Denoised CBCT image]
{\includegraphics[width=0.213\textwidth]{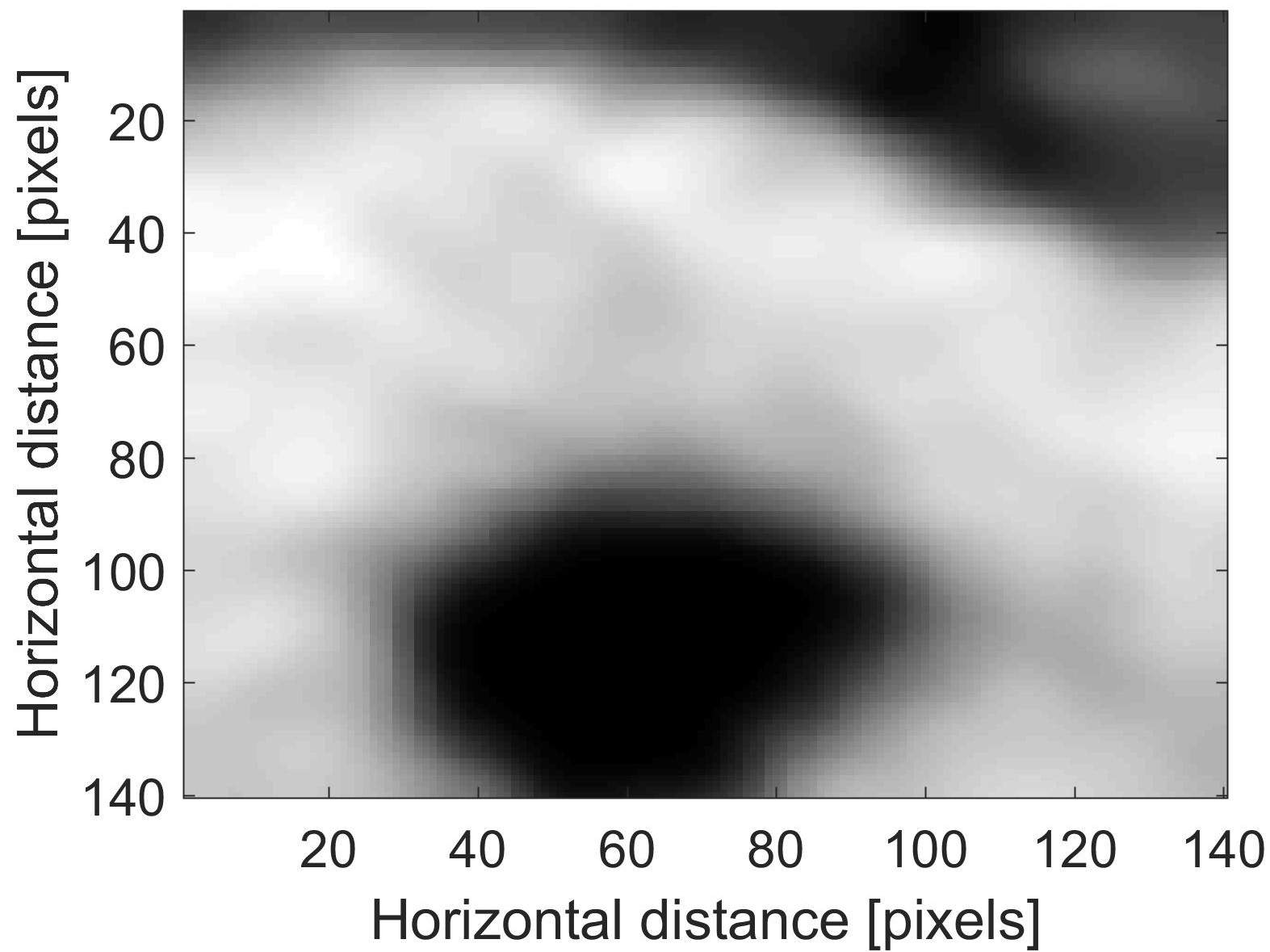}}

  \caption{Result of the denoising algorithm for a biomedical image: (a) Clean CBCT dental image, (b) Noisy CBCT dental image, (c) CBCT dental image after denoising considering the hyperparameters $\hbar^2/2m = 0.5$, $\sigma^2 = 20$, $\rho = 1$ and $s = 3000$.}
\label{fig:xray}

\end{figure*}

Three quantitative measurements were used to evaluate the denoised images: the signal to noise ratio (SNR), the peak signal to noise ratios (PSNR) and the structure similarity (SSIM) \cite{Wang04}. All the quantitative results are regrouped in Table~\ref{tab_gp} where the best and the second best values have been highlighted by red and blue colors respectively for each dataset. Note that VST is only used for data-dependent noise, whereas GSP and NLM is used only for Gaussian noise. Moreover, VST, GSP,  NLM, and DL were only tested for images, as initially suggested by the seminal papers. Illustrative results for Fruits image (Fig.~\ref{fig:2Dsamples} (e)) corrupted by Gaussian noise, Moon image (Fig.~\ref{fig:2Dsamples} (f)) with Poisson noise and Lena image (Fig.~\ref{fig:2Dsamples} (d)) with speckle noise are shown respectively in Fig.~\ref{fig:stillResults}, \ref{fig:moonResults} and \ref{fig:lenaResults}. All these results allow us to draw some conclusions. First, one may remark that in almost all the cases, regardless of the noise nature and the image, the proposed method is one of the two best ones. This proves its adaptability to different scenarios {and general applicability} which can be considered a strong point in number of practical applications. Second, we may remark that for the synthetic signal and image, our method outperforms all the others. The main reason is that the synthetic signal and image were generated to \SD{provide a best case} for the proposed decomposition, that keeps preferentially higher frequencies for low gray levels and lower frequencies for high gray levels. For such images or signals, the proposed method is very efficient. On the contrary, TV and DL, for example, fail in these cases because of the non piece-wise constant nature of the synthetic data. Finally, we remark that the proposed denoising algorithm provides competitive results compared to DL that learns the redundant dictionary from a database of clean images. Of course the proposed method does not need such a database. \SD{In summary, the results show that while our method is clearly the best for some specific types of signals or images for which it is well-adapted, it is also competitive for general types of images, being in almost all cases one of the two best methods. This indicates that the algorithm we propose can be used reliably for denoising applications in a variety of contexts.}

\subsection{Application to CBCT dental image denoising}
\label{sec:medicalimage}

This section illustrates the ability of the proposed method to denoise real medical images. In particular, the application considered in this work for illustration purpose is CBCT dental imaging. CBCT is a medical imaging modality that allows tooth visualization with low radiation doses, and is thus suitable for dental applications. However, the low radiation prevents the current scanners to provide images with high SNR. In \cite{michetti2015cone}, 
the quality of CBCT dental image within phantom and \textit{in vivo} data  were evaluated. Fig.~\ref{fig:xray} shows a noisy image resulting from that study, as well as the denoised images with the proposed approach.
The region of interest in this image is the dark region in the middle of the tooth, that represents the canal root.  The results displayed show that the method has some practical applications in this field. \SD{For a quantitative analysis the contrast-to-noise ratio (CNR) computed between the dark region representing the canal root and the bright region representing the dentine, and the SSIM values comparing the noisy and the denoised image to the clean one, are presented in Table \ref{tab_teeth}. They clearly show the ability of the proposed method to enhance the noisy CBCT image.}

\begin{table}[h!]
\begin{center}
	\caption{\SD{Quantitative results for CBCT image.}}
	\label{tab_teeth}
	\begin{tabular}{c c c }
	\hline
\SD{	Sample} & \SD{CNR (dB)} & \SD{SSIM} \\
	\hline
\SD{Noisy CBCT image} 		& \SD{23.89} & \SD{0.66} \\
\SD{Denoised CBCT image}		& \SD{25.26} & \SD{0.75} \\

	\hline
\end{tabular}\end{center}
\end{table}

\section{Conclusions}
\label{sec:conclusion}

We investigated in this paper an original approach of constructing an adaptive transform in the context of signal and image processing based on the resolution of a quantum mechanical problem. More precisely, the signal or image is used as the potential in a quantum problem, the resolution of which gives as eigenvectors the proposed adaptive basis. The basis vectors automatically use a different range of frequencies to explore low potential valued regions compare to the regions corresponding to the high potential values. Therefore, thresholding the coefficients of the signal or image expanded in this basis will process differently high and low values of the signal or image. This framework has been illustrated through denoising applications on different signals and images in presence of Gaussian, Poisson and speckle noise. We have performed a detailed investigation of the impact of the hyperparameters. We have also presented a quantitative comparison of the denoising efficiency of the proposed adaptive method compared to state-of-the-art methods on synthetic signals and standard images. The results of our investigation show that our method has interesting potential to denoise signals and images, especially for Poisson and speckle noise to which it is well adapted; indeed, as a vector in the adaptive basis naturally uses higher frequencies for low values of the signal compared to low values, the thresholding process keeps more frequencies for low values than for high values. Our results show that our denoising procedure outperforms standard methods in specific cases, and ranks among the best methods in most cases. In general, the method should be optimal for signals or images with large contrasts in presence of Poisson-like noise. Our study of the hyperparameters shows that they cannot be chosen at random, but that the range of optimality is large enough to allow to set them beforehand independently of the signal or image, although the choice may be modified according to the type of noise present in the application.

The computational time of the eigenvectors of the Hamiltonian operator is the major drawback of this method, which can be tackled by more refined algorithms \SD{or by adapting the patch-based processing to the proposed framework, using for example the theory of multiple-particle quantum mechanics.} It should be also noted that in many applications the computational efficiency of the algorithm, while important, is less crucial than the efficiency to denoise the signal or image considered.
\SD{Using more complex quantum mechanical tools, such as the time-dependent Schrodinger equation, \textit{i.e.,} the wave functions and the potential change with time, gives a very fascinating direction for further research.}
As another future perspective of this study, it would be very interesting to extend this framework \SD{to three dimensional data or color images. It could be also extended} to other reconstruction applications available in the literature, such as deconvolution, super-resolution or compressed sensing.

\section{Acknowledgments}
\label{sec:Acknowledge}
We thank Raphael Smith who participated in a preliminary version of this work. We also thank CNRS for funding through the 80 prime program.

\bibliographystyle{IEEEbib}
\bibliography{Quantumdenoising}

\begin{thebibliography}{10}

\bibitem{donoho1994ideal}
David~L Donoho and Jain~M Johnstone,
\newblock ``Ideal spatial adaptation by wavelet shrinkage,''
\newblock {\em biometrika}, vol. 81, no. 3, pp. 425--455, 1994.

\bibitem{donoho1995wavelet}
David~L Donoho, Iain~M Johnstone, G{\'e}rard Kerkyacharian, and Dominique
  Picard,
\newblock ``Wavelet shrinkage: asymptopia?,''
\newblock {\em Journal of the Royal Statistical Society: Series B
  (Methodological)}, vol. 57, no. 2, pp. 301--337, 1995.

\bibitem{Aharon06}
M.~Aharon, M.~Elad, and A.~Bruckstein,
\newblock ``$rm k$-svd: An algorithm for designing overcomplete dictionaries
  for sparse representation,''
\newblock {\em IEEE Transactions on Signal Processing}, vol. 54, no. 11, pp.
  4311--4322, Nov 2006.

\bibitem{Elad06}
M.~Elad and M.~Aharon,
\newblock ``Image denoising via sparse and redundant representations over
  learned dictionaries,''
\newblock {\em IEEE Transactions on Image Processing}, vol. 15, no. 12, pp.
  3736--3745, Dec 2006.

\bibitem{smith2018adaptive}
Raphael Smith, Adrian Basarab, Bertrand Georgeot, and Denis Kouam{\'e},
\newblock ``Adaptive transform via quantum signal processing: application to
  signal and image denoising,''
\newblock in {\em 2018 25th IEEE International Conference on Image Processing
  (ICIP)}. IEEE, 2018, pp. 1523--1527.

\bibitem{eldar2002quantum}
Yonina~C Eldar and Alan~V Oppenheim,
\newblock ``Quantum signal processing,''
\newblock {\em IEEE Signal Processing Magazine}, vol. 19, no. 6, pp. 12--32,
  2002.

\bibitem{Gabbouj2013}
C.~Aytekin, S.~Kiranyaz, and M.~Gabbouj,
\newblock ``Quantum mechanics in computer vision: Automatic object
  extraction,''
\newblock in {\em IEEE International Conference on Image Processing}, Sept
  2013, pp. 2489--2493.

\bibitem{youssry2015quantum}
Akram Youssry, Ahmed El-Rafei, and Salwa Elramly,
\newblock ``A quantum mechanics-based framework for image processing and its
  application to image segmentation,''
\newblock {\em Quantum Information Processing}, vol. 14, no. 10, pp.
  3613--3638, 2015.

\bibitem{Laleg2013}
Taous-Meriem Laleg-Kirati, Emmanuelle Cr{\'e}peau, and Michel Sorine,
\newblock ``Semi-classical signal analysis,''
\newblock {\em Mathematics of Control, Signals, and Systems}, vol. 25, no. 1,
  pp. 37--61, Mar 2013.

\bibitem{LalegKirati16}
Taous-Meriem Laleg-Kirati, Jiayu Zhang, Eric Achten, and Hacene Serrai,
\newblock ``Spectral data de-noising using semi-classical signal analysis:
  application to localized mrs,''
\newblock {\em NMR in Biomedicine}, vol. 29, no. 10, pp. 1477--1485, 2016.

\bibitem{iliyasu2013towards}
Abdullah~M Iliyasu,
\newblock ``Towards realising secure and efficient image and video processing
  applications on quantum computers,''
\newblock {\em Entropy}, vol. 15, no. 8, pp. 2874--2974, 2013.

\bibitem{zhang2013neqr}
Yi~Zhang, Kai Lu, Yinghui Gao, and Mo~Wang,
\newblock ``Neqr: a novel enhanced quantum representation of digital images,''
\newblock {\em Quantum Information Processing}, vol. 12, no. 8, pp. 2833--2860,
  2013.

\bibitem{kaisserli2014image}
Zineb Kaisserli and Taous-Meriem Laleg-Kirati,
\newblock ``Image representation and denoising using squared eigenfunctions of
  schrodinger operator,''
\newblock {\em arXiv preprint arXiv:1409.3720}, 2014.

\bibitem{chahid2018new}
Abderrazak Chahid, Hacene Serrai, Eric Achten, and Taous-Meriem Laleg-Kirati,
\newblock ``A new roi-based performance evaluation method for image denoising
  using the squared eigenfunctions of the schr{\"o}dinger operator,''
\newblock in {\em 2018 40th Annual International Conference of the IEEE
  Engineering in Medicine and Biology Society (EMBC)}. IEEE, 2018, pp.
  5579--5582.

\bibitem{meyer2014perturbation}
Fran{\c{c}}ois~G Meyer and Xilin Shen,
\newblock ``Perturbation of the eigenvectors of the graph laplacian:
  Application to image denoising,''
\newblock {\em Applied and Computational Harmonic Analysis}, vol. 36, no. 2,
  pp. 326--334, 2014.

\bibitem{Ortega2018Graph}
A.~{Ortega}, P.~{Frossard}, J.~{Kovačević}, J.~M.~F. {Moura}, and
  P.~{Vandergheynst},
\newblock ``Graph signal processing: Overview, challenges, and applications,''
\newblock {\em Proceedings of the IEEE}, vol. 106, no. 5, pp. 808--828, 2018.

\bibitem{Cheung2018Graph}
G.~{Cheung}, E.~{Magli}, Y.~{Tanaka}, and M.~K. {Ng},
\newblock ``Graph spectral image processing,''
\newblock {\em Proceedings of the IEEE}, vol. 106, no. 5, pp. 907--930, 2018.

\bibitem{Pang2017Graph}
J.~{Pang} and G.~{Cheung},
\newblock ``Graph laplacian regularization for image denoising: Analysis in the
  continuous domain,''
\newblock {\em IEEE Transactions on Image Processing}, vol. 26, no. 4, pp.
  1770--1785, 2017.

\bibitem{Shekkizhar2020Efficient}
S.~{Shekkizhar} and A.~{Ortega},
\newblock ``Efficient graph construction for image representation,''
\newblock in {\em 2020 IEEE International Conference on Image Processing
  (ICIP)}, 2020, pp. 1956--1960.

\bibitem{feynman}
Richard~Phillips Feynman, Robert~Benjamin Leighton, and Matthew Sands,
\newblock {\em The Feynman lectures on physics; New millennium ed.},
\newblock Basic Books, New York, NY, 2010,
\newblock Originally published 1963-1965.

\bibitem{landau}
L.~D. Landau and L.~M. Lifshitz,
\newblock {\em Quantum Mechanics Non-Relativistic Theory, Third Edition},
\newblock Butterworth-Heinemann, 1981.

\bibitem{cohen}
Claude Cohen-Tannoudji, Bernard Diu, and Franck Laloë,
\newblock {\em Quantum mechanics; 1st ed.},
\newblock Wiley, New York, NY, 1977,
\newblock Trans. of : Mécanique quantique. Paris : Hermann, 1973.

\bibitem{schrodinger1926undulatory}
Erwin Schr{\"o}dinger,
\newblock ``An undulatory theory of the mechanics of atoms and molecules,''
\newblock {\em Physical review}, vol. 28, no. 6, pp. 1049, 1926.

\bibitem{Anderson1958}
P.W. Anderson,
\newblock ``Absence of diffusion in certain random lattices,''
\newblock {\em Physical Review}, vol. 109, pp. 1492--1505, 1958.

\bibitem{makitalo2011closed}
Markku Makitalo and Alessandro Foi,
\newblock ``A closed-form approximation of the exact unbiased inverse of the
  anscombe variance-stabilizing transformation,''
\newblock {\em IEEE transactions on image processing}, vol. 20, no. 9, pp.
  2697--2698, 2011.

\bibitem{Figueiredo10}
M.~A.~T. Figueiredo and J.~M. Bioucas-Dias,
\newblock ``Restoration of poissonian images using alternating direction
  optimization,''
\newblock {\em IEEE Transactions on Image Processing}, vol. 19, no. 12, pp.
  3133--3145, Dec 2010.

\bibitem{rudin1994total}
Leonid~I Rudin and Stanley Osher,
\newblock ``Total variation based image restoration with free local
  constraints,''
\newblock in {\em Proceedings of 1st International Conference on Image
  Processing}. IEEE, 1994, vol.~1, pp. 31--35.

\bibitem{tasdizen2009principal}
T.~{Tasdizen},
\newblock ``Principal neighborhood dictionaries for nonlocal means image
  denoising,''
\newblock {\em IEEE Transactions on Image Processing}, vol. 18, no. 12, pp.
  2649--2660, 2009.

\bibitem{elad2006image}
Michael Elad and Michal Aharon,
\newblock ``Image denoising via sparse and redundant representations over
  learned dictionaries,''
\newblock {\em IEEE Transactions on Image processing}, vol. 15, no. 12, pp.
  3736--3745, 2006.

\bibitem{NumericalTours}
Gabriel Peyré,
\newblock ``The numerical tours of signal processing,''
\newblock {\em Computing in Science \& Engineering}, vol. 13, no. 4, pp.
  94--97, 2011.

\bibitem{Wang04}
Zhou Wang, A.~C. Bovik, H.~R. Sheikh, and E.~P. Simoncelli,
\newblock ``Image quality assessment: from error visibility to structural
  similarity,''
\newblock {\em IEEE Transactions on Image Processing}, vol. 13, no. 4, pp.
  600--612, April 2004.

\bibitem{michetti2015cone}
J{\'e}r{\^o}me Michetti, Adrian Basarab, Michel Tran, Franck Diemer, and Denis
  Kouam{\'e},
\newblock ``Cone-beam computed tomography contrast validation of an artificial
  periodontal phantom for use in endodontics,''
\newblock in {\em 2015 37th Annual International Conference of the IEEE
  Engineering in Medicine and Biology Society (EMBC)}. IEEE, 2015, pp.
  7905--7908.

\end{thebibliography}

\begin{IEEEbiography}[
{\includegraphics[width=1in,height=1.25in,clip,keepaspectratio]{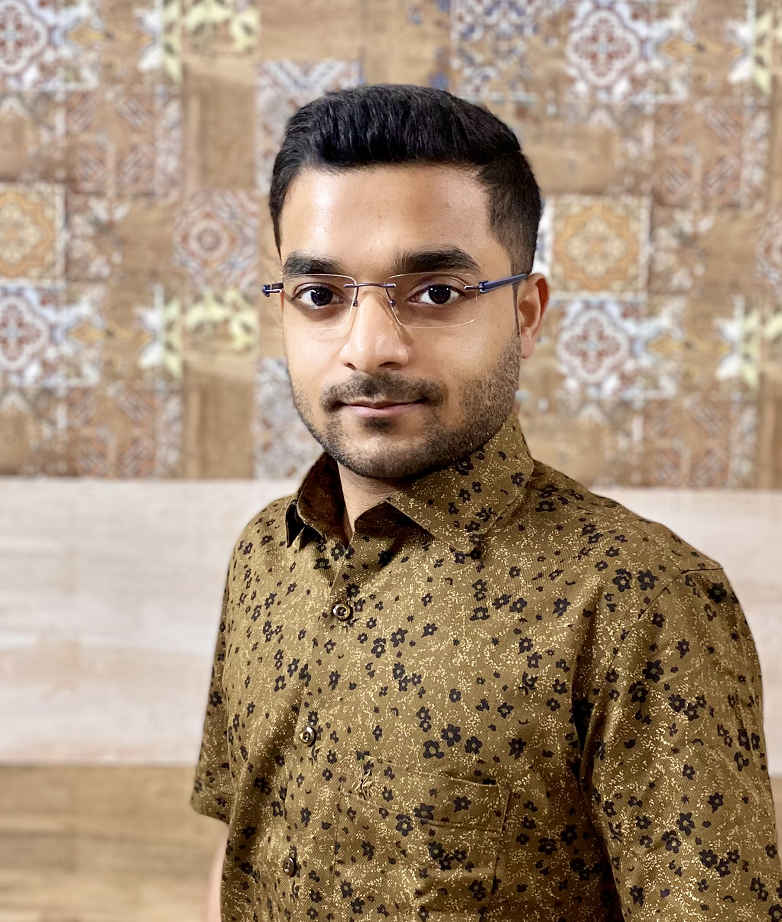}}]
{Sayantan Dutta} received the B.Sc. degree in Mathematics from the University of Burdwan, India, in 2016, followed by M.Sc. degree in Applied Mathematics from the Visva-Bharati University, India, in 2018, and M.S. in Fundamental Physics from the University of Tours, France, in 2019. He is currently pursuing the Ph.D. degree at the University Paul Sabatier Toulouse 3, France, working at the Institut de Recherche en Informatique de Toulouse (IRIT) and Laboratoire de Physique Th\'eorique de Toulouse (LPT) laboratories. His research interests include quantum image processing and inverse problems, particularly denoising, deblurring, and compressed sensing.
\end{IEEEbiography}

\begin{IEEEbiography}[{\includegraphics[width=1in,height=1.25in,clip,keepaspectratio]{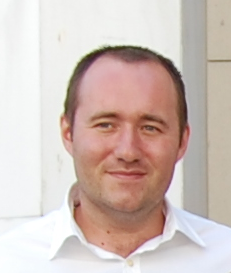}}]
{Adrian Basarab} (Senior Member, IEEE) received the M.S. and Ph.D. degrees in signal and image processing from the National Institute for Applied Sciences, Lyon, France, in 2005 and 2008, respectively. Since 2009 (respectively 2016), he has been an Assistant (respectively Associate) Professor with the University Paul Sabatier Toulouse 3 and a member of IRIT Laboratory (UMR CNRS 5505). His research interests include computational medical imaging and more particularly inverse problems (deconvolution, super-resolution, compressive sampling, beamforming, image registration and fusion) applied to ultrasound image formation, ultrasound elastography, cardiac ultrasound, quantitative acoustic microscopy, computed tomography and magnetic resonance imaging. He is currently an Associate Editor for Digital Signal Processing and was a member of the French National Council of Universities Section 61 – Computer sciences, Automatic Control and Signal Processing from 2010 to 2015. In 2017, he was a Guest Co-Editor for the IEEE TUFFC special issue on “Sparsity driven methods in medical ultrasound.” Since 2018, he has been the head of “Computational Imaging and Vision” Group, IRIT Laboratory. Since 2019, he has been a member of the EURASIP Technical Area Committee Biomedical Image \& Signal Analytics. Since 2020, he has been a member of the IEEE Ultrasonics Symposium TPC.
\end{IEEEbiography}

\begin{IEEEbiography}[{\includegraphics[width=1in,height=1.25in,clip,keepaspectratio]{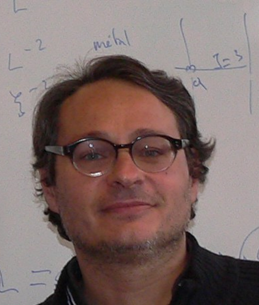}}]
{Bertrand Georgeot} is Directeur de Recherche (group leader) at CNRS in Laboratoire de Physique Th\'eorique, Universit\'e Paul Sabatier, Toulouse. After studies at the Ecole Polytechnique, he obtained his PhD in Orsay in 1993 then was postdoctoral associate at the University of Maryland (USA) and Nils Bohr Institute (Denmark). He is CNRS researcher in Toulouse since 1996. Since 2016 he is head of the laboratoire de physique th\'eorique.

He has authored or coauthored more than 90 publications in peer-reviewed journals, mostly in quantum physics (quantum chaos, Anderson localization, cold atom physics, multifractal quantum states) but also in interdisciplinary research in dynamical systems, astrophysics, classical and quantum computer sciences, network theory.
\end{IEEEbiography}

\begin{IEEEbiography}[{\includegraphics[width=1in,height=1.25in,clip,keepaspectratio]{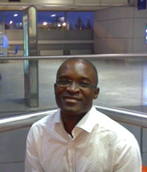}}]
{Denis Kouam\'e} is a professor in medical imaging and signal processing at Paul Sabatier University of Toulouse, France, since 2008. From 1998 to 2008 he was assistant then associate professor at University of Tours. From 1996 to 1998 he was senior engineer at GIP Tours, France. He received the M.Sc, the Ph.D. and the habilitation to supervise research works(HDR) in signal processing and medical ultrasound imaging from the University of Tours in 1993, 1996 and 2004 respectively.

He was head of signal and image processing group, and then head of Ultrasound imaging group at Ultrasound and Signal Lab at University of Tours respectively from 2000 to 2006 and from 2006 to 2008. From 2009-2015 He was head of Health and Information Technology (HIT) strategic field at the Institut de Recherche en Informatique IRIT Laboratory, Toulouse. He currently leads the Signals and Image department at IRIT. His research interests cover the following areas: Medical imaging, ultrasound imaging, high resolution imaging, Doppler signal processing, Multidimensional biomedical Signal and image analysis including parametric modeling, spectral analysis and application to flow estimation, Sparse representation, Inverse problems.

He was nvited for talks or in charge of different invited special sessions or tutorials at several IEEE conferences : ICASSP, ISBI, ISSPIT, ICIP. He has served on several international conferences technical programme committees in signal, image processing or medical imaging, and also chaired various sessions at different international conferences. He was/is/ invited for talks in different universities inside and outside France. He was/is involved, as principal investigator or as member, in different European or French research projets (ANR,FUI, INSERM,…).

He is an Associate Editor for the IEEE Transactions on Ultrasonics Ferroelectrics and Frequency Control and for for the IEEE Transactions on Image Processing.
\end{IEEEbiography}

\end{document}